\documentclass[seceq]{ptptex}

\usepackage{amsmath}
\usepackage{amsfonts}
\usepackage{latexsym}
\usepackage{amssymb} 
\usepackage{verbatim}
\usepackage{amsthm}
\usepackage{graphicx}
\usepackage{wrapft}

\title{
Consistency of Equations in the Second-order Gauge-invariant
Cosmological Perturbation Theory
}

\author{
  Kouji \textsc{Nakamura}%
}

\inst{
  Department of Astronomical Science,
  the Graduate University for Advanced Studies,
  Mitaka, Tokyo 181-8588, Japan.
}

\abst{
  Along the general framework of the gauge-invariant
  perturbation theory developed in the papers [K.~Nakamura,
  Prog.~Theor.~Phys.\ {\bf 110} (2003), 723; {\it ibid},
  {\bf 113} (2005), 481.], we re-derive the second-order Einstein
  equations on four-dimensional homogeneous isotropic
  background universe in gauge-invariant manner without ignoring
  any mode of perturbations.
  We consider the perturbations both in the universe dominated
  by the single perfect fluid and in that dominated by the
  single scalar field.
  We also confirmed the consistency of all equations of the
  second-order Einstein equation and the equations of motion for
  matter fields which are derived in the paper [K.~Nakamura,
  arXiv:0804.3840 [gr-qc]].
  This confirmation implies that the all derived equations of the
  second order are self-consistent and these equations are
  correct in this sense.
}

\begin{document}

\maketitle

\section{Introduction}
\label{sec:intro}


The general relativistic second-order cosmological perturbation
theory is one of topical subjects in the recent cosmology. 
By the recent observation\cite{WMAP}, the first order
approximation of the fluctuations of our universe from a
homogeneous isotropic one was revealed, the cosmological
parameters are accurately measured, we have obtained the
standard cosmological model, and so-called ``the precision
cosmology'' has begun. 
The observational results also suggest that the fluctuations of
our universe are adiabatic and Gaussian at least in the first
order approximation.
We are now on the stage to discuss the deviation from this first
order approximation from the
observational\cite{Non-Gaussianity-observation-WMAP} and the 
theoretical
side\cite{Non-Gaussianity-inflation,Non-Gaussianity-in-CMB}
through the non-Gaussianity, the non-adiabaticity, and so on.
To carry out this, some analyses beyond linear order are
required.
The second-order cosmological perturbation theory is one of
such perturbation theories beyond linear order.


Although the second-order perturbation theory in general
relativity is old topics, a general framework of the
gauge-invariant formulation of the second-order general
relativistic perturbation are recently proposed by the present
author\cite{kouchan-gauge-inv,kouchan-second}.
We refer these works as KN2003\cite{kouchan-gauge-inv} and
KN2005\cite{kouchan-second} in this paper. 
Further, this general framework was also applied to cosmological
perturbations.
We demonstrated the derivation of the second-order perturbation
of the Einstein equation in gauge-invariant
manner without any gauge fixing\cite{kouchan-second-cosmo}.
We also showed that the above general framework is also
applicable when we discuss the second-order perturbations of the
equation of motion for matter
fields\cite{kouchan-second-cosmo-matter}. 
We also refer these works as KN2007\cite{kouchan-second-cosmo}
and KN2008\cite{kouchan-second-cosmo-matter}.
This gauge-invariant formulation of second-order
cosmological perturbations is a natural extension of the
first-order gauge-invariant cosmological perturbation
theory\cite{Bardeen-1980,Kodama-Sasaki-1984,Mukhanov-Feldman-Brandenberger-1992}.


In this paper, we re-derive all components of the second-order
perturbations of the Einstein equations without ignoring any
modes of perturbations.
In the first-order cosmological perturbation theory, the
perturbations are classified into three types which are called
the scalar-, the vector-, and the tensor-mode, respectively. 
In KN2007, we ignore the vector- and tensor-modes of the first
order when we derive the second-order perturbations of the
Einstein equation.
These modes are taken into account in this paper and we show the
precise mode-coupling of the these three types of the
first-order perturbations in the case of the universe filled
with a perfect fluid and with a scalar field, respectively.
In particular, in the case of a perfect fluid, we see that the 
any types of mode-coupling appear in the second-order
perturbations of the Einstein equations, in principle.


We also confirm the consistency of all equations of the
second-order Einstein equation and the equations of motion for
matter fields which are derived in
KN2008\cite{kouchan-second-cosmo-matter}.  
Further, due to the fact that the Einstein equations are the
first class constrained system, we have initial value
constraints in the Einstein equations.
Moreover, since the Einstein equations include the equation of
motion for matter fields, the second-order perturbations of the
equations of motion for matter fields are not independent
equations of the second-order perturbation of the Einstein
equations.
Through these facts, we can check whether the derived equations
of the second order are consistent or not.
In this paper, we do check this consistency.
Namely, we show that the second-order perturbations of the
equations of motion for the matter field are consistent with
the second-order perturbations of the Einstein equations through
the background and the first-order perturbations of the Einstein
equations.
This confirmation implies that the all derived equations of the
second order are self-consistent and these equations are
correct in this sense.


The explicit derivations of the second-order perturbations of
the Einstein equations without ignoring any mode are not only for
the cosmological perturbations but also for the post-Minkowski
expansion for a binary system\cite{Bel-Damour-Deruelle-Ibanez-Martine-1981}.
At least in the cosmological perturbations with the single
matter field, it is well-known that the vector-mode of the
first-order perturbation is a just decaying mode.
Further, the tensor-mode whose wavelength is shorter than the
Hubble horizon size also decays due to the expansion of the
universe, though the tensor-mode whose wavelength is longer than the
horizon size is frozen.
Therefore, in many situations in cosmology, we may neglect these
modes, safely.
However, in this paper, we dare to include these modes in our
considerations in spite that the expressions of the second-order
perturbations of the Einstein equations become very complicated.
The explicit expressions of the second-order perturbations of the
Einstein equations are reduced to those for Minkowski background
spacetime if we neglect the expansion of universe.
Thus, formulae to derive the explicit Einstein equations of the
second order will be also useful to reconsider the post-Minkowski
expansion for a binary system in gauge-invariant manner.
Of course, we have to reconsider the regularization procedure to
treat the self-gravity of the point particles within a
gauge-invariant manner to complete discussions of the
post-Minkowski expansion for a binary system.


The organization of this paper is as follows.
In \S\ref{sec:Second-order-cosmological-perturbatios}, 
we briefly review the definitions of the gauge-invariant
variables for the second-order perturbation which was defined by
KN2007\cite{kouchan-second-cosmo}.
We also summarize the components of the gauge-invariant part of
the first- and the second-order perturbations of the
Einstein tensor in the cosmological perturbations in this
section.
We did not ignore any modes in these formulae.
Further, we briefly review the background Einstein equations and
the equations of motion for the matter field in
\S\ref{sec:Background-equations} and its first-order
perturbations in
\S\ref{sec:Consistency-of-the-first-order-perturbations}.
The ingredients of these sections are used in
\S\ref{sec:Consistency-of-the-second-order-perturbations}.
In \S\ref{sec:Consistency-of-the-second-order-perturbations}, 
we show the explicit expression of the second-order perturbation
of the Einstein equations without ignoring any modes of
perturbations and check the consistency with the second-order
perturbations of the equations of motion for matter field.
The final section, \S\ref{sec:summary}, is devoted to the
summary and discussions.


We employ the notation of our series of papers
KN2003\cite{kouchan-gauge-inv}, KN2005\cite{kouchan-second},
KN2007\cite{kouchan-second-cosmo}, and
KN2008\cite{kouchan-second-cosmo-matter} and use the abstract
index notation\cite{Wald-book}.
We also employ the natural unit in which the light velocity is
denoted by $c=1$ and denote the Newton's gravitational constant
by $G$.


\section{Perturbations of Einstein tensor in terms of gauge-invariant variables}
\label{sec:Second-order-cosmological-perturbatios}


\subsection{Gauge invariant variables}
\label{sec:Second-order-cosmological-perturbatios-Definitions}


In any perturbation theory, we always treat two spacetime
manifolds.
One is the physical spacetime ${\cal M}={\cal M}_{\lambda}$ and
the other is the background spacetime ${\cal M}_{0}$.
Since these two spacetime manifolds are distinct from each
other, we have to introduce a point-identification map 
${\cal X}_{\lambda}: {\cal M}_{0} \rightarrow {\cal M}_{\lambda}$.
This point-identification map ${\cal X}$ called a gauge choice
in perturbation theories.
Through the pull-back ${\cal X}^{*}_{\lambda}$ of the gauge choice
${\cal X}_{\lambda}$, any physical variable $\hat{Q}_{\lambda}$
on the physical manifold ${\cal M}_{\lambda}$ is pulled back to 
a representation ${\cal X}^{*}_{\lambda}\hat{Q}_{\lambda}$ on
the background spacetime ${\cal M}_{0}$.
It is important to note that the gauge choice 
${\cal X}_{\lambda}$ is not unique by virtue of general
covariance in general relativity.
When we have two different gauge choices ${\cal Y}_{\lambda}$ and
${\cal X}_{\lambda}$, we can consider the gauge transformation
rule from a gauge choice ${\cal X}_{\lambda}$ to another one
${\cal Y}_{\lambda}$ through the diffeomorphism 
$\Phi_{\lambda}:=({\cal X}_{\lambda})^{-1}\circ {\cal Y}_{\lambda}$.
The pull-back $\Phi_{\lambda}^{*}$ of the diffeomorphism
$\Phi_{\lambda}$ does change the representation 
${\cal X}^{*}_{\lambda}\hat{Q}_{\lambda}$ of the physical
variable $\hat{Q}_{\lambda}$ to another representation as 
${\cal Y}^{*}_{\lambda}\hat{Q}_{\lambda}=\Phi^{*}_{\lambda}{\cal
  X}_{\lambda}^{*}\hat{Q}_{\lambda}$.


The pull-back ${\cal X}^{*}_{\lambda}\hat{Q}_{\lambda}$ is
expanded as 
\begin{equation}
  {\cal X}^{*}_{\lambda}\hat{Q}_{\lambda}
  =
  Q_{0}
  + \lambda {}^{(1)}_{{\cal X}}\!Q
  + \frac{1}{2} \lambda^{2} {}^{(2)}\!_{{\cal X}}Q
  + O(\lambda^{3})
  .
  \label{eq:perturbative-expansion-def}
\end{equation}
The first- and the second-order perturbations 
${}^{(1)}_{{\cal X}}\!Q$ and ${}^{(2)}\!_{{\cal X}}Q$ are
defined by this equation (\ref{eq:perturbative-expansion-def}).
For example, we expand the pulled-back 
${\cal X}^{*}_{\lambda}\bar{g}_{ab}$ of the metric
$\bar{g}_{ab}$ on the physical spacetime ${\cal M}_{\lambda}$ by
the gauge choice ${\cal X}_{\lambda}$: 
\begin{eqnarray}
  {\cal X}^{*}_{\lambda}\bar{g}_{ab}
  &=&
  g_{ab} + \lambda {}_{{\cal X}}\!h_{ab} 
  + \frac{\lambda^{2}}{2} {}_{{\cal X}}\!l_{ab}
  + O(\lambda^{3}),
  \label{eq:metric-expansion}
\end{eqnarray}
where $g_{ab}$ is the metric on the background spacetime 
${\cal M}_{0}$.
In the case of the cosmological perturbations,
we consider the homogeneous isotropic spacetime whose metric is
given by  
\begin{eqnarray}
  \label{eq:background-metric}
  g_{ab} = a^{2}\left(
    - (d\eta)_{a}(d\eta)_{b}
    + \gamma_{ij} (dx^{i})_{a} (dx^{j})_{b}
  \right),
\end{eqnarray}
where $\gamma_{ab} := \gamma_{ij} (dx^{i})_{a} (dx^{j})_{b}$ is
the metric on the maximally symmetric three space and the
indices $i,j,k,...$ for the spatial components run from 1 to 3.
Henceforth, we do not explicitly express the index of the gauge
choice ${\cal X}_{\lambda}$ in expressions if there is no 
possibility of confusion.


From the generic form of the Taylor expansion of the pull-back 
$\Phi_{\lambda}^{*}:={\cal Y}_{\lambda}^{*}\circ({\cal X}_{\lambda}^{*})^{-1}$,
we can easily derive the gauge transformation rule of each
order: 
\begin{eqnarray}
  \label{eq:gauge-trans-first-order-generic}
  {}^{(1)}_{\;\cal Y}\!Q - {}^{(1)}_{\;\cal X}\!Q &=& 
  {\pounds}_{\xi_{(1)}} Q_{0} 
  ,
  \\
  \label{eq:gauge-trans-second-order-generic}
  {}^{(2)}_{\;\cal Y}\!Q - {}^{(2)}_{\;\cal X}\!Q &=& 
  2 {\pounds}_{\xi_{(1)}} {}^{(1)}_{\;\cal X}\!Q 
  +\left\{{\pounds}_{\xi_{(2)}} +
  {\pounds}_{\xi_{(1)}}^{2}\right\} Q_{0},
\end{eqnarray}
where $\xi_{1}^{a}$ and $\xi_{2}^{a}$ are generators of the
diffeomorphism $\Phi^{*}$.
Inspecting these gauge transformation rules
(\ref{eq:gauge-trans-first-order-generic}) and
(\ref{eq:gauge-trans-second-order-generic}), we consider the
notion of the order by order gauge invariance.
We call the $p$th-order perturbation ${}^{(p)}_{\;\cal X}\!Q$ is
gauge invariant iff
\begin{equation}
  {}^{(p)}_{\;\cal Y}\!Q = {}^{(p)}_{\;\cal X}\!Q
\end{equation}
for any gauge choice ${\cal X}_{\lambda}$ and
${\cal Y}_{\lambda}$.
Employing this idea of order by order gauge invariance for
each-order perturbations, we proposed a procedure to construct
gauge invariant variables of higher-order perturbations in
KN2003\cite{kouchan-gauge-inv}.


In the cosmological perturbation case, we can show that the
first-order metric perturbation $h_{ab}$ is decomposed as 
\begin{eqnarray}
  h_{ab} =: {\cal H}_{ab} + {\pounds}_{X}g_{ab},
  \label{eq:linear-metric-decomp}
\end{eqnarray}
where ${\cal H}_{ab}$ and $X^{a}$ are the gauge-invariant and
gauge-variant parts of the linear-order metric
perturbations\cite{kouchan-gauge-inv}, i.e., under the gauge
transformation (\ref{eq:gauge-trans-first-order-generic}), these
are transformed as 
\begin{equation}
  {}_{{\cal Y}}\!{\cal H}_{ab} - {}_{{\cal X}}\!{\cal H}_{ab} =  0, 
  \quad
  {}_{\quad{\cal Y}}\!X^{a} - {}_{{\cal X}}\!X^{a} = \xi^{a}_{(1)}. 
  \label{eq:linear-metric-decomp-gauge-trans}
\end{equation}
As shown in KN2007, the decomposition
(\ref{eq:linear-metric-decomp}) is accomplished if we assume the
existence of the Green functions $\Delta^{-1}:=(D^{i}D_{i})^{-1}$, 
$(\Delta + 2 K)^{-1}$, and $(\Delta + 3 K)^{-1}$, where $D_{i}$
is the covariant derivative associated with the metric
$\gamma_{ij}$ on the maximally symmetric three space and $K$ is
the curvature constant of this maximally symmetric three space.
Further, we may choose the components of the gauge-invariant
part ${\cal H}_{ab}$ of the first-order metric perturbation as
\begin{eqnarray}
  \label{eq:components-calHab}
  {\cal H}_{ab}
  &=& 
  a^{2} \left\{
    - 2 \stackrel{(1)}{\Phi} (d\eta)_{a}(d\eta)_{b}
    + 2 \stackrel{(1)}{\nu}_{i} (d\eta)_{(a}(dx^{i})_{b)}
  \right.
  \nonumber\\
  && \quad\quad\quad
  \left.
    + \left( - 2 \stackrel{(1)}{\Psi} \gamma_{ij} 
      + \stackrel{(1)}{\chi}_{ij} \right)
    (dx^{i})_{a}(dx^{j})_{b}
  \right\},
\end{eqnarray}
where $\stackrel{(1)}{\nu}_{i}$ and $\stackrel{(1)}{\chi_{ij}}$
satisfy the properties
\begin{eqnarray}
  D^{i}\stackrel{(1)}{\nu}_{i} =
  \gamma^{ij}D_{i}\stackrel{(1)}{\nu}_{j} = 0, \quad
  \stackrel{(1)}{\chi^{i}_{\;\;i}} = 0, \quad
   D^{i}\stackrel{(1)}{\chi}_{ij} = 0,
\end{eqnarray}
where $\gamma^{kj}$ is the inverse of the metric $\gamma_{ij}$.


If the decomposition (\ref{eq:linear-metric-decomp}) is true, we
can easily show that the second-order metric perturbation
$l_{ab}$ is also decomposed as
\begin{eqnarray}
  \label{eq:second-metric-decomp}
  l_{ab}
  =:
  {\cal L}_{ab} + 2 {\pounds}_{X} h_{ab}
  + \left(
      {\pounds}_{Y}
    - {\pounds}_{X}^{2}
  \right)
  g_{ab},
\end{eqnarray}
where ${\cal L}_{ab}$ and $Y^{a}$ are the gauge-invariant and
gauge-variant parts of the second-order metric perturbations,
i.e.,
\begin{eqnarray}
  {}_{{\cal Y}}\!{\cal L}_{ab} - {}_{{\cal X}}\!{\cal L}_{ab} = 0,
  \quad
  {}_{{\cal Y}}\!Y^{a} - {}_{{\cal X}}\!Y^{a}
  = \xi_{(2)}^{a} + [\xi_{(1)},X]^{a}.
  \label{eq:second-metric-decomp-gauge-trans}
\end{eqnarray}
Further, in the cosmological perturbation case, we may also
choose the components of the gauge-invariant part 
${\cal L}_{ab}$ of the second-order metric perturbation as
\begin{eqnarray}
  \label{eq:components-calLab}
  {\cal L}_{ab}
  &=& 
  a^{2} \left\{
    - 2 \stackrel{(2)}{\Phi} (d\eta)_{a}(d\eta)_{b}
    + 2 \stackrel{(2)}{\nu}_{i} (d\eta)_{(a}(dx^{i})_{b)}
  \right.
  \nonumber\\
  && \quad\quad\quad
  \left.
    + \left( - 2 \stackrel{(2)}{\Psi} \gamma_{ij} 
      + \stackrel{(2)}{\chi}_{ij} \right)
    (dx^{i})_{a}(dx^{j})_{b}
  \right\},
\end{eqnarray}
where $\stackrel{(2)}{\nu}_{i}$ and $\stackrel{(2)}{\chi_{ij}}$ satisfy
the properties
\begin{eqnarray}
  D^{i}\stackrel{(2)}{\nu}_{i} =
  \gamma^{ij}D_{i}\stackrel{(2)}{\nu}_{j} = 0, \quad
  \stackrel{(2)}{\chi^{i}_{\;\;i}} = 0, \quad
   D^{i}\stackrel{(2)}{\chi}_{ij} = 0.
\end{eqnarray}


\subsection{Components of the Perturbative Einstein tensor}
\label{sec:Perturbations-of-the-Einstein-tensor}


As shown in KN2003, through the above first- and the
second-order gauge-variant parts, $X^{a}$ and $Y^{a}$, of the
metric perturbations, we can define the gauge-invariant
variables for an arbitrary field $Q$ other than the metric. 
The definitions in KN2003 imply that the first- and the
second-order perturbation ${}^{(1)}\!Q$ and ${}^{(2)}\!Q$ are
always decomposed into gauge-invariant part and gauge-variant
part as 
\begin{eqnarray}
  \label{eq:matter-gauge-inv-decomp-1.0}
  {}^{(1)}\!Q &=:& {}^{(1)}\!{\cal Q} + {\pounds}_{X}Q_{0}
  , \\ 
  \label{eq:matter-gauge-inv-decomp-2.0}
  {}^{(2)}\!Q  &=:& {}^{(2)}\!{\cal Q} + 2 {\pounds}_{X} {}^{(1)}Q 
  + \left\{ {\pounds}_{Y} - {\pounds}_{X}^{2} \right\} Q_{0}
  ,
\end{eqnarray}
respectively.
Here, ${}^{(1)}\!{\cal Q}$ and ${}^{(2)}\!{\cal Q}$ are
gauge-invariant parts of the first- and the second-order
perturbations of ${}^{(1)}\!Q$ and ${}^{(2)}\!Q$ in the sense of
the above ``order by order gauge invariance'', respectively.


To evaluate the perturbations of the Einstein equation, we
expand the Einstein tensor $\bar{G}_{a}^{\;\;b}$ on the physical
spacetime ${\cal M}$ so that
\begin{equation}
  {\cal X}^{*}\bar{G}_{a}^{\;\;b} = G_{a}^{\;\;b} 
  + \lambda {}^{(1)}\!\bar{G}_{a}^{\;\;b}
  + \frac{1}{2} \lambda^{2} {}^{(2)}\!\bar{G}_{a}^{\;\;b}
  + O(\lambda^{3}).
\end{equation}
As shown in KN2005\cite{kouchan-second}, the first- and 
second-order perturbations of the Einstein tensor are given in
the same form as (\ref{eq:matter-gauge-inv-decomp-1.0}) and
(\ref{eq:matter-gauge-inv-decomp-2.0}):
\begin{eqnarray}
  \label{eq:first-Einstein-generic}
  {}^{(1)}\!\bar{G}_{a}^{\;\;b}
  &=& 
  {}^{(1)}{\cal G}_{a}^{\;\;b}\left[{\cal H}\right]
  + {\pounds}_{X} G_{a}^{\;\;b}
  , \\
  \label{eq:second-Einstein-generic}
  {}^{(2)}\!\bar{G}_{a}^{\;\;b}
  &=& 
  {}^{(1)}{\cal G}_{a}^{\;\;b}\left[{\cal L}\right]
  + {}^{(2)}{\cal G}_{a}^{\;\;b} \left[{\cal H}, {\cal H}\right]
  + 2 {\pounds}_{X} {}^{(1)}\!\bar{G}_{a}^{\;\;b}
  + \left\{
    {\pounds}_{Y} - {\pounds}_{X}^{2}
  \right\} G_{a}^{\;\;b}
  ,
\end{eqnarray}
where
\begin{eqnarray}
  {}^{(1)}{\cal G}_{a}^{\;\;b}\left[A\right]
  &:=&
  - 2 \nabla_{[a}^{}H_{d]}^{\;\;\;bd}\left[A\right]
  - A^{cb} R_{ac}
  \nonumber\\
  && \quad
  + \frac{1}{2} \delta_{a}^{\;\;b}
  \left(
    2 \nabla_{[e}^{}H_{d]}^{\;\;\;ed}[A]
    + R_{ed} A^{ed}
  \right)
  \label{eq:cal-G-def-linear}
  , \\
  {}^{(2)}\!{\cal G}_{a}^{\;\;b}[A,A]
  &:=& \Sigma_{a}^{\;\;b}[A]
  - \frac{1}{2} \delta_{a}^{\;\;b} \Sigma_{c}^{\;\;c}[A],
  \label{eq:calG-Sigam-relation}
  \\
  \Sigma_{a}^{\;\;b}[A]
  &:=& 
    2 R_{ad} A_{c}^{\;\;b} A^{dc}
  + 4 H_{[a}^{\;\;\;de}[A] H_{d]\;\;e}^{\;\;\;b}[A]
  \nonumber\\
  && \quad
  + 4 A_{e}^{\;\;d} \nabla_{[a}H_{d]}^{\;\;\;be}[A]
  + 4 A_{c}^{\;\;b} \nabla_{[a}H_{d]}^{\;\;\;cd}[A]
  \label{eq:Sigma-ab-def}
\end{eqnarray}
and 
\begin{eqnarray}
  H_{abc}[A] &:=&
  \nabla_{(a}A_{b)c} - \frac{1}{2} \nabla_{c}A_{ab} 
  ,
  \nonumber\\
  H_{a}^{\;\;bc}[A] &:=& g^{be}g^{cd} H_{aed}[A]
  , \quad
  H_{a\;\;c}^{\;\;b}[A] := g^{be} H_{aec}[A]
  \label{eq:Habc-def}
  ,
\end{eqnarray}
for arbitrary tensor $A_{ab}$ of the second rank. 
The terms ${}^{(1)}{\cal G}_{a}^{\;\;b}\left[*\right]$ in
Eqs.~(\ref{eq:first-Einstein-generic}) and
(\ref{eq:second-Einstein-generic}) are the gauge-invariant parts
of the perturbative Einstein tensors, which consists of the
linear combinations of the gauge-invariant variables for the
metric perturbations of the first- (${\cal H}_{ab}$) or the
second-order (${\cal L}_{ab}$).
The term ${}^{(2)}{\cal G}_{a}^{\;\;b}[{\cal H},{\cal H}]$ in
the second-order perturbation (\ref{eq:second-Einstein-generic})
of the Einstein tensor consists of the quadratic terms of the
gauge-invariant part of the first-order metric perturbation.


\subsubsection{Components of ${}^{(1)}{\cal G}_{b}^{\;\;a}$}
\label{sec:Components-of-1-cal-G}


As shown in KN2007, the components of 
${}^{(1)}{\cal G}_{a}^{\;\;b}
= {}^{(1)}{\cal G}_{a}^{\;\;b}\left[{\cal H}\right]$ in  
Eq.~(\ref{eq:first-Einstein-generic}) are given in terms of the
gauge-invariant variables defined in
Eq.~(\ref{eq:components-calHab}) as follows: 
\begin{eqnarray}
  {}^{(1)}\!{\cal G}_{\eta}^{\;\;\eta}
  &=&
  - \frac{1}{a^{2}} \left\{
    \left(
      - 6 {\cal H} \partial_{\eta}
      + 2 \Delta
      + 6 K
    \right) \stackrel{(1)}{\Psi}
    - 6 {\cal H}^{2} \stackrel{(1)}{\Phi}
  \right\}
  \label{eq:kouchan-15.32-linear}
  , \\
  {}^{(1)}\!{\cal G}_{i}^{\;\;\eta}
  &=&
  - \frac{1}{a^{2}}
  \left\{
    2 \partial_{\eta} D_{i} \stackrel{(1)}{\Psi}
    + 2 {\cal H} D_{i} \stackrel{(1)}{\Phi}
    - \frac{1}{2} \left(
      \Delta
      + 2 K
    \right)
    \stackrel{(1)}{\nu_{i}}
  \right\}
  \label{eq:kouchan-15.33-linear}
  , \\
  {}^{(1)}\!{\cal G}_{\eta}^{\;\;i}
  &=&
  \frac{1}{a^{2}} \left\{
    2 \partial_{\eta} D^{i} \stackrel{(1)}{\Psi}
    + 2 {\cal H} D^{i} \stackrel{(1)}{\Phi}
    + \frac{1}{2} \left(
      - \Delta
      + 2 K
      + 4 {\cal H}^{2}
      - 4 \partial_{\eta}{\cal H}
    \right)
    \stackrel{(1)}{\nu^{i}}
  \right\}
  \label{eq:kouchan-15.34-linear}
  , \\
  {}^{(1)}\!{\cal G}_{i}^{\;\;j}
  &=& 
  \frac{1}{a^{2}} \left[
    - D_{i} D^{j} \stackrel{(1)}{\Phi}
    + D_{i} D^{j} \stackrel{(1)}{\Psi}
    + 
    \left\{
      \left(
        -   \Delta
        + 2 \partial_{\eta}^{2} 
        + 4 {\cal H} \partial_{\eta}
        - 2 K
      \right)
      \stackrel{(1)}{\Psi}
    \right.
  \right.
  \nonumber\\
  && \quad\quad\quad\quad\quad\quad\quad\quad\quad\quad\quad\quad\quad
  \left.
    \left.
      + \left(
          2 {\cal H} \partial_{\eta}
        + 4 \partial_{\eta}{\cal H}
        + 2 {\cal H}^{2}
        + \Delta
      \right)
      \stackrel{(1)}{\Phi}
    \right\}
    \gamma_{i}^{\;\;j}
  \right.
  \nonumber\\
  && \quad\quad
  \left.
    - \frac{1}{2} \partial_{\eta} \left( 
      D_{i} \stackrel{(1)}{\nu^{j}} + D^{j} \stackrel{(1)}{\nu_{i}}
    \right)
    - {\cal H} \left( 
      D_{i} \stackrel{(1)}{\nu^{j}} + D^{j} \stackrel{(1)}{\nu_{i}}
    \right)
  \right.
  \nonumber\\
  && \quad\quad
  \left.
    + \frac{1}{2} \left(
      \partial_{\eta}^{2}
      + 2 {\cal H} \partial_{\eta}
      + 2 K
      - \Delta
    \right) \stackrel{(1)}{\chi_{i}^{\;\;j}}
  \right]
  \label{eq:kouchan-15.35-linear}
  ,
\end{eqnarray}
where ${\cal H} := \partial_{\eta}a/a$ and
$\gamma_{i}^{\;\;j}:=\gamma_{ik}\gamma^{kj}$ is the
three-dimensional Kronecker's delta. 
The components of 
${}^{(1)}{\cal G}_{a}^{\;\;b}\left[{\cal L}\right]$ in
Eq.~(\ref{eq:second-Einstein-generic}) in terms of the
gauge-invariant variables defined in
Eq.~(\ref{eq:components-calLab}) are given by the replacement of
the variables
\begin{eqnarray}
  \stackrel{(1)}{\Phi} \rightarrow \stackrel{(2)}{\Phi}, \quad
  \stackrel{(1)}{\nu_{i}} \rightarrow \stackrel{(2)}{\nu}_{i}, \quad
  \stackrel{(1)}{\Psi} \rightarrow \stackrel{(2)}{\Psi}, \quad
  \stackrel{(1)}{\chi_{ij}} \rightarrow \stackrel{(2)}{\chi}_{ij},
  \label{eq:replacement-linear-to-second-order}
\end{eqnarray}
in the equations
(\ref{eq:kouchan-15.32-linear})--(\ref{eq:kouchan-15.35-linear}).


\subsubsection{Components of ${}^{(2)}{\cal G}_{b}^{\;\;a}$}
\label{sec:Components-of-2-cal-G}


From Eq.~(\ref{eq:calG-Sigam-relation}) and the components of
the gauge-invariant parts (\ref{eq:components-calHab}) of the 
first-order metric perturbation, and the components of tensors 
$H_{a\;\;c}^{\;\;b}[{\cal H}]$ and $H_{a}^{\;\;bc}[{\cal H}]$,
which are summarized in Appendix A in the paper
KN2007\cite{kouchan-second-cosmo}, we can derive the components
of ${}^{(2)}\!{\cal G}_{a}^{\;\;b} = {}^{(2)}\!{\cal
  G}_{a}^{\;\;b}[{\cal H},{\cal H}]$ in a straightforward
manner. These components are summarized as follows:
\begin{eqnarray}
  {}^{(2)}\!{\cal G}_{\eta}^{\;\;\eta}
  &=&
  \frac{2}{a^{2}}
  \left[
    -          3  D_{k}\stackrel{(1)}{\Psi} D^{k}\stackrel{(1)}{\Psi}
    -          8  \stackrel{(1)}{\Psi} \Delta\stackrel{(1)}{\Psi}
    -          3  \left(\partial_{\eta}\stackrel{(1)}{\Psi}\right)^{2}
    -         12  K \left(\stackrel{(1)}{\Psi}\right)^{2}
    -         12  {\cal H}^{2} \left(\stackrel{(1)}{\Phi}\right)^{2}
  \right.
  \nonumber\\
  && \quad\quad\quad\quad
  \left.
    -         12  {\cal H} \left(
      \stackrel{(1)}{\Phi} - \stackrel{(1)}{\Psi} 
    \right) \partial_{\eta}\stackrel{(1)}{\Psi}
  \right.
  \nonumber\\
  && \quad\quad
  \left.
    -          2  D^{k}\left\{
      \partial_{\eta}\stackrel{(1)}{\Psi}
      + {\cal H} \left(
        \stackrel{(1)}{\Phi} + \stackrel{(1)}{\Psi}
      \right)
    \right\} \stackrel{(1)}{\nu^{k}}
  \right.
  \nonumber\\
  && \quad\quad
  \left.
    + \frac{1}{2} D_{k}\stackrel{(1)}{\nu_{l}} D^{(k}\stackrel{(1)}{\nu^{l)}}
    + \frac{1}{2} \stackrel{(1)}{\nu_{k}} \left(
      \Delta + 2 K + 6 {\cal H}^{2} 
    \right) \stackrel{(1)}{\nu^{k}}
  \right.
  \nonumber\\
  && \quad\quad
  \left.
    +             D_{l}D_{k}\stackrel{(1)}{\Psi} \stackrel{(1)}{\chi^{lk}}
  \right.
  \nonumber\\
  && \quad\quad
  \left.
    - \frac{1}{2} D^{k}\stackrel{(1)}{\nu^{l}} \left(
      \partial_{\eta} + 4 {\cal H}
    \right) \stackrel{(1)}{\chi_{lk}}
  \right.
  \nonumber\\
  && \quad\quad
  \left.
    + \frac{1}{8} \partial_{\eta}\stackrel{(1)}{\chi^{kl}}
    \left(
      \partial_{\eta}
      + 8 {\cal H}
    \right) \stackrel{(1)}{\chi_{kl}}
    + \frac{1}{2} D_{k}\stackrel{(1)}{\chi_{lm}} D^{[l}\stackrel{(1)}{\chi^{k]m}}
    - \frac{1}{8} D_{k}\stackrel{(1)}{\chi_{lm}} D^{k}\stackrel{(1)}{\chi^{ml}}
  \right.
  \nonumber\\
  && \quad\quad\quad\quad
  \left.
    - \frac{1}{2} \stackrel{(1)}{\chi^{lm}} \left(
      \Delta - K
    \right) \stackrel{(1)}{\chi_{lm}}
  \right]
  ,
  \label{eq:generic-2-calG-eta-eta}
  \\
  {}^{(2)}\!{\cal G}_{\eta}^{\;\;i}
  &=&
  \frac{2}{a^{2}}
  \left[
               4  {\cal H} \left(
      \stackrel{(1)}{\Psi} - \stackrel{(1)}{\Phi} 
    \right) D^{i}\stackrel{(1)}{\Phi}
    +          2  \partial_{\eta}\stackrel{(1)}{\Psi}
    D^{i}\left(
      2 \stackrel{(1)}{\Psi} - \stackrel{(1)}{\Phi}
    \right)
    +          8  \stackrel{(1)}{\Psi} \partial_{\eta}D^{i}\stackrel{(1)}{\Psi}
  \right.
  \nonumber\\
  && \quad\quad
  \left.
    + 2 \left(
        2 \partial_{\eta}{\cal H}
      - 2 {\cal H}^{2}
      +   {\cal H} \partial_{\eta}
    \right)
    \left(
      \stackrel{(1)}{\Phi} - \stackrel{(1)}{\Psi} 
    \right) \stackrel{(1)}{\nu^{i}}
    + D_{j}\stackrel{(1)}{\Phi} D^{(i}\stackrel{(1)}{\nu^{j)}}
    + D_{j}\stackrel{(1)}{\Psi} D^{[i}\stackrel{(1)}{\nu^{j]}}
  \right.
  \nonumber\\
  && \quad\quad\quad\quad
  \left.
    +   \left(
        2 \partial_{\eta}^{2}\stackrel{(1)}{\Psi}
      - 2 \stackrel{(1)}{\Psi} \Delta
      +   \Delta\stackrel{(1)}{\Phi}
      + 4 K \stackrel{(1)}{\Psi}
    \right) \stackrel{(1)}{\nu^{i}}
    -   D_{j}D^{i}\stackrel{(1)}{\Phi} \stackrel{(1)}{\nu^{j}}
  \right.
  \nonumber\\
  && \quad\quad
  \left.
    +   \left(
        2 {\cal H} D^{[i}\stackrel{(1)}{\nu^{j]}}
      -   \partial_{\eta}D^{(i}\stackrel{(1)}{\nu^{j)}}
    \right) \stackrel{(1)}{\nu_{j}}
  \right.
  \nonumber\\
  && \quad\quad
  \left.
    - \frac{1}{2} D_{j}\left(
      \stackrel{(1)}{\Phi} + \stackrel{(1)}{\Psi}
    \right) \partial_{\eta}\stackrel{(1)}{\chi^{ji}}
    -   \left(
      \partial_{\eta}D_{j}\stackrel{(1)}{\Psi}
      + 2 {\cal H} D_{j}\stackrel{(1)}{\Phi}
    \right) \stackrel{(1)}{\chi^{ij}}
  \right.
  \nonumber\\
  && \quad\quad
  \left.
    + \frac{1}{2} \stackrel{(1)}{\nu_{j}} \left(
          \partial_{\eta}^{2}
      + 2 {\cal H} \partial_{\eta}
      + 4 \partial_{\eta}{\cal H}
      - 4 {\cal H}^{2}
      - 2 K
    \right) \stackrel{(1)}{\chi^{ji}}
    + D_{k}\stackrel{(1)}{\nu_{j}} D^{[k}\stackrel{(1)}{\chi^{j]i}}
  \right.
  \nonumber\\
  && \quad\quad\quad\quad
  \left.
    + D^{l}D^{[k}\stackrel{(1)}{\nu^{i]}} \stackrel{(1)}{\chi_{kl}}
    + \frac{1}{2} \Delta\stackrel{(1)}{\nu_{j}} \stackrel{(1)}{\chi^{ij}}
  \right.
  \nonumber\\
  && \quad\quad
  \left.
    + \frac{1}{4} \partial_{\eta}\stackrel{(1)}{\chi_{jk}} D^{i}\stackrel{(1)}{\chi^{kj}}
    + \stackrel{(1)}{\chi_{kl}} \partial_{\eta}D^{[i}\stackrel{(1)}{\chi^{k]l}}
  \right]
  ,
  \label{eq:generic-2-calG-eta-i}
  \\
  {}^{(2)}\!{\cal G}_{i}^{\;\;\eta}
  &=&
  \frac{2}{a^{2}}
  \left[
      8 {\cal H} \stackrel{(1)}{\Phi} D_{i}\stackrel{(1)}{\Phi}
    + 2 D_{i}\left(
      \stackrel{(1)}{\Phi} - 2 \stackrel{(1)}{\Psi}
    \right) \partial_{\eta}\stackrel{(1)}{\Psi}
    + 4 \left(
      \stackrel{(1)}{\Phi} - \stackrel{(1)}{\Psi}
    \right) \partial_{\eta}D_{i}\stackrel{(1)}{\Psi}
  \right.
  \nonumber\\
  && \quad\quad
  \left.
    -   D^{j}\stackrel{(1)}{\Phi} D_{(i}\stackrel{(1)}{\nu_{j)}}
    +   D^{j}\stackrel{(1)}{\Psi} D_{[j}\stackrel{(1)}{\nu_{i]}}
    + \left(
      \stackrel{(1)}{\Psi} - \stackrel{(1)}{\Phi}
    \right)
    \left(
      \Delta + 2 K
    \right)
    \stackrel{(1)}{\nu_{i}}
  \right.
  \nonumber\\
  && \quad\quad\quad\quad
  \left.
    + \stackrel{(1)}{\nu^{j}} \left(
          D_{i}D_{j}
      +   \gamma_{ij} \Delta
    \right)\stackrel{(1)}{\Psi}
  \right.
  \nonumber\\
  && \quad\quad
  \left.
    -          2  {\cal H} \stackrel{(1)}{\nu^{j}} D_{i}\stackrel{(1)}{\nu_{j}}
  \right.
  \nonumber\\
  && \quad\quad
  \left.
    + \frac{1}{2} D^{j}\left(
      \stackrel{(1)}{\Phi} + \stackrel{(1)}{\Psi}
    \right) \partial_{\eta}\stackrel{(1)}{\chi_{ij}}
    -             \partial_{\eta}D^{j}\stackrel{(1)}{\Psi} \stackrel{(1)}{\chi_{ij}}
  \right.
  \nonumber\\
  && \quad\quad
  \left.
    +   D_{k}D_{[i}\stackrel{(1)}{\nu_{j]}} \stackrel{(1)}{\chi^{kj}}
    +   D^{[k}\stackrel{(1)}{\nu^{j]}} D_{j}\stackrel{(1)}{\chi_{ik}}
    - \frac{1}{2} \stackrel{(1)}{\nu^{j}} \left(
         \Delta
      - 2 K
    \right) \stackrel{(1)}{\chi_{ji}}
  \right.
  \nonumber\\
  && \quad\quad
  \left.
    - \frac{1}{4} \partial_{\eta}\stackrel{(1)}{\chi^{kj}} D_{i}\stackrel{(1)}{\chi_{kj}}
    +   \stackrel{(1)}{\chi^{kj}} \partial_{\eta}D_{[j}\stackrel{(1)}{\chi_{i]k}}
  \right]
  ,
  \label{eq:generic-2-calG-i-eta}
  \\
  {}^{(2)}\!{\cal G}_{i}^{\;\;j}
  &=&
  \frac{2}{a^{2}}
  \left[
    D_{i}\stackrel{(1)}{\Phi} D^{j}\left(
        \stackrel{(1)}{\Phi}
      - \stackrel{(1)}{\Psi}
    \right)
    - D_{i}\stackrel{(1)}{\Psi} D^{j}\left(
          \stackrel{(1)}{\Phi}
      - 3 \stackrel{(1)}{\Psi}
    \right)
    +          4  \stackrel{(1)}{\Psi} D_{i}D^{j}\stackrel{(1)}{\Psi}
  \right.
  \nonumber\\
  && \quad\quad
  \left.
    + 2 \left(
      \stackrel{(1)}{\Phi} - \stackrel{(1)}{\Psi}
    \right) D_{i}D^{j}\stackrel{(1)}{\Phi}
  \right.
  \nonumber\\
  && \quad\quad
  \left.
    + \left\{
      -             D_{k}\stackrel{(1)}{\Phi} D^{k}\stackrel{(1)}{\Phi}
      -          2  D_{k}\stackrel{(1)}{\Psi} D^{k}\stackrel{(1)}{\Psi}
      - 2 \left(
        \stackrel{(1)}{\Phi} - \stackrel{(1)}{\Psi}
      \right) \Delta\stackrel{(1)}{\Phi}
    \right.
  \right.
  \nonumber\\
  && \quad\quad\quad\quad
  \left.
    \left.
      - 4 \stackrel{(1)}{\Psi} \left(
        \Delta + K
      \right) \stackrel{(1)}{\Psi}
      + \partial_{\eta}\stackrel{(1)}{\Psi} \partial_{\eta}\left(
            \stackrel{(1)}{\Psi}
        - 2 \stackrel{(1)}{\Phi}
      \right)
      + 4 \left(
            \stackrel{(1)}{\Psi}
        -   \stackrel{(1)}{\Phi}
      \right) \partial_{\eta}^{2}\stackrel{(1)}{\Psi}
    \right.
  \right.
  \nonumber\\
  && \quad\quad\quad\quad
  \left.
    \left.
      -          8  {\cal H} \stackrel{(1)}{\Phi} \partial_{\eta}\stackrel{(1)}{\Phi}
      - 8 {\cal H} \left(
        \stackrel{(1)}{\Phi} - \stackrel{(1)}{\Psi}
      \right) \partial_{\eta}\stackrel{(1)}{\Psi}
    \right.
  \right.
  \nonumber\\
  && \quad\quad\quad\quad
  \left.
    \left.
      - 4 \left(
          2 \partial_{\eta}{\cal H}
        +   {\cal H}^{2}
      \right) \left(\stackrel{(1)}{\Phi}\right)^{2}
    \right\} \gamma_{i}^{\;\;j}
  \right.
  \nonumber\\
  && \quad\quad
  \left.
    + \frac{1}{2} \partial_{\eta}\left(\stackrel{(1)}{\Psi} + \stackrel{(1)}{\Phi}\right) \left(
      D_{i}\stackrel{(1)}{\nu^{j}} + D^{j}\stackrel{(1)}{\nu_{i}}
    \right)
    -             D^{j}\stackrel{(1)}{\Psi} \partial_{\eta}\stackrel{(1)}{\nu_{i}}
    +             \partial_{\eta}D_{i}\stackrel{(1)}{\Psi} \stackrel{(1)}{\nu^{j}}
  \right.
  \nonumber\\
  && \quad\quad\quad
  \left.
    + \left(
      \stackrel{(1)}{\Phi} - \stackrel{(1)}{\Psi}
    \right)
    \left(
      \partial_{\eta} + 2 {\cal H} 
    \right)
    \left(
      D_{i}\stackrel{(1)}{\nu^{j}} + D^{j}\stackrel{(1)}{\nu_{i}}
    \right)
    - D_{i}\stackrel{(1)}{\Psi} \left(
      \partial_{\eta} + 2 {\cal H}
    \right) \stackrel{(1)}{\nu^{j}}
  \right.
  \nonumber\\
  && \quad\quad\quad
  \left.
    - \stackrel{(1)}{\nu_{i}} \left(
      \partial_{\eta} + 2 {\cal H}
    \right) D^{j}\stackrel{(1)}{\Psi}
    + 2 {\cal H} D_{i}\stackrel{(1)}{\Phi} \stackrel{(1)}{\nu^{j}}
    - 2 D_{k}\left(
          \partial_{\eta}\stackrel{(1)}{\Psi}
      +   {\cal H} \stackrel{(1)}{\Phi}
    \right) \stackrel{(1)}{\nu^{k}} \gamma_{i}^{\;\;j}
  \right.
  \nonumber\\
  && \quad\quad
  \left.
    + \frac{1}{2} \stackrel{(1)}{\nu^{k}} D_{k}\left(
      D^{j}\stackrel{(1)}{\nu_{i}} + D_{i}\stackrel{(1)}{\nu^{j}}
    \right)
    -             \stackrel{(1)}{\nu_{k}} D_{i}D^{j}\stackrel{(1)}{\nu^{k}}
    - \frac{1}{2} D_{i}\stackrel{(1)}{\nu_{k}} D^{j}\stackrel{(1)}{\nu^{k}}
  \right.
  \nonumber\\
  && \quad\quad\quad
  \left.
    - \frac{1}{2} D^{k}\stackrel{(1)}{\nu_{i}} D_{k}\stackrel{(1)}{\nu^{j}}
    - \frac{1}{2} \stackrel{(1)}{\nu^{j}} \left(
      \Delta + 2 K
    \right) \stackrel{(1)}{\nu_{i}}
  \right.
  \nonumber\\
  && \quad\quad\quad
  \left.
    + \left\{
      \stackrel{(1)}{\nu_{k}} \left(
          2 {\cal H} \partial_{\eta}
        +   \Delta
        + 2 \partial_{\eta}{\cal H}
        +   {\cal H}^{2}
      \right) \stackrel{(1)}{\nu^{k}}
    \right.
  \right.
  \nonumber\\
  && \quad\quad\quad\quad\quad
  \left.
    \left.
      + \frac{1}{2} D_{k}\stackrel{(1)}{\nu_{l}} \left(
        D^{[k}\stackrel{(1)}{\nu^{l]}} + D^{k}\stackrel{(1)}{\nu^{l}}
      \right)
    \right\} \gamma_{i}^{\;\;j}
  \right.
  \nonumber\\
  && \quad\quad
  \left.
    - \left(
      \stackrel{(1)}{\Phi} - \stackrel{(1)}{\Psi}
    \right)
    \left(
          \partial_{\eta}^{2}
      + 2 {\cal H} \partial_{\eta}
    \right) \stackrel{(1)}{\chi_{i}^{\;\;j}}
    - \frac{1}{2} \partial_{\eta}\stackrel{(1)}{\chi_{i}^{\;\;j}} \partial_{\eta}\left(
      \stackrel{(1)}{\Phi} - \stackrel{(1)}{\Psi}
    \right)
  \right.
  \nonumber\\
  && \quad\quad\quad
  \left.
    + \stackrel{(1)}{\chi_{i}^{\;\;j}} \left(
          \partial_{\eta}^{2}
      + 2 {\cal H} \partial_{\eta}
    \right) \stackrel{(1)}{\Psi}
    + \frac{1}{2} D_{k}\left(
      \stackrel{(1)}{\Phi} + \stackrel{(1)}{\Psi}
    \right)
    \left(
      D_{i}\stackrel{(1)}{\chi^{jk}} + D^{j}\stackrel{(1)}{\chi_{ik}}
    \right)
  \right.
  \nonumber\\
  && \quad\quad\quad
  \left.
    - \frac{1}{2} D^{k}\left(
      \stackrel{(1)}{\Phi} + 3 \stackrel{(1)}{\Psi}
    \right) D_{k}\stackrel{(1)}{\chi_{i}^{\;\;j}}
    - 2 \stackrel{(1)}{\Psi} \left(
      \Delta - 2 K
    \right) \stackrel{(1)}{\chi_{i}^{\;\;j}}
    -             \Delta \stackrel{(1)}{\Psi} \stackrel{(1)}{\chi_{i}^{\;\;j}}
  \right.
  \nonumber\\
  && \quad\quad\quad
  \left.
    +             D_{k}D_{i}\stackrel{(1)}{\Phi} \stackrel{(1)}{\chi^{jk}}
    +             D^{m}D^{j}\stackrel{(1)}{\Psi} \stackrel{(1)}{\chi_{im}}
    -             D_{l}D_{k}\stackrel{(1)}{\Phi} \stackrel{(1)}{\chi^{lk}} \gamma_{i}^{\;\;j}
  \right.
  \nonumber\\
  && \quad\quad
  \left.
    + \frac{1}{2} \left(\partial_{\eta} + 2 {\cal H} \right) \left(
        \stackrel{(1)}{\nu_{k}} D_{i}\stackrel{(1)}{\chi^{kj}}
      + \stackrel{(1)}{\nu^{k}} D^{j}\stackrel{(1)}{\chi_{ki}}
      - \stackrel{(1)}{\nu^{k}} D_{k}\stackrel{(1)}{\chi_{i}^{\;\;j}}
    \right)
  \right.
  \nonumber\\
  && \quad\quad\quad
  \left.
    + \frac{1}{2} D^{k}\stackrel{(1)}{\nu^{j}} \partial_{\eta}\stackrel{(1)}{\chi_{ik}}
    + \frac{1}{2} D_{k}\stackrel{(1)}{\nu_{i}} \partial_{\eta}\stackrel{(1)}{\chi^{jk}}
    + \stackrel{(1)}{\chi^{jk}} \left(
      \partial_{\eta} + 2 {\cal H}
    \right) D_{(i}\stackrel{(1)}{\nu_{k)}}
  \right.
  \nonumber\\
  && \quad\quad\quad
  \left.
    - \frac{1}{2} \stackrel{(1)}{\nu^{k}} \partial_{\eta}D_{k}\stackrel{(1)}{\chi_{i}^{\;\;j}}
    - \left\{
        \stackrel{(1)}{\chi_{lk}} \partial_{\eta}D^{k}\stackrel{(1)}{\nu^{l}}
      + \frac{1}{2} D^{k}\stackrel{(1)}{\nu^{l}} \left(
        \partial_{\eta} + 4 {\cal H}
      \right) \stackrel{(1)}{\chi_{lk}}
    \right\} \gamma_{i}^{\;\;j}
  \right.
  \nonumber\\
  && \quad\quad
  \left.
    - \frac{1}{2} \partial_{\eta}\stackrel{(1)}{\chi_{ik}} \partial_{\eta}\stackrel{(1)}{\chi^{kj}}
    + D_{k}\stackrel{(1)}{\chi_{il}} D^{[k}\stackrel{(1)}{\chi^{l]j}}
    + \frac{1}{4} D^{j}\stackrel{(1)}{\chi_{lk}} D_{i}\stackrel{(1)}{\chi^{lk}}
  \right.
  \nonumber\\
  && \quad\quad\quad
  \left.
    + \frac{1}{2} \stackrel{(1)}{\chi_{lm}} D_{i}D^{j}\stackrel{(1)}{\chi^{ml}}
    - \frac{1}{2} \stackrel{(1)}{\chi_{lm}} D^{l}D_{i}\stackrel{(1)}{\chi^{mj}}
    - \frac{1}{2} \stackrel{(1)}{\chi^{lm}} D_{l}D^{j}\stackrel{(1)}{\chi_{mi}}
  \right.
  \nonumber\\
  && \quad\quad\quad
  \left.
    + \frac{1}{2} \stackrel{(1)}{\chi^{lm}} D_{m}D_{l}\stackrel{(1)}{\chi_{i}^{\;\;j}}
    - \frac{1}{2} \stackrel{(1)}{\chi^{jk}}  \left(
          \partial_{\eta}^{2}
      + 2 {\cal H} \partial_{\eta}
      -   \Delta
      + 2 K
    \right) \stackrel{(1)}{\chi_{ik}}
  \right.
  \nonumber\\
  && \quad\quad\quad
  \left.
    + \frac{1}{2} \left\{
        \frac{3}{4} \partial_{\eta}\stackrel{(1)}{\chi_{lk}} \partial_{\eta}\stackrel{(1)}{\chi^{kl}}
      - \frac{1}{4} D_{k}\stackrel{(1)}{\chi_{lm}} D^{k}\stackrel{(1)}{\chi^{ml}}
      +             D_{k}\stackrel{(1)}{\chi_{lm}} D^{[l}\stackrel{(1)}{\chi^{k]m}}
    \right.
  \right.
  \nonumber\\
  && \quad\quad\quad\quad\quad\quad
  \left.
    \left.
      +   \stackrel{(1)}{\chi_{kl}} \left(
            \partial_{\eta}^{2}
        + 2 {\cal H} \partial_{\eta}
        -   \Delta
        +   K
      \right) \stackrel{(1)}{\chi^{lk}}
    \right\} \gamma_{i}^{\;\;j}
  \right]
  .
  \label{eq:generic-2-calG-i-j}
\end{eqnarray}
In these components, the terms of different types of the
mode-coupling are written in the different lines to show the
different types of mode-coupling, manifestly.
For example, in the expression of the component 
${}^{(2)}\!{\cal G}_{\eta}^{\;\;\eta}$ in
Eq.~(\ref{eq:generic-2-calG-eta-eta}), the first two lines show
the mode-coupling of the scalar-scalar type, the third line
shows the scalar-vector mode-coupling, the fourth line shows the
vector-vector type, the fifth line shows the scalar-tensor type,
the sixth line shows vector-tensor type, and the seventh and
eighth lines show the mode-coupling of the tensor-tensor type.


We have checked the perturbations of the Bianchi identity for
the components
(\ref{eq:kouchan-15.32-linear})--(\ref{eq:kouchan-15.35-linear}) and 
(\ref{eq:generic-2-calG-eta-eta})--(\ref{eq:generic-2-calG-i-j})
of the gauge-invariant parts 
${}^{(1)}\!{\cal G}_{a}^{\;\;b}[{\cal H}]$,
${}^{(1)}\!{\cal G}_{a}^{\;\;b}[{\cal L}]$, and 
${}^{(2)}\!{\cal G}_{a}^{\;\;b}[{\cal H}, {\cal H}]$.
As shown in KN2005\cite{kouchan-second}, the first- and the
second-order perturbations of the Bianchi identity
$\bar{\nabla}_{a}\bar{G}_{a}^{\;\;b}=0$ give the identities
\begin{eqnarray}
  \nabla_{a}{}^{(1)}\!{\cal G}_{b}^{\;\;a}\left[{\cal H}\right]
  &=& 
  - H_{ca}^{\;\;\;\;a}\left[{\cal H}\right] G_{b}^{\;\;c}
  + H_{ba}^{\;\;\;\;c}\left[{\cal H}\right] G_{c}^{\;\;a},
  \label{eq:kouchan-second-3.95}
  \\
  \nabla_{a}{}^{(2)}{\cal G}_{b}^{\;\;a}[{\cal H},{\cal H}] 
  &=& 
  - 2 H_{ca}^{\;\;\;\;a}[{\cal H}]
  \; {}^{(1)}\!{\cal G}_{b}^{\;\;c}[{\cal H}]
  + 2 H_{ba}^{\;\;\;\;e}[{\cal H}]
  \; {}^{(1)}\!{\cal G}_{e}^{\;\;a}[{\cal H}]
  \nonumber\\
  &&
  - 2 H_{bad}[{\cal H}]\; {\cal H}^{dc} G_{c}^{\;\;a}
  + 2 H_{cad}[{\cal H}]\; {\cal H}^{ad} G_{b}^{\;\;c}.
  \label{eq:second-div-of-calGab-1,1-2}
\end{eqnarray}
Although we can check these identities without the explicit
components of the gauge-invariant part of the metric
perturbations as shown in KN2005, we can also check these
identities (\ref{eq:kouchan-second-3.95}) and
(\ref{eq:second-div-of-calGab-1,1-2}) in the case of
cosmological perturbation through the explicit components of the
gauge-invariant parts ${\cal H}_{ab}$ and ${\cal L}_{ab}$ in
Eqs.~(\ref{eq:components-calHab}) and
(\ref{eq:components-calLab}) of the metric perturbations.
The actual calculations to confirm the identities
(\ref{eq:kouchan-second-3.95}) and
(\ref{eq:second-div-of-calGab-1,1-2}) are straightforward.
These checks of these Bianchi identities guarantee that the
expressions
(\ref{eq:kouchan-15.32-linear})--(\ref{eq:kouchan-15.35-linear})
of the components of ${}^{(1)}\!{\cal G}_{a}^{\;\;b}$ and the
expressions 
(\ref{eq:generic-2-calG-eta-eta})--(\ref{eq:generic-2-calG-i-j})
of the components of ${}^{(2)}\!{\cal G}_{a}^{\;\;b}$ are
self-consistent. 
In this sense, we may say that the formulae 
(\ref{eq:kouchan-15.32-linear})--(\ref{eq:kouchan-15.35-linear})
and
(\ref{eq:generic-2-calG-eta-eta})--(\ref{eq:generic-2-calG-i-j})
are correct.


\section{Consistency of the background equations}
\label{sec:Background-equations}


In this section, we briefly review the Einstein equations and
the equations for the matter field on a four-dimensional
homogeneous isotropic universe whose metric is given by
(\ref{eq:background-metric}). 
Further, we show the consistency between the equations of motion
for matter field and the Einstein equations.
These equations are used throughout this paper.
As the matter contents, we consider a perfect fluid and a scalar
field, respectively.


\subsection{Perfect fluid case}
\label{sec:Background-Equations-perfect-fluid}


The energy momentum tensor for a perfect fluid on the background
spacetime, whose metric is given by
(\ref{eq:background-metric}), is given by
Eq.~(\ref{eq:energy-momentum-perfect-fluid}) (or
Eq.~(\ref{eq:energy-momentum-perfect-fluid-homogeneous})). 
The Einstein equations $G_{a}^{\;\;b} = 8\pi G T_{a}^{\;\;b}$
for this background spacetime filled with a perfect fluid are
given by 
\begin{eqnarray}
  \label{eq:background-Einstein-equation-1}
  \stackrel{(0)}{{}^{(p)}E_{(1)}}
  &:=& {\cal H}^{2} + K - \frac{8 \pi G}{3} a^{2} \epsilon = 0
  , \\
  \label{eq:background-Einstein-equation-2}
  \stackrel{(0)}{{}^{(p)}E_{(2)}}
  &:=& 2 \partial_{\eta}{\cal H} + {\cal H}^{2} + K + 8 \pi G a^{2}p
  = 0
  .
\end{eqnarray}
To consider the consistency of the perturbative equations, the 
equation
\begin{eqnarray}
  \label{eq:background-Einstein-equations-3}
  \frac{1}{2} \left(
    3 \stackrel{(0)}{{}^{(p)}E_{(1)}}
    - \stackrel{(0)}{{}^{(p)}E_{(2)}}
  \right)
  &=&
  {\cal H}^{2} + K - \partial_{\eta}{\cal H} - 4 \pi G a^{2} (\epsilon + p) = 0
\end{eqnarray}
is also useful.


The divergence of the energy momentum tensor
(\ref{eq:energy-momentum-perfect-fluid}) gives the two
equations, which are well-known as the energy continuity equation
and the Euler equation.
The Euler equation is trivial due to the fact that the pressure
$p$ is homogeneous (i.e., $p=p(\eta)$) and the integral curves
of the fluid four-velocity $u^{a}=g^{ab}u_{b}$ with the
component (\ref{eq:kouchan-17.378}) are geodesics on the
background spacetime with the metric (\ref{eq:background-metric}). 
On the other hand, the energy continuity equation is given by  
\begin{eqnarray}
  a \stackrel{(0)}{C_{0}^{(p)}}
  :=
  \partial_{\eta}\epsilon
  +
  3 {\cal H} \left(\epsilon + p\right)
  =
  0
  .
  \label{eq:kouchan-19.41}
\end{eqnarray}
Through
Eqs.~(\ref{eq:background-Einstein-equation-1})--(\ref{eq:background-Einstein-equations-3}),
we easily verify that 
\begin{eqnarray}
  8 \pi G a^{3} \stackrel{(0)}{C_{0}^{(p)}}
  =
  - 3 \partial_{\eta}\stackrel{(0)}{{}^{(p)}E_{(1)}}
  - 3 {\cal H} \left(
    \stackrel{(0)}{{}^{(p)}E_{(1)}} - \stackrel{(0)}{{}^{(p)}E_{(2)}}
  \right)
  .
  \label{eq:consistency-background-fluid}
\end{eqnarray}
We may say that the energy continuity equation for the
background spacetime is consistent with the Einstein equations 
(\ref{eq:background-Einstein-equation-1}) and
(\ref{eq:background-Einstein-equation-2}).
We also note that the relation
(\ref{eq:consistency-background-fluid}) has nothing to do with
the equation of state of the perfect fluid.
This is a well-known fact and is just due to the Bianchi
identity of the background spacetime.
However, in
\S\S\ref{sec:Consistency-of-the-first-order-perturbations} and
\ref{sec:Consistency-of-the-second-order-perturbations},
we will show this kind consistency between the Einstein
equations and the equations of motion for matter field in the
case of the first- and the second-order perturbations.
We use this kind of the relations to check the consistency of
the set of perturbative equations.


\subsection{Scalar field case}
\label{sec:Background-Equations-scalar-field}


In this paper, we also consider the universe filled with a
single scalar field.
The energy momentum tensor for a scalar field on the
homogeneous isotropic universe is given by
Eq.~(\ref{eq:energy-momentum-single-scalar-homogeneous}) with
the homogeneous condition (\ref{eq:kouchan-19.181}).
Through this energy momentum tensor for a scalar field, the
Einstein equations $G_{a}^{\;\;b} = 8\pi G T_{a}^{\;\;b}$ for
this background spacetime filled with a scalar field are given
by
\begin{eqnarray}
  \label{eq:background-Einstein-equations-scalar-1}
  \stackrel{(0)}{{}^{(s)}E_{(1)}}
  &:=& {\cal H}^{2} + K - \frac{8 \pi G}{3} \left(
    \frac{1}{2} (\partial_{\eta}\varphi)^{2} + a^{2} V(\varphi)
  \right)
  = 0
  ,\\
  \label{eq:background-Einstein-equations-scalar-2}
  \stackrel{(0)}{{}^{(s)}E_{(2)}}
  &:=&
  2 \partial_{\eta}{\cal H} + {\cal H}^{2} + K + 8 \pi G
  \left(\frac{1}{2} (\partial_{\eta}\varphi)^{2} - a^{2} V(\varphi)\right)
  = 0.
\end{eqnarray}
We also note that these equations
(\ref{eq:background-Einstein-equations-scalar-1}) and
(\ref{eq:background-Einstein-equations-scalar-2}) lead
\begin{eqnarray}
  \label{eq:background-Einstein-equations-scalar-3}
  \frac{1}{2} \left(
    3 \stackrel{(0)}{{}^{(s)}E_{(1)}}
    - \stackrel{(0)}{{}^{(s)}E_{(2)}}
  \right)
  &=& {\cal H}^{2} + K - \partial_{\eta}{\cal H}
  - 4 \pi G (\partial_{\eta}\varphi)^{2}
  = 0
  ,\\
  \label{eq:background-Einstein-equations-scalar-4}
  \frac{1}{2}
  \left(
    3 \stackrel{(0)}{{}^{(s)}E_{(1)}}
    + \stackrel{(0)}{{}^{(s)}E_{(2)}}
  \right)
  &=&
    2 {\cal H}^{2}
  + 2 K
  +   \partial_{\eta}{\cal H}
  - 8 \pi G a^{2} V(\varphi)
  = 0.
\end{eqnarray}
The equations (\ref{eq:background-Einstein-equations-scalar-3})
and (\ref{eq:background-Einstein-equations-scalar-4}) are also
useful when we check the consistency of equations for the first-
and the second-order perturbations, respectively.


The divergence of the energy momentum tensor
(\ref{eq:energy-momentum-single-scalar}) gives the Klein-Gordon
equation on the background spacetime.
Further, the Klein-Gordon equation is also consistent with the
background Einstein equations
(\ref{eq:background-Einstein-equations-scalar-1}) and
(\ref{eq:background-Einstein-equations-scalar-2}).
This can be easily seen from the relation
\begin{eqnarray}
  \frac{8 \pi G}{3} \partial_{\eta}\varphi \left(
    a^{2} \stackrel{(0)}{C_{K}}
  \right)
  &=&
  - \frac{8 \pi G}{3} \partial_{\eta}\varphi \left(
    \partial_{\eta}^{2}\varphi
    + 2 {\cal H} \partial_{\eta}\varphi
    +   a^{2} \frac{\partial V}{\partial\varphi}
  \right)
  \label{eq:background-Klein-Gordon-eq}
  \\
  &=&
    \partial_{\eta}\stackrel{(0)}{{}^{(s)}E_{(1)}}
  + {\cal H} \left(
    \stackrel{(0)}{{}^{(s)}E_{(1)}}
    - \stackrel{(0)}{{}^{(s)}E_{(2)}}
  \right)
  \label{eq:background-Klein-Gordon-consistency}
  .
\end{eqnarray}
Thus, the Klein-Gordon equation 
\begin{eqnarray}
  \label{eq:background-Klein-Gordon-eq-explicit}
  \stackrel{(0)}{C_{K}}=0
\end{eqnarray}
for the background spacetime is consistent with the Einstein 
equations (\ref{eq:background-Einstein-equations-scalar-1}) and 
(\ref{eq:background-Einstein-equations-scalar-2}).
Of course, this is well-known fact and is just due to the Bianchi
identity of the background spacetime as in the case of a perfect
fluid.
However, as we noted in the case of the perfect fluid, this kind
of relations are useful to check whether the derived system of
equations are consistent or not.


\section{Consistency of the first-order equations}
\label{sec:Consistency-of-the-first-order-perturbations}


Here, we consider the consistency of the first-order
perturbations of the Einstein equations and equations of matter
fields.
The essence of this consistency check is same as those for the
background equations.
However, these consistency checks for the first-order
perturbations are similar to those for the second-order
perturbations.
Therefore, the consistency check for the set of the equations of
the first order is instructive when we consider the set of
equations for the second-order perturbations.


As shown in KN2007\cite{kouchan-second-cosmo}, we can derive the
components of the first-order perturbation of the Einstein
equation 
\begin{eqnarray}
  {}^{(1)}\!{\cal G}_{a}^{\;\;b}\left[{\cal H}\right]
  =
  8 \pi G \; {}^{(1)}\!{\cal T}_{a}^{\;\;b}
  \label{eq:kouchan-first-order-einstein-general}
\end{eqnarray}
and the components of the equations of motion for matter field, 
which are derived in KN2008\cite{kouchan-second-cosmo-matter}. 
As in the case of the background equations, we consider the cases
for a perfect fluid
(\S\ref{sec:Fist-order-perturbations-perfect-fluid}) and a
scalar field
(\S\ref{sec:Fist-order-perturbations-scalar-field}).


\subsection{Perfect fluid case}
\label{sec:Fist-order-perturbations-perfect-fluid}


Through the components of the linearized Einstein tensor
Eqs.~(\ref{eq:kouchan-15.32-linear})--(\ref{eq:kouchan-15.35-linear})
and the components of the first-order perturbation of the energy
momentum tensor for a perfect fluid
[Eqs.~(\ref{eq:kouchan-19.22})--(\ref{eq:kouchan-19.25})], the 
components of the first-order Einstein equation 
(\ref{eq:kouchan-first-order-einstein-general}) are summarized
as 
\begin{eqnarray}
  \stackrel{(1)}{{}^{(p)}E_{(1)}}
  &:=&
  \left(
    - 3 {\cal H} \partial_{\eta}
    +   \Delta
    + 3 K
  \right) \stackrel{(1)}{\Psi}
  - 3 {\cal H}^{2} \stackrel{(1)}{\Phi}
  - 4 \pi G a^{2} \stackrel{(1)}{{\cal E}}
  = 0
  \label{eq:kouchan-19.71}
  , \\
  \stackrel{(1)}{{}^{(p)}E_{(2)}}
  &:=&
  \left(
                  \partial_{\eta}^{2} 
    +          2  {\cal H} \partial_{\eta}
    -             K
    - \frac{1}{3} \Delta
  \right)
  \stackrel{(1)}{\Psi}
  + \left(
                  {\cal H} \partial_{\eta}
    +          2  \partial_{\eta}{\cal H}
    +             {\cal H}^{2}
    + \frac{1}{3} \Delta
  \right)
  \stackrel{(1)}{\Phi}
  \nonumber\\
  &&
  - 4 \pi G a^{2} \stackrel{(1)}{{\cal P}}
  = 0
  ,
  \label{eq:kouchan-19.65}
  \\
  \stackrel{(1)}{{}^{(p)}E_{(3)}}
  &:=&
  \stackrel{(1)}{\Psi} - \stackrel{(1)}{\Phi} = 0
  ,
  \label{eq:kouchan-19.66}
  \\
  \stackrel{(1)}{{}^{(p)}E_{(4)i}}
  &:=&
    \partial_{\eta}D_{i}\stackrel{(1)}{\Psi}
  + {\cal H} D_{i}\stackrel{(1)}{\Phi}
  + 4 \pi G \left( \epsilon + p \right) a^{2} D_{i}\stackrel{(1)}{v}
  = 0
  \label{eq:kouchan-19.69}
  ,\\
  \stackrel{(1)}{{}^{(p)}E_{(5)i}}
  &:=&
  \left( \Delta + 2 K \right) \stackrel{(1)}{\nu_{i}}
  - 16 \pi G \left( \epsilon + p \right) a^{2} \stackrel{(1)}{{\cal V}_{i}}
  = 0
  \label{eq:kouchan-19.70}
  , \\ 
  \stackrel{(1)}{{}^{(p)}E_{(6)i}}
  &:=&
  \partial_{\eta}\left(
    a^{2}  \stackrel{(1)}{\nu_{i}}
  \right)
  =
  0
  ,
  \label{eq:kouchan-19.67}
  \\
  \stackrel{(1)}{{}^{(p)}E_{(7)ij}}
  &:=&
  \left(
    \partial_{\eta}^{2}
    + 2 {\cal H} \partial_{\eta}
    + 2 K
    - \Delta
  \right) \stackrel{(1)}{\chi_{ij}}
  =
  0
  .
  \label{eq:kouchan-19.68}
\end{eqnarray}


\subsubsection{Continuity equations and Euler equations}
\label{sec:Fist-order-perturbations-perfect-fluid-continuity-Euler}


As shown in KN2008\cite{kouchan-second-cosmo-matter}, the
first-order perturbation of the energy continuity equation in
terms of the gauge-invariant variables is given by
\begin{eqnarray}
  a {}^{(1)}\!{\cal C}_{0}^{(p)}
  &=&
      \partial_{\eta}\stackrel{(1)}{{\cal E}}
  + 3 {\cal H} \left(
      \stackrel{(1)}{{\cal E}}
    + \stackrel{(1)}{{\cal P}}
  \right)
  + \left(\epsilon + p\right) \left(
                  \Delta\stackrel{(1)}{v} 
    -          3  \partial_{\eta}\stackrel{(1)}{\Phi}
  \right)
  =
  0
  \label{eq:kouchan-19.81}
  .
\end{eqnarray}
Further, in terms of the gauge-invariant variables, the
first-order perturbation of the Euler equation is given by 
\begin{eqnarray}
  {}^{(1)}\!{\cal C}_{i}^{(p)}
  &=&
  \left( \epsilon + p \right) \left\{
      \left(\partial_{\eta} + {\cal H} \right) \left(
      D_{i}\stackrel{(1)}{v} 
      + \stackrel{(1)}{{\cal V}_{i}}
    \right)
    + D_{i}\stackrel{(1)}{\Phi}
  \right\}
  + D_{i}\stackrel{(1)}{{\cal P}}
  + \partial_{\eta}p \left(
    D_{i}\stackrel{(1)}{v}
      + \stackrel{(1)}{{\cal V}_{i}}
  \right)
  \nonumber\\
  &=& 
  0,
  \label{eq:kouchan-19.82}
\end{eqnarray}
and this equation is decomposed into the scalar- and the
vector-parts: 
\begin{eqnarray}
  {}^{(1)}\!{\cal C}_{i}^{(pS)}
  &:=&
  D_{i}\Delta^{-1}D^{j}{}^{(1)}\!{\cal C}_{j}^{(p)}
  \nonumber\\
  &=&
  \left( \epsilon + p \right) \left\{
      \left(\partial_{\eta} + {\cal H} \right) D_{i}\stackrel{(1)}{v} 
    + D_{i}\stackrel{(1)}{\Phi}
  \right\}
  + D_{i}\stackrel{(1)}{{\cal P}}
  + \partial_{\eta}p D_{i}\stackrel{(1)}{v}
  \nonumber\\
  &=&
  0,
  \label{eq:kouchan-19.82-2-1}
  \\
  {}^{(1)}\!{\cal C}_{i}^{(pV)}
  &:=&
  {}^{(1)}\!{\cal C}_{i}^{(p)}
  - D_{i}\Delta^{-1}D^{j}{}^{(1)}\!{\cal C}_{j}^{(p)}
  \nonumber\\
  &=&
    \left(\epsilon + p\right) \left(\partial_{\eta} + {\cal H} \right)\stackrel{(1)}{{\cal V}_{i}}
  + \partial_{\eta}p \stackrel{(1)}{{\cal V}_{i}}
  \nonumber\\
  &=&
  0.
  \label{eq:kouchan-19.82-2-2}
\end{eqnarray}


Now, we check the consistency of Eqs.~(\ref{eq:kouchan-19.81}),
(\ref{eq:kouchan-19.82-2-1}), and (\ref{eq:kouchan-19.82-2-2})
for a perfect fluid and the Einstein equations 
(\ref{eq:kouchan-19.71})--(\ref{eq:kouchan-19.68}). 
First, we consider the perturbation of the continuity equation
(\ref{eq:kouchan-19.81}).
Substituting (\ref{eq:background-Einstein-equations-3}),
(\ref{eq:kouchan-19.71}), (\ref{eq:kouchan-19.65}), and
(\ref{eq:kouchan-19.69}) into (\ref{eq:kouchan-19.81}), we can
easily verify the relation
\begin{eqnarray}
  4 \pi G a^{3} {}^{(1)}\!{\cal C}_{0}^{(p)}
  &=&
  - \left(
    \partial_{\eta} - 2 {\cal H}
  \right) \stackrel{(1)}{{}^{(p)}E_{(1)}}
  + 3 {\cal H} \left(
    - \stackrel{(1)}{{}^{(p)}E_{(1)}}
    - \stackrel{(1)}{{}^{(p)}E_{(2)}}
  \right)
  \nonumber\\
  &&
  +   D^{i}\stackrel{(1)}{{}^{(p)}E_{(4)i}}
  + \frac{3}{2} \left(
      3 \stackrel{(0)}{{}^{(p)}E_{(1)}}
    -   \stackrel{(0)}{{}^{(p)}E_{(2)}}
  \right) \partial_{\eta}\stackrel{(1)}{\Psi}
  .
  \label{eq:kouchan-19.81-2}
\end{eqnarray}
Therefore, the first-order perturbation of the energy continuity
equation is consistent with the background Einstein equations
and the first-order perturbation of the Einstein equation.


Next, we consider the first-order perturbation
(\ref{eq:kouchan-19.82}) of the Euler equation.
Through
Eqs.~(\ref{eq:kouchan-19.65})--(\ref{eq:kouchan-19.69}),
(\ref{eq:kouchan-19.41}), and
(\ref{eq:background-Einstein-equations-3}), we can easily derive
the equation
\begin{eqnarray}
  4 \pi G a^{2} {}^{(1)}\!{\cal C}_{i}^{(pS)}
  &=&
    \partial_{\eta}\stackrel{(1)}{{}^{(p)}E_{(4)i}}
  + 2 {\cal H} \stackrel{(1)}{{}^{(p)}E_{(4)i}}
  - D_{i}\stackrel{(1)}{{}^{(p)}E_{(2)}}
  - \frac{1}{3} D_{i}\left(
    \Delta + 3 K
  \right) \stackrel{(1)}{{}^{(p)}E_{(3)}}
  \nonumber\\
  &&
  - 4 \pi G a^{3} \stackrel{(0)}{C_{0}^{(p)}} D_{i}\stackrel{(1)}{v}
  - \frac{1}{2} \left(
    3 \stackrel{(0)}{{}^{(p)}E_{(1)}}
    - \stackrel{(0)}{{}^{(p)}E_{(2)}}
  \right) D_{i}\stackrel{(1)}{\Phi}
  .
  \label{eq:Euler-consistency-linear-scalar-fluid}
\end{eqnarray}
Further, through Eqs.~(\ref{eq:kouchan-19.70}),
(\ref{eq:kouchan-19.67}), and (\ref{eq:kouchan-19.41}), we can
easily see that 
\begin{eqnarray}
  4 \pi G a^{2} {}^{(1)}\!{\cal C}_{i}^{(pV)}
  &=&
  \frac{1}{4a^{2}} \left\{
    -   \partial_{\eta}\left(a^{2} \stackrel{(1)}{{}^{(p)}E_{(5)i}}\right)
    +   \left( \Delta + 2 K \right) \stackrel{(1)}{{}^{(p)}E_{(6)i}}
  \right\}
  \nonumber\\
  &&
  - 4 \pi G a^{3} \stackrel{(0)}{C_{0}^{(p)}} \stackrel{(1)}{{\cal V}_{i}}
  .
  \label{eq:Euler-consistency-linear-vector-fluid}
\end{eqnarray}
Since, the background energy continuity equation
(\ref{eq:kouchan-19.41}) is consistent with the background
Einstein equation, 
Eqs.~(\ref{eq:Euler-consistency-linear-scalar-fluid}) and 
(\ref{eq:Euler-consistency-linear-vector-fluid}) show that the
first-order perturbations (\ref{eq:kouchan-19.82}) of the Euler
equation are consistent with the set of the background and the
first-order Einstein equations.


\subsection{Scalar field case}
\label{sec:Fist-order-perturbations-scalar-field}


Through
Eqs.~(\ref{eq:kouchan-15.32-linear})--(\ref{eq:kouchan-15.35-linear})
and Eqs.~(\ref{eq:kouchan-19.196})--(\ref{eq:kouchan-19.199}),
the linear-order Einstein equations are given as follows:  
For the scalar-mode, we have four equations:
\begin{eqnarray}
  \stackrel{(1)}{{}^{(s)}E_{(1)}}
  &:=&
                \partial_{\eta}^{2}\stackrel{(1)}{\Phi}
  + 2 \left(
    {\cal H}
    - \frac{\partial_{\eta}^{2}\varphi}{\partial_{\eta}\varphi}
  \right) \partial_{\eta}\stackrel{(1)}{\Phi}
  -             \Delta\stackrel{(1)}{\Phi}
  + 2 \left(
      \partial_{\eta}{\cal H}
    - {\cal H} \frac{\partial_{\eta}^{2}\varphi}{\partial_{\eta}\varphi}
    -          2  K
  \right) \stackrel{(1)}{\Phi}
  \nonumber\\
  &=&
  0
  ,
  \label{eq:kouchan-19.262}
  \\
  \stackrel{(1)}{{}^{(s)}E_{(2)}}
  &:=&
  \partial_{\eta} \stackrel{(1)}{\Phi}
  + {\cal H} \stackrel{(1)}{\Phi}
  - 4 \pi G \partial_{\eta}\varphi \varphi_{1}
  =
  0
  ,
  \label{eq:kouchan-19.263}
  \\
  \stackrel{(1)}{{}^{(s)}E_{(3)}}
  &:=& 
  \left(
    -             \partial_{\eta}^{2} 
    -          6  {\cal H} \partial_{\eta}
    +             \Delta
    -          2  \partial_{\eta}{\cal H}
    -          4  {\cal H}^{2}
    +          4  K
  \right) \stackrel{(1)}{\Phi}
  - 8 \pi G a^{2} \frac{\partial V}{\partial\varphi} \varphi_{1}
  = 
  0
  ,
  \label{eq:kouchan-19.264}
  \\
  \stackrel{(1)}{{}^{(s)}E_{(4)}}
  &:=&
  \stackrel{(1)}{\Psi} - \stackrel{(1)}{\Phi} = 0.
  \label{eq:kouchan-19.265}
\end{eqnarray}
For the vector-mode, we obtain 
\begin{eqnarray}
  \stackrel{(1)}{{}^{(s)}E_{(5)i}} :=
  \stackrel{(1)}{\nu^{i}} = 0.
  \label{eq:kouchan-19.266}
\end{eqnarray}
For the tensor-mode, we obtain 
\begin{eqnarray}
  \stackrel{(1)}{{}^{(s)}E_{(6)ij}}
  &:=&
  \left(
    \partial_{\eta}^{2}
    + 2 {\cal H} \partial_{\eta}
    + 2 K
    - \Delta
  \right) \stackrel{(1)}{\chi_{ij}}
  =
  0
  .
  \label{eq:kouchan-19.267}
\end{eqnarray}
These equations are also summarized in
KN2007\cite{kouchan-second-cosmo}.


In Eqs.~(\ref{eq:kouchan-19.262})--(\ref{eq:kouchan-19.264}) for
the scalar-modes, Eq.~(\ref{eq:kouchan-19.262}) determines the
evolution of the scalar potential $\stackrel{(1)}{\Phi}$ with
appropriate boundary conditions.
Equation (\ref{eq:kouchan-19.263}) determines the behavior of
the first-order perturbation $\varphi_{1}$ of the scalar field
through the scalar potential $\stackrel{(1)}{\Phi}$.
Therefore, Eq.~(\ref{eq:kouchan-19.264}) is not necessary
to solve this system.
However, we can easily see that Eq.~(\ref{eq:kouchan-19.264}) is
consistent with the set of two equations
(\ref{eq:kouchan-19.262}) and (\ref{eq:kouchan-19.263}).
Actually, from Eqs.~(\ref{eq:kouchan-19.262}),
(\ref{eq:kouchan-19.263}), and
(\ref{eq:background-Klein-Gordon-eq}),
Eq.~(\ref{eq:kouchan-19.264}) is given by 
\begin{eqnarray}
  \stackrel{(1)}{{}^{(s)}E_{(3)}}
  &=&
  - \stackrel{(1)}{{}^{(s)}E_{(1)}}
  - 2 \left(
        \frac{\partial_{\eta}^{2}\varphi}{\partial_{\eta}\varphi}
    + 2 {\cal H}
  \right) \stackrel{(1)}{{}^{(s)}E_{(2)}}
  +  8 \pi G \varphi_{1} a^{2} \stackrel{(0)}{C_{K}}
  .
  \label{eq:kouchan-19.264-consistency}
\end{eqnarray}
This shows that the Einstein equation (\ref{eq:kouchan-19.264})
is not independent of the set of the equations
Eqs.~(\ref{eq:kouchan-19.262}), (\ref{eq:kouchan-19.263}) and
the background Klein-Gordon equation
(\ref{eq:background-Klein-Gordon-eq-explicit}).
Since the background Klein-Gordon equation is derived from the
set of the background Einstein equations as in 
Eq.~(\ref{eq:background-Klein-Gordon-consistency}), the equation
(\ref{eq:kouchan-19.264}) is satisfied through the background
Einstein equations 
(\ref{eq:background-Einstein-equations-scalar-1}) and
(\ref{eq:background-Einstein-equations-scalar-2}), the
first-order perturbations (\ref{eq:kouchan-19.262}) and
(\ref{eq:kouchan-19.263}) of the Einstein equations.
Therefore to solve this system of the scalar-mode, we first
solve Eq.~(\ref{eq:kouchan-19.262}) with an appropriate initial
condition.
These initial conditions gives the first-order perturbation of
the scalar field $\varphi_{1}$ through the momentum constraint
(\ref{eq:kouchan-19.263}).
Further, after these initial states,
Eq.~(\ref{eq:kouchan-19.263}) gives the first-order perturbation
$\varphi_{1}$ of the scalar field in terms of the scalar
potential $\stackrel{(1)}{\Phi}$ at any time.
Thus, the free initial values for the linear-order perturbation
of scalar-modes are only $\stackrel{(1)}{\Phi}$ and
$\partial_{\eta}\stackrel{(1)}{\Phi}$ on the initial surface.
There is no other degree of freedom for initial value in the
scalar-mode.


\subsubsection{Klein-Gordon equation}
\label{sec:Fist-order-scalar-field-Klein-Gordon}


Here, we consider the first-order perturbation of the
Klein-Gordon equation which given in
KN2008\cite{kouchan-second-cosmo-matter}: 
\begin{eqnarray}
  a^{2} \stackrel{(1)}{{\cal C}_{(K)}}
  &=&
  -   \partial_{\eta}^{2}\varphi_{1}
  - 2 {\cal H} \partial_{\eta}\varphi_{1}
  +   \Delta\varphi_{1}
  +   \partial_{\eta}\stackrel{(1)}{\Phi} \partial_{\eta}\varphi
  + 3 \partial_{\eta}\stackrel{(1)}{\Psi} \partial_{\eta}\varphi
  \nonumber\\
  &&
  + 4 {\cal H} \stackrel{(1)}{\Phi} \partial_{\eta}\varphi
  + 2 \stackrel{(1)}{\Phi} \partial_{\eta}^{2}\varphi
  -   a^{2}\varphi_{1} \frac{\partial^{2}V}{\partial\bar{\varphi}^{2}}(\varphi)
  \label{eq:kouchan-17.806-first-explicit}
  \\
  &=&
    3 \partial_{\eta}\stackrel{(1)}{{}^{(s)}E_{(4)}} \partial_{\eta}\varphi
  - 2 a^{2} \stackrel{(1)}{\Phi} \stackrel{(0)}{C_{K}}
  \nonumber\\
  &&
  -   \partial_{\eta}^{2}\varphi_{1}
  - 2 {\cal H} \partial_{\eta}\varphi_{1}
  +   \Delta\varphi_{1}
  + 4 \partial_{\eta}\stackrel{(1)}{\Phi} \partial_{\eta}\varphi
  \nonumber\\
  && \quad\quad
  - 2 a^{2} \stackrel{(1)}{\Phi} \frac{\partial V}{\partial\varphi}
  -   a^{2} \varphi_{1} \frac{\partial^{2}V}{\partial\varphi^{2}}
  \label{eq:kouchan-19.270}
  \\
  &=&
  0
  \label{eq:kouchan-19.270-eq}
  ,
\end{eqnarray}
where we have used the component (\ref{eq:kouchan-19.265}) of
the linearized Einstein equation and the background Klein-Gordon
equations Eqs.~(\ref{eq:background-Klein-Gordon-eq}).


Now, we check the consistency of the first-order perturbation
(\ref{eq:kouchan-19.270}) of the Klein-Gordon equation with the
first-order perturbations
(\ref{eq:kouchan-19.262})--(\ref{eq:kouchan-19.264}) of the
Einstein equations.
We can easily derive the relation as follows:
\begin{eqnarray}
  &&
  8 \pi G (\partial_{\eta}\varphi) a^{2} {}^{(1)}\!{\cal C}_{(K)}
  \nonumber\\
  &=& 
  - 2 \left[
    \partial_{\eta} + {\cal H}
  \right] \stackrel{(1)}{{}^{(s)}E_{(1)}}
  \nonumber\\
  && 
  + 2 \left[
        \partial_{\eta}^{2}
    - 2 \left(
      \frac{\partial_{\eta}^{2}\varphi}{\partial_{\eta}\varphi} - {\cal H} 
    \right) \partial_{\eta}
    -   \Delta
    - 2 \frac{\partial_{\eta}^{3}\varphi}{\partial_{\eta}\varphi}
    - 2 \partial_{\eta}{\cal H}
    + 4 {\cal H}^{2}
  \right.
  \nonumber\\
  && \quad\quad\quad
  \left.
    + 2 \frac{\partial_{\eta}^{2}\varphi}{\partial_{\eta}\varphi} \left(
      \frac{\partial_{\eta}^{2}\varphi}{\partial_{\eta}\varphi}
      - {\cal H} 
    \right)
  \right] \stackrel{(1)}{{}^{(s)}E_{(2)}}
  \nonumber\\
  && 
  + \left[
    \left(
           \stackrel{(1)}{\Phi}
      - 3 {\cal H} \frac{\varphi_{1}}{\partial_{\eta}\varphi}
    \right) \partial_{\eta}
    + 2 \left(
      2 \partial_{\eta}\stackrel{(1)}{\Phi}
      - {\cal H} \stackrel{(1)}{\Phi}
    \right)
  \right.
  \nonumber\\
  && \quad\quad\quad
  \left.
    + 3 \frac{\varphi_{1}}{\partial_{\eta}\varphi} \left(
        2 {\cal H}^{2} 
      -   \partial_{\eta}{\cal H} 
      +   {\cal H} \frac{\partial_{\eta}^{2}\varphi}{\partial_{\eta}\varphi}
    \right)
  \right]
  \left(
        \stackrel{(0)}{{}^{(s)}E_{(2)}}
    - 3 \stackrel{(0)}{{}^{(s)}E_{(1)}}
  \right)
  \nonumber\\
  && 
  + 3 \left[
    \frac{\varphi_{1}}{\partial_{\eta}\varphi} \left\{
         \partial_{\eta}^{2}
      -  \left(
           \frac{\partial_{\eta}^{2}\varphi}{\partial_{\eta}\varphi}
        + 4 {\cal H} 
      \right) \partial_{\eta}
      - 2 \partial_{\eta}{\cal H}
      + 4 {\cal H}^{2}
      + 2 {\cal H} \frac{\partial_{\eta}^{2}\varphi}{\partial_{\eta}\varphi}
    \right\}
  \right.
  \nonumber\\
  && \quad\quad\quad
  \left.
    + 2 \stackrel{(1)}{\Phi} \left(
        \partial_{\eta}
      - 2 {\cal H}
    \right)
  \right] \stackrel{(0)}{{}^{(s)}E_{(1)}}
  \label{eq:kouchan-19.272}
  .
\end{eqnarray}
Through the background Einstein equations
(\ref{eq:background-Einstein-equations-scalar-1}),
(\ref{eq:background-Einstein-equations-scalar-2}), and the
first-order perturbations (\ref{eq:kouchan-19.262}) and
(\ref{eq:kouchan-19.263}) of the Einstein equation, we have seen
that 
\begin{eqnarray}
  {}^{(1)}\!{\cal C}_{(K)} = 0.
\end{eqnarray}
Hence, the first-order perturbation of the Klein-Gordon equation
is not independent equation of the background and the
first-order perturbation of the Einstein equation.
Therefore, from the point of view of the Cauchy problem, any
information which are obtained from the first-order
perturbation of the Klein-Gordon equation should be also
obtained from the set of the background Einstein equations and
the first-order perturbations of the Einstein equation, in
principle.


\section{Consistency of the second-order equations}
\label{sec:Consistency-of-the-second-order-perturbations}


Now, we consider the consistency of the second-order
perturbations of the Einstein equations
\begin{eqnarray}
  {}^{(1)}\!{\cal G}_{a}^{\;\;b}[{\cal L}]
  + {}^{(2)}\!{\cal G}_{a}^{\;\;b}[{\cal H},{\cal H}]
  = 8 \pi G {}^{(2)}\!{\cal T}_{a}^{\;\;b},
  \label{eq:second-order-Einstein-general}
\end{eqnarray}
and equations of motion for matter fields derived in
KN2008\cite{kouchan-second-cosmo-matter}. 
The essence of this consistency check is same as those for the
background equations and for the equations of the first-order 
perturbations as mentioned in the last section.
However, these consistency checks for the second-order
perturbations are necessary to guarantee that the derived
equations are correct, since the second-order Einstein equations
have complicated forms due to the quadratic terms of the
linear-order perturbations which arise from the non-linear
effects of the Einstein equations.


\subsection{Perfect fluid case}


First, we consider the second-order Einstein equation
(\ref{eq:second-order-Einstein-general}) in the case of a
perfect fluid. 
The components of 
${}^{(1)}\!{\cal G}_{a}^{\;\;b}\left[{\cal L}\right]$ are given
by
Eqs.~(\ref{eq:kouchan-15.32-linear})--(\ref{eq:kouchan-15.35-linear})
and the replacement
(\ref{eq:replacement-linear-to-second-order}) and the components
of the second term in the right hand side of 
Eq.~(\ref{eq:second-order-Einstein-general}) are given by 
Eqs.~(\ref{eq:generic-2-calG-eta-eta})--(\ref{eq:generic-2-calG-i-j}).
Further, the components of the right hand side of
Eq.~(\ref{eq:second-order-Einstein-general}) are given by
Eqs.~(\ref{eq:kouchan-19.34})--(\ref{eq:kouchan-19.37}). 
So, we can write down the components of the second-order
perturbations of the Einstein equation. 
The resulting equations are completely same as those obtained in
KN2007 except for the definitions of $\Gamma_{0}$, $\Gamma_{i}$, and
$\Gamma_{ij}$.
Therefore, the same calculations as those in
KN2007\cite{kouchan-second-cosmo} lead
\begin{eqnarray}
  \stackrel{(2)}{{}^{(p)}E_{(1)}}
  &:=&
  \left(
    - 3 {\cal H} \partial_{\eta}
    +   \Delta
    + 3 K
  \right) \stackrel{(2)}{\Psi}
  - 3 {\cal H}^{2} \stackrel{(2)}{\Phi}
  - 4\pi G a^{2} \stackrel{(2)}{{\cal E}}
  - \Gamma_{0}
  \nonumber\\
  &=& 0
  ,
  \label{eq:kouchan-19.120}
  \\
  \stackrel{(2)}{{}^{(p)}E_{(2)}}
  &:=&
  \left(
                  \partial_{\eta}^{2} 
    +          2  {\cal H} \partial_{\eta}
    -             K
    - \frac{1}{3} \Delta
  \right)
  \stackrel{(2)}{\Psi}
  + \left(
                  {\cal H} \partial_{\eta}
    +          2  \partial_{\eta}{\cal H}
    +             {\cal H}^{2}
    + \frac{1}{3} \Delta
  \right)
  \stackrel{(2)}{\Phi}
  \nonumber\\
  && \quad\quad
  - 4 \pi G a^{2} \stackrel{(2)}{{\cal P}}
  - \frac{1}{6} \Gamma_{k}^{\;\;k}
  \nonumber\\
  &=& 0
  ,
  \label{eq:kouchan-19.121}
  \\
  \stackrel{(2)}{{}^{(p)}E_{(3)}}
  &:=&
  \stackrel{(2)}{\Psi} - \stackrel{(2)}{\Phi}
  -
  \frac{3}{2}
  \left( \Delta + 3 K \right)^{-1}
  \left(
    \Delta^{-1} D^{i}D^{j}\Gamma_{ij}
    - \frac{1}{3} \Gamma_{k}^{\;\;k}
  \right)
  \nonumber\\
  &=& 0
  ,
  \label{eq:kouchan-19.122}
  \\
  \stackrel{(2)}{{}^{(p)}E_{(4)i}}
  &:=&
                \partial_{\eta}D_{i}\stackrel{(2)}{\Psi}
  +             {\cal H} D_{i}\stackrel{(2)}{\Phi}
  - \frac{1}{2} D_{i}\Delta^{-1}D^{k}\Gamma_{k}
  +          4  \pi G a^{2} (\epsilon + p) D_{i}\stackrel{(2)}{v} 
  \nonumber\\
  &=& 0
  ,
  \label{eq:kouchan-19.125}
  \\
  \stackrel{(2)}{{}^{(p)}E_{(5)i}}
  &:=&
  \left( \Delta + 2 K \right) \stackrel{(2)}{\nu_{i}}
  + 2 \left(\Gamma_{i} - D_{i}\Delta^{-1}D^{k}\Gamma_{k}\right)
  - 16 \pi G a^{2} (\epsilon + p) \stackrel{(2)}{{\cal V}_{i}}
  \nonumber\\
  &=& 0
  ,
  \label{eq:kouchan-19.126}
  \\
  \stackrel{(2)}{{}^{(p)}E_{(6)i}}
  &:=&
  \partial_{\eta}\left(
    a^{2}  \stackrel{(2)}{\nu_{i}}
  \right)
  - 2 a^{2} \left(\Delta+2K\right)^{-1} \left\{
    D_{i}\Delta^{-1}D^{k}D^{l}\Gamma_{kl} - D^{k}\Gamma_{ik}
  \right\}
  \nonumber\\
  &=& 0
  ,
  \label{eq:kouchan-19.123}
  \\
  \stackrel{(2)}{{}^{(p)}E_{(7)ij}}
  &:=&
  \left(
    \partial_{\eta}^{2}
    + 2 {\cal H} \partial_{\eta}
    + 2 K
    - \Delta
  \right) \stackrel{(2)}{\chi_{ij}}
  - 2 \Gamma_{ij}
  + \frac{2}{3} \gamma_{ij} \Gamma_{k}^{\;\;k}
  \nonumber\\
  &&
  + 3
  \left(
    D_{i}D_{j} - \frac{1}{3} \gamma_{ij} \Delta
  \right) 
  \left( \Delta + 3 K \right)^{-1}
  \left(
    \Delta^{-1} D^{k}D^{l}\Gamma_{kl}
    - \frac{1}{3} \Gamma_{k}^{\;\;k}
  \right)
  \nonumber\\
  &&
  - 4
  \left( 
    D_{(i}\left(\Delta+2K\right)^{-1}D_{j)}\Delta^{-1}D^{l}D^{k}\Gamma_{lk}
    - D_{(i}\left(\Delta+2K\right)^{-1}D^{k}\Gamma_{j)k}
  \right)
  \nonumber\\
  &=& 0
  .
  \label{eq:kouchan-19.124}
\end{eqnarray}
Here, the definitions of $\Gamma_{0}$, $\Gamma_{i}$, and
$\Gamma_{ij}$ are as follows:
\begin{eqnarray}
  \Gamma_{0}
  &:=&
     8 \pi G a^{2} \left(\epsilon+p\right) D_{i}\stackrel{(1)}{v} D^{i}\stackrel{(1)}{v} 
  -  3 D_{k}\stackrel{(1)}{\Phi} D^{k}\stackrel{(1)}{\Phi}
  -  8 \stackrel{(1)}{\Phi} \Delta\stackrel{(1)}{\Phi}
  -  3 \left(\partial_{\eta}\stackrel{(1)}{\Phi}\right)^{2}
  \nonumber\\
  && \quad\quad
  - 12 K \left(\stackrel{(1)}{\Phi}\right)^{2}
  - 12 {\cal H}^{2} \left(\stackrel{(1)}{\Phi}\right)^{2}
  \nonumber\\
  && 
  - 4 \left(
    \partial_{\eta}D_{i}\stackrel{(1)}{\Phi}+{\cal H} D_{i}\stackrel{(1)}{\Phi}
  \right) \stackrel{(1)}{{\cal V}^{i}} 
  - 2 {\cal H} D_{k}\stackrel{(1)}{\Phi} \stackrel{(1)}{\nu^{k}}
  \nonumber\\
  && 
  + 8  \pi G a^{2} \left(\epsilon+p\right) \stackrel{(1)}{{\cal V}_{i}} \stackrel{(1)}{{\cal V}^{i}}
  + \frac{1}{2} D_{k}\stackrel{(1)}{\nu_{l}} D^{(k}\stackrel{(1)}{\nu^{l)}}
  + 3  {\cal H}^{2} \stackrel{(1)}{\nu^{k}} \stackrel{(1)}{\nu_{k}}
  \nonumber\\
  && 
  +             D_{l}D_{k}\stackrel{(1)}{\Phi} \stackrel{(1)}{\chi^{lk}}
  \nonumber\\
  && 
  - 2 {\cal H} D^{k}\stackrel{(1)}{\nu^{l}} \stackrel{(1)}{\chi_{kl}}
  -\frac{1}{2}D^{k}\stackrel{(1)}{\nu^{l}}\partial_{\eta}\stackrel{(1)}{\chi_{lk}}
  \nonumber\\
  && 
  + \frac{1}{8} \partial_{\eta}\stackrel{(1)}{\chi_{lk}} \partial_{\eta}\stackrel{(1)}{\chi^{kl}}
  + {\cal H} \stackrel{(1)}{\chi_{kl}} \partial_{\eta}\stackrel{(1)}{\chi^{lk}}
  - \frac{1}{8} D_{k}\stackrel{(1)}{\chi_{lm}} D^{k}\stackrel{(1)}{\chi^{ml}}
  \nonumber\\
  && \quad\quad
  + \frac{1}{2} D_{k}\stackrel{(1)}{\chi_{lm}} D^{[l}\stackrel{(1)}{\chi^{k]m}}
  - \frac{1}{2} \stackrel{(1)}{\chi^{lm}} \left(\Delta-K\right)\stackrel{(1)}{\chi_{lm}}
  ,
  \label{eq:kouchan-19.117}
  \\
  \Gamma_{i}
  &:=&
  - 16 \pi G a^{2} \left(
    \stackrel{(1)}{{\cal E}} + \stackrel{(1)}{{\cal P}} 
  \right)
  D_{i}\stackrel{(1)}{v}
  +         12  {\cal H} \stackrel{(1)}{\Phi} D_{i}\stackrel{(1)}{\Phi}
  \nonumber\\
  && \quad\quad
  -          4  \stackrel{(1)}{\Phi} \partial_{\eta}D_{i}\stackrel{(1)}{\Phi}
  -          4  \partial_{\eta}\stackrel{(1)}{\Phi} D_{i}\stackrel{(1)}{\Phi}
  \nonumber\\
  &&
  - 16 \pi G a^{2} \left(
    \stackrel{(1)}{{\cal E}} + \stackrel{(1)}{{\cal P}} 
  \right)
  \stackrel{(1)}{{\cal V}_{i}}
  -          2  D^{j}\stackrel{(1)}{\Phi} D_{i}\stackrel{(1)}{\nu_{j}}
  +          2  D_{i}D^{j}\stackrel{(1)}{\Phi} \stackrel{(1)}{\nu_{j}}
  +          2  \Delta\stackrel{(1)}{\Phi} \stackrel{(1)}{\nu_{i}}
  \nonumber\\
  && \quad\quad
  +             \stackrel{(1)}{\Phi} \Delta\stackrel{(1)}{\nu_{i}}
  +          2  K \stackrel{(1)}{\Phi} \stackrel{(1)}{\nu_{i}}
  \nonumber\\
  &&
  -          4  {\cal H} \stackrel{(1)}{\nu^{j}} D_{i}\stackrel{(1)}{\nu_{j}}
  \nonumber\\
  &&
  +          2  D^{j}\stackrel{(1)}{\Phi} \partial_{\eta}\stackrel{(1)}{\chi_{ji}}
  -          2  \partial_{\eta}D^{j}\stackrel{(1)}{\Psi} \stackrel{(1)}{\chi_{ij}}
  \nonumber\\
  &&
  +          2  D_{k}D_{[i}\stackrel{(1)}{\nu_{m]}} \stackrel{(1)}{\chi^{km}}
  +          2  D^{[k}\stackrel{(1)}{\nu^{j]}} D_{j}\stackrel{(1)}{\chi_{ki}}
  +          2  K \stackrel{(1)}{\nu^{j}} \stackrel{(1)}{\chi_{ij}}
  -             \stackrel{(1)}{\nu^{j}} \Delta\stackrel{(1)}{\chi_{ji}}
  \nonumber\\
  &&
  - \frac{1}{2} \partial_{\eta}\stackrel{(1)}{\chi^{jk}} D_{i}\stackrel{(1)}{\chi_{kj}}
  +          2  \stackrel{(1)}{\chi^{kj}} \partial_{\eta}D_{[j}\stackrel{(1)}{\chi_{i]k}}
  ,
  \label{eq:kouchan-19.118}
  \\
  \Gamma_{ij}
  &:=&
    16 \pi G a^{2} \left( \epsilon + p \right) D_{i}\stackrel{(1)}{v} D_{j}\stackrel{(1)}{v}
  -  4 D_{i}\stackrel{(1)}{\Phi} D_{j}\stackrel{(1)}{\Phi}
  -  8 \stackrel{(1)}{\Phi} D_{i}D_{j}\stackrel{(1)}{\Phi}
  \nonumber\\
  && \quad\quad
  +  2 \left\{
       3 D_{k}\stackrel{(1)}{\Phi} D^{k}\stackrel{(1)}{\Phi}
    +  4 \stackrel{(1)}{\Phi} \Delta\stackrel{(1)}{\Phi}
    +    \left(\partial_{\eta}\stackrel{(1)}{\Phi}\right)^{2}
    +  8 {\cal H} \stackrel{(1)}{\Phi} \partial_{\eta}\stackrel{(1)}{\Phi}
  \right.
  \nonumber\\
  && \quad\quad\quad\quad\quad
  \left.
    + 4 \left(
      2 \partial_{\eta}{\cal H} + K + {\cal H}^{2}
    \right)
    \left(\stackrel{(1)}{\Phi}\right)^{2}
  \right\} \gamma_{ij}
  \nonumber\\
  && 
  + 32 \pi G a^{2} \left( \epsilon + p \right) D_{(i}\stackrel{(1)}{v} \stackrel{(1)}{{\cal V}_{j)}}
  -  4 \partial_{\eta}\stackrel{(1)}{\Phi} D_{(i}\stackrel{(1)}{\nu_{j)}}
  +  4 \partial_{\eta}D_{(i}\stackrel{(1)}{\Phi} \stackrel{(1)}{\nu_{j)}}
  \nonumber\\
  && \quad\quad
  + 4 \left(
       \partial_{\eta}D_{k}\stackrel{(1)}{\Phi} \stackrel{(1)}{\nu^{k}}
    +  {\cal H} D_{k}\stackrel{(1)}{\Phi} \stackrel{(1)}{\nu^{k}}
  \right) \gamma_{ij}
  \nonumber\\
  && 
  + 16 \pi G a^{2} \left( \epsilon + p \right) \stackrel{(1)}{{\cal V}_{i}} \stackrel{(1)}{{\cal V}_{j}}
  -  2 \stackrel{(1)}{\nu^{k}} D_{k}D_{(i}\stackrel{(1)}{\nu_{j)}}
  +  2 \stackrel{(1)}{\nu_{k}} D_{i}D_{j}\stackrel{(1)}{\nu^{k}}
  \nonumber\\
  && \quad\quad
  +    D_{i}\stackrel{(1)}{\nu^{k}} D_{j}\stackrel{(1)}{\nu_{k}}
  +    D^{k}\stackrel{(1)}{\nu_{i}} D_{k}\stackrel{(1)}{\nu_{j}}
  \nonumber\\
  && \quad\quad
  - \left\{
         D_{k}\stackrel{(1)}{\nu_{l}} D^{[k}\stackrel{(1)}{\nu^{l]}}
    +    D_{k}\stackrel{(1)}{\nu_{l}} D^{k}\stackrel{(1)}{\nu^{l}}
    +  2 \stackrel{(1)}{\nu^{k}} \left(
      \Delta
      +  2 \partial_{\eta}{\cal H}
      -  3 {\cal H}^{2}
    \right) \stackrel{(1)}{\nu_{k}}
  \right\} \gamma_{ij}
  \nonumber\\
  && 
  -  4 {\cal H} \partial_{\eta}\stackrel{(1)}{\Phi} \stackrel{(1)}{\chi_{ij}}
  -  2 \partial_{\eta}^{2}\stackrel{(1)}{\Phi} \stackrel{(1)}{\chi_{ij}}
  -  4 D^{k}\stackrel{(1)}{\Phi} D_{(i}\stackrel{(1)}{\chi_{j)k}}
  +  4 D^{k}\stackrel{(1)}{\Phi} D_{k}\stackrel{(1)}{\chi_{ij}}
  -  8 K \stackrel{(1)}{\Phi} \stackrel{(1)}{\chi_{ij}}
  \nonumber\\
  && \quad\quad
  +  4 \stackrel{(1)}{\Phi} \Delta\stackrel{(1)}{\chi_{ij}}
  -  4 D^{k}D_{(i}\stackrel{(1)}{\Phi} \stackrel{(1)}{\chi_{j)k}}
  +  2 \Delta \stackrel{(1)}{\Phi} \stackrel{(1)}{\chi_{ij}}
  +  2 D_{l}D_{k}\stackrel{(1)}{\Phi} \stackrel{(1)}{\chi^{lk}} \gamma_{ij}
  \nonumber\\
  && 
  -  2 D^{k}\stackrel{(1)}{\nu_{(i}} \partial_{\eta}\stackrel{(1)}{\chi_{j)k}}
  -  2 \stackrel{(1)}{\nu^{k}} \partial_{\eta}D_{(i}\stackrel{(1)}{\chi_{j)k}}
  +  2 \stackrel{(1)}{\nu^{k}} \partial_{\eta}D_{k}\stackrel{(1)}{\chi_{ij}}
  +    D^{k}\stackrel{(1)}{\nu^{l}} \partial_{\eta}\stackrel{(1)}{\chi_{lk}} \gamma_{ij}
  \nonumber\\
  && 
  +             \partial_{\eta}\stackrel{(1)}{\chi_{ik}} \partial_{\eta}\stackrel{(1)}{\chi_{j}^{\;\;k}}
  +          2  D_{[l}\stackrel{(1)}{\chi_{k]i}} D^{k}\stackrel{(1)}{\chi_{j}^{\;\;l}}
  - \frac{1}{2} D_{j}\stackrel{(1)}{\chi_{lk}} D_{i}\stackrel{(1)}{\chi^{lk}}
  \nonumber\\
  && \quad\quad
  -             \stackrel{(1)}{\chi^{lm}} D_{i}D_{j}\stackrel{(1)}{\chi_{ml}}
  +          2  \stackrel{(1)}{\chi^{lm}} D_{l}D_{(i}\stackrel{(1)}{\chi_{j)m}}
  -             \stackrel{(1)}{\chi^{lm}} D_{m}D_{l}\stackrel{(1)}{\chi_{ij}}
  \nonumber\\
  && \quad\quad
  - \frac{1}{4} \left(
       3 \partial_{\eta}\stackrel{(1)}{\chi_{lk}} \partial_{\eta}\stackrel{(1)}{\chi^{kl}}
    - 3 D_{k}\stackrel{(1)}{\chi_{lm}} D^{k}\stackrel{(1)}{\chi^{ml}}
  \right.
  \nonumber\\
  && \quad\quad\quad\quad\quad
  \left.
    + 2 D_{k}\stackrel{(1)}{\chi_{lm}} D^{l}\stackrel{(1)}{\chi^{mk}}
    - 4 K \stackrel{(1)}{\chi_{lm}} \stackrel{(1)}{\chi^{lm}}
  \right) \gamma_{ij}
  \label{eq:kouchan-19.119}
  ,
\end{eqnarray}
and we denote $\Gamma_{i}^{\;\;j} = \gamma^{jk}\Gamma_{ik}$.
The definitions
(\ref{eq:kouchan-19.117})--(\ref{eq:kouchan-19.119}) of the
source terms $\Gamma_{0}$, $\Gamma_{i}$, and $\Gamma_{ij}$ in
the second-order perturbations of the Einstein equations
represent the precise mode-coupling:
For example, the first two lines in
Eq.~(\ref{eq:kouchan-19.117}) represent the scalar-scalar
mode-coupling; the third line in Eq.~(\ref{eq:kouchan-19.117}) 
represents the scalar-vector mode-coupling; the fourth line in
Eq.~(\ref{eq:kouchan-19.117}) represents the vector-vector
mode-coupling; the fifth line in Eq.~(\ref{eq:kouchan-19.117})
shows the scalar-tensor mode-coupling; the sixth line in
Eq.~(\ref{eq:kouchan-19.117}) is the vector-tensor mode
coupling; and the last two lines in
Eq.~(\ref{eq:kouchan-19.117}) corresponds to the tensor-tensor
mode coupling.
Thus, in the second-order perturbations of the Einstein
equations, any types of mode-coupling appear due to the
non-linear effects of the Einstein equations.
Actually, the definitions
(\ref{eq:kouchan-19.117})--(\ref{eq:kouchan-19.119}) of the
source terms $\Gamma_{0}$, $\Gamma_{i}$, and $\Gamma_{ij}$ do
shows these any types of mode-coupling occur in
the second-order perturbations of the Einstein equations.


\subsubsection{Consistency with the continuity equation}
\label{sec:Consystency-with-the-continuity-equation-second}


Now, we consider the second-order perturbation of the energy
continuity equation which is the second-order perturbation of
$\bar{u}^{a}\bar{\nabla}^{b}\bar{T}_{a}^{\;\;b} = 0$.
As shown in KN2008\cite{kouchan-second-cosmo-matter}, the
second-order perturbation of the energy continuity equation is
given by 
\begin{eqnarray}
  a {}^{(2)}\!{\cal C}_{0}^{(p)}
  &:=&
      \partial_{\eta}\stackrel{(2)}{{\cal E}} 
  + 3 {\cal H}
  \left(
      \stackrel{(2)}{{\cal E}} 
    + \stackrel{(2)}{{\cal P}}
  \right)
  + \left(\epsilon + p\right) \left(
         \Delta \stackrel{(2)}{v}
    -  3 \partial_{\eta}\stackrel{(2)}{\Psi}
  \right)
  - \Xi_{0}
  \nonumber\\
  &=& 0
  \label{eq:kouchan-19.132}
  .
\end{eqnarray}
Here, $\Xi_{0}$ consists of the quadratic terms of the linear
order perturbations defined by 
\begin{eqnarray}
  \Xi_{0}
  &:=&
  2 \left[
       6 \stackrel{(1)}{\Psi} \partial_{\eta}\stackrel{(1)}{\Psi}
    -  2 \stackrel{(1)}{\Psi} \Delta\stackrel{(1)}{v}
    -    \stackrel{(1)}{\Phi} \Delta\stackrel{(1)}{v} 
    -    \stackrel{(1)}{\nu^{k}} D_{k}\stackrel{(1)}{\Psi}
  \right.
  \nonumber\\
  && \quad\quad\quad\quad
  \left.
    +   \left(
        D^{k}\stackrel{(1)}{v}
      + \stackrel{(1)}{{\cal V}^{k}}
    \right)
    \left(
          D_{k}\stackrel{(1)}{\Psi}
      -   D_{k}\stackrel{(1)}{\Phi}
      -   \partial_{\eta}D_{k}\stackrel{(1)}{v}
      -   \partial_{\eta}\stackrel{(1)}{{\cal V}_{k}}
    \right)
  \right.
  \nonumber\\
  && \quad\quad\quad\quad
  \left.
    +   \stackrel{(1)}{\chi^{ik}}
    \left(
                    D_{i}D_{k}\stackrel{(1)}{v} 
      +             D_{i}\stackrel{(1)}{{\cal V}_{k}}
      -             D_{i}\stackrel{(1)}{\nu_{k}}
      + \frac{1}{2} \partial_{\eta}\stackrel{(1)}{\chi_{ik}}
    \right)
  \right]
  \left(\epsilon + p\right)
  \nonumber\\
  && \quad
  - 2 \left(
      D^{i}\stackrel{(1)}{v}
    + \stackrel{(1)}{{\cal V}^{i}}
    - \stackrel{(1)}{\nu^{i}}
  \right)
  D_{i}\stackrel{(1)}{{\cal E}} 
  - 2 \left(
                  \Delta\stackrel{(1)}{v} 
    -          3  \partial_{\eta}\stackrel{(1)}{\Psi}
  \right)
  \left(
    \stackrel{(1)}{{\cal E}} + \stackrel{(1)}{{\cal P}}
  \right)
  .
  \label{eq:kouchan-19.133}
\end{eqnarray}


To confirm the consistency of the background and the
perturbations of the Einstein equation and the energy continuity
equation (\ref{eq:kouchan-19.132}), we first substitute the
second-order Einstein equations
(\ref{eq:kouchan-19.120})--(\ref{eq:kouchan-19.125}) into
Eq.~(\ref{eq:kouchan-19.132}) as in the case of the first-order
perturbations in
\S\ref{sec:Fist-order-perturbations-perfect-fluid-continuity-Euler}.
For simplicity, we first impose Eq.~(\ref{eq:kouchan-19.66}) on
all equations.
Then, we obtain
\begin{eqnarray}
  4 \pi G a^{3} {}^{(2)}\!{\cal C}_{0}^{(p)}
  &=&
  -             \partial_{\eta}\stackrel{(2)}{{}^{(p)}E_{(1)}}
  -             {\cal H} \stackrel{(2)}{{}^{(p)}E_{(1)}}
  -          3  {\cal H} \stackrel{(2)}{{}^{(p)}E_{(2)}}
  +             D^{i}\stackrel{(2)}{{}^{(p)}E_{(4)i}}
  \nonumber\\
  &&
  + \frac{3}{2} \left(
    3 \stackrel{(0)}{{}^{(p)}E_{(1)}}
    - \stackrel{(0)}{{}^{(p)}E_{(2)}}
  \right) \partial_{\eta}\stackrel{(2)}{\Psi}
  \nonumber\\
  &&
  -             \partial_{\eta}\Gamma_{0}
  -             {\cal H} \Gamma_{0}
  - \frac{1}{2} {\cal H} \Gamma_{k}^{\;\;k}
  + \frac{1}{2} D^{k}\Gamma_{k}
  - 4 \pi G a^{2} \Xi_{0}
  .
\end{eqnarray}
Imposing the second-order Einstein equations
(\ref{eq:kouchan-19.120}), (\ref{eq:kouchan-19.121}),
(\ref{eq:kouchan-19.125}), and the background Einstein equation
(\ref{eq:background-Einstein-equations-3}), we see that the
second-order perturbation (\ref{eq:kouchan-19.132}) of the
energy continuity equation is consistent with the second-order
and the background Einstein equations if the equation 
\begin{eqnarray}
    4 \pi G a^{2} \Xi_{0}
  +   \left(
    \partial_{\eta}
    +   {\cal H}
  \right) \Gamma_{0}
  + \frac{1}{2} {\cal H} \Gamma_{k}^{\;\;k}
  - \frac{1}{2} D^{k}\Gamma_{k}
  &=&
  0
  \label{eq:kouchan-19.135}
\end{eqnarray}
is satisfied under the background, the first-order Einstein
equations.
Actually, through the background Einstein equation
(\ref{eq:background-Einstein-equations-3}) (or equivalently
Eqs.~(\ref{eq:background-Einstein-equation-1}) and
(\ref{eq:background-Einstein-equation-2})) and the first-order
perturbations of the Einstein equations
(\ref{eq:kouchan-19.71})--(\ref{eq:kouchan-19.68}), we can
easily see that
\begin{eqnarray}
  &&
    4 \pi G a^{2} \Xi_{0}
  + \partial_{\eta}\Gamma_{0}
  + {\cal H} \Gamma_{0}
  + \frac{1}{2} {\cal H} \Gamma_{k}^{\;\;k}
  - \frac{1}{2} D^{k}\Gamma_{k}
  \nonumber\\
  &=&
  - \left[
     6  \stackrel{(1)}{\Phi} \partial_{\eta}\stackrel{(1)}{\Phi}
    + D^{i}\stackrel{(1)}{\Phi} D_{i}\stackrel{(1)}{v}
  \right]
  \left(
    3 \stackrel{(0)}{{}^{(p)}E_{(1)}}
    - \stackrel{(0)}{{}^{(p)}E_{(2)}}
  \right) 
  \nonumber\\
  && \quad\quad
  + 2 \left(
    D^{i}\stackrel{(1)}{v} \partial_{\eta}
    + 2 {\cal H} D^{i}\stackrel{(1)}{v} 
    - D^{i}\stackrel{(1)}{\Phi}
    - 3 \stackrel{(1)}{\Phi} D^{i}
  \right)\stackrel{(1)}{{}^{(p)}E_{(4)i}}
  \nonumber\\
  && \quad\quad
  - 6 \partial_{\eta}\stackrel{(1)}{\Phi} \stackrel{(1)}{{}^{(p)}E_{(1)}}
  - 2 \left(
    3 \partial_{\eta}\stackrel{(1)}{\Phi}
    +   D_{i}\stackrel{(1)}{v} D^{i}
  \right) \stackrel{(1)}{{}^{(p)}E_{(2)}}
  \nonumber\\
  &&
  + D_{i}\stackrel{(1)}{\Phi} \left(
    \stackrel{(1)}{\nu^{i}}
    - \stackrel{(1)}{{\cal V}^{i}}
  \right)
  \left(
    3 \stackrel{(0)}{{}^{(p)}E_{(1)}}
    - \stackrel{(0)}{{}^{(p)}E_{(2)}}
  \right)
  - 2 \stackrel{(1)}{\nu^{i}} D_{i}\stackrel{(1)}{{}^{(p)}E_{(1)}}
  - 2 \stackrel{(1)}{{\cal V}^{i}} D_{i}\stackrel{(1)}{{}^{(p)}E_{(2)}}
  \nonumber\\
  && \quad\quad
  + 2 \stackrel{(1)}{{\cal V}^{i}} \left(
    \partial_{\eta}  + 2 {\cal H}
  \right) \stackrel{(1)}{{}^{(p)}E_{(4)i}}
  - \frac{1}{2} \left[
    D^{i}\stackrel{(1)}{v} \left(
      \partial_{\eta} + 2 {\cal H}
    \right)
    - D^{i}\stackrel{(1)}{\Phi}
  \right] \stackrel{(1)}{{}^{(p)}E_{(5)i}}
  \nonumber\\
  && \quad\quad
  + \frac{1}{2a^{2}} \left[
    D^{k}\stackrel{(1)}{v} \left( \Delta + 2 K \right)
    - 4 {\cal H} D^{k}\stackrel{(1)}{\Phi}
  \right]\stackrel{(1)}{{}^{(p)}E_{(6)k}}
  \nonumber\\
  &&
  - \frac{1}{2} \stackrel{(1)}{{\cal V}^{i}} \left(\partial_{\eta} + 2 {\cal H}\right)\stackrel{(1)}{{}^{(p)}E_{(5)i}}
  + \frac{1}{2a^{2}} \left[
         \stackrel{(1)}{{\cal V}^{k}} \left(\Delta + 2 K \right)
    +  2 D^{(l}\stackrel{(1)}{\nu^{k)}} D_{l}
    + 12 {\cal H}^{2} \stackrel{(1)}{\nu^{k}}
  \right] \stackrel{(1)}{{}^{(p)}E_{(6)k}}
  \nonumber\\
  &&
  +          2  \stackrel{(1)}{\chi^{ik}} D_{k}\stackrel{(1)}{{}^{(p)}E_{(4)i}}
  \nonumber\\
  &&
  + \left(
    3 \stackrel{(0)}{{}^{(p)}E_{(1)}} - \stackrel{(0)}{{}^{(p)}E_{(2)}}
  \right) D_{i}\stackrel{(1)}{\nu_{k}} \stackrel{(1)}{\chi^{ik}}
  - \frac{1}{2} \stackrel{(1)}{\chi^{ik}} D_{i}\stackrel{(1)}{{}^{(p)}E_{(5)k}}
  \nonumber\\
  && \quad\quad
  - \frac{1}{2a^{2}} \left(
    \partial_{\eta} + 4 {\cal H}
  \right)\stackrel{(1)}{\chi^{lk}} D_{k}\stackrel{(1)}{{}^{(p)}E_{(6)l}}
  - \frac{1}{2} D^{k}\stackrel{(1)}{\nu^{l}} \stackrel{(1)}{{}^{(p)}E_{(7)lk}}
  \nonumber\\
  &&
  - \frac{1}{2} \stackrel{(1)}{\chi^{ik}} \partial_{\eta}\stackrel{(1)}{\chi_{ik}} \left(
    3 \stackrel{(0)}{{}^{(p)}E_{(1)}}
    - \stackrel{(0)}{{}^{(p)}E_{(2)}}
  \right)
  + \frac{1}{4} \left(
    \partial_{\eta} + 4 {\cal H} 
  \right) \stackrel{(1)}{\chi^{lk}} \stackrel{(1)}{{}^{(p)}E_{(7)lk}}
  \label{eq:kouchan-19.140}
  .
\end{eqnarray}
This shows that Eq.~(\ref{eq:kouchan-19.135}) is satisfied under
the Einstein equations of the background and of the first order.
Thus, this implies that the second-order perturbation
(\ref{eq:kouchan-19.132}) of the energy continuity equation
together with the source term (\ref{eq:kouchan-19.133}) is
consistent with the Einstein equations of the background, the
first- and the second-order, i.e.,
Eqs.~(\ref{eq:background-Einstein-equations-3}),
(\ref{eq:kouchan-19.71})--(\ref{eq:kouchan-19.68}), 
(\ref{eq:kouchan-19.120}), (\ref{eq:kouchan-19.121}), and
(\ref{eq:kouchan-19.125}).


\subsubsection{Consistency with the Euler equation}
\label{sec:Consystency-with-the-continuity-equation}


Next, we consider the second-order perturbations of the Euler
equations.
For simplicity, we first impose Eq.~(\ref{eq:kouchan-19.66}) on
all equations, again.
As shown in KN2008\cite{kouchan-second-cosmo-matter}, the
second-order perturbation of the Euler equation for a single
perfect fluid is given in terms of gauge-invariant form as 
\begin{eqnarray}
  {}^{(2)}\!{\cal C}_{i}^{(p)}
  &=&
  \left( \epsilon + p \right) \left\{
    \left(
      \partial_{\eta} + {\cal H}
    \right)
    \left(
      D_{i}\stackrel{(2)}{v}
      + \stackrel{(2)}{{\cal V}_{i}}
    \right)
    + D_{i}\stackrel{(2)}{\Phi}
  \right\}
  \nonumber\\
  &&
  + D_{i}\stackrel{(2)}{{\cal P}}
  + \partial_{\eta}p \left(
      D_{i}\stackrel{(2)}{v}
    + \stackrel{(2)}{{\cal V}_{i}}
  \right)
  - \Xi_{i}^{(p)}
  = 0,
  \label{eq:kouchan-17.680}
\end{eqnarray}
where $\Xi_{i}^{(p)}$ is defined by 
\begin{eqnarray}
  \Xi_{i}^{(p)}
  &:=&
  - 2 \stackrel{(1)}{\Phi}
  D_{i}\left\{
      \stackrel{(1)}{{\cal P}}
    - \left( \epsilon + p \right) \stackrel{(1)}{\Phi}
  \right\}
  \nonumber\\
  &&
  - 2 \left( \epsilon + p \right) \left(
      \stackrel{(1)}{\nu^{j}}
    - D^{j}\stackrel{(1)}{v} 
    - \stackrel{(1)}{{\cal V}^{j}}
  \right)
  \left\{
      D_{i}\stackrel{(1)}{\nu_{j}}
    - D_{j}\left(
        D_{i}\stackrel{(1)}{v} 
      + \stackrel{(1)}{{\cal V}_{i}}
    \right)
  \right\}
  \nonumber\\
  &&
  - 2 \left(
      \stackrel{(1)}{{\cal E}}
    + \stackrel{(1)}{{\cal P}}
  \right)
  \left\{
      D_{i}\stackrel{(1)}{\Phi}
    + \partial_{\eta}\left(
      D_{i}\stackrel{(1)}{v} 
      + \stackrel{(1)}{{\cal V}_{i}}
    \right)
    + {\cal H} \left(
      D_{i}\stackrel{(1)}{v}
      + \stackrel{(1)}{{\cal V}_{i}}
    \right)
  \right\}
  \nonumber\\
  &&
  - 2 \left(
      D_{i}\stackrel{(1)}{v}
    + \stackrel{(1)}{{\cal V}_{i}}
  \right) \partial_{\eta}\stackrel{(1)}{{\cal P}}
  \label{eq:kouchan-17.682}
  .
\end{eqnarray}
This $\Xi_{i}^{(p)}$ is the collection of the quadratic terms of
the linear-order perturbations in the second-order perturbation
of the Euler equation.
As in the case of the first-order perturbations of the Euler
equation, the equation (\ref{eq:kouchan-17.680}) is decomposed
into the scalar- and the vector-parts as 
\begin{eqnarray}
  {}^{(2)}\!{\cal C}_{i}^{(pS)}
  &:=&
  D_{i}\Delta^{-1}D^{j}{}^{(2)}\!{\cal C}_{j}^{(p)}
  \nonumber\\
  &=&
  \left( \epsilon + p \right) \left\{
    \left( \partial_{\eta} + {\cal H} \right) D_{i}\stackrel{(2)}{v}
    + D_{i}\stackrel{(2)}{\Phi}
  \right\}
  + D_{i}\stackrel{(2)}{{\cal P}}
  + \partial_{\eta}p D_{i}\stackrel{(2)}{v}
  \nonumber\\
  &&
  - D_{i} \Delta^{-1} D^{j}\Xi_{j}^{(p)} 
  \nonumber\\
  &=& 0
  \label{eq:kouchan-17.684}
  , \\
  {}^{(2)}\!{\cal C}_{i}^{(pV)}
  &:=&
  {}^{(2)}\!{\cal C}_{i}^{(p)}
  - D_{i}\Delta^{-1}D^{j}{}^{(2)}\!{\cal C}_{j}^{(p)}
  \nonumber\\
  &=&
  \left( \epsilon + p \right) \left(
    \partial_{\eta} + {\cal H}
  \right)
    \stackrel{(2)}{{\cal V}_{i}}
  + \partial_{\eta}p \stackrel{(2)}{{\cal V}_{i}}
  \nonumber\\
  &&
  - \Xi_{i}^{(p)} + D_{i} \Delta^{-1} D^{j}\Xi_{j}^{(p)}
  \nonumber\\
  &=& 0
  \label{eq:kouchan-17.685}
  .
\end{eqnarray}


First, we consider the scalar-part (\ref{eq:kouchan-17.684})
of the Euler equation.
As in the case of the first-order perturbation of the Euler
equation, through the background Einstein equation
(\ref{eq:background-Einstein-equations-3}), the background
energy continuity equation (\ref{eq:kouchan-19.41}), and the
Einstein equations of the second order
(\ref{eq:kouchan-19.121})--(\ref{eq:kouchan-19.125}), we can 
obtain
\begin{eqnarray}
  4 \pi G a^{2} {}^{(2)}\!{\cal C}_{i}^{(pS)}
  &=&
  -  4 \pi G a^{3} D_{i}\stackrel{(2)}{v} \stackrel{(0)}{C_{0}^{(p)}}
  - \frac{1}{2} D_{i}\stackrel{(2)}{\Phi} \left(
    3 \stackrel{(0)}{{}^{(p)}E_{(1)}}
    - \stackrel{(0)}{{}^{(p)}E_{(2)}}
  \right)
  -    D_{i}\stackrel{(2)}{{}^{(p)}E_{(2)}}
  \nonumber\\
  &&
  - \frac{1}{3} D_{i}\left(
    \Delta + 3 K
  \right) \stackrel{(2)}{{}^{(p)}E_{(3)}}
  + \left(
    \partial_{\eta}
    + 2 {\cal H}
  \right) \stackrel{(2)}{{}^{(p)}E_{(4)i}}
  \nonumber\\
  &&
  - \frac{1}{2} D_{i}\Delta^{-1}D^{j}{\cal J}_{j}
  ,
  \label{eq:kouchan-19.150}
\end{eqnarray}
where we defined
\begin{eqnarray}
  {\cal J}_{j} &:=&
     8 \pi G a^{2} \Xi_{j}^{(p)} 
  - \left(
    \partial_{\eta} + 2 {\cal H}
  \right) \Gamma_{j}
  +    D^{l}\Gamma_{jl}
  \label{eq:kouchan-19.151}
  .
\end{eqnarray}
On the other hand, through the background energy continuity
equation (\ref{eq:kouchan-19.41}) and the Einstein equation of
the second order (\ref{eq:kouchan-19.126}) and
(\ref{eq:kouchan-19.123}), the vector-part
(\ref{eq:kouchan-17.685}) of the Euler equation
(\ref{eq:kouchan-17.680}) is given by
\begin{eqnarray}
  8 \pi G a^{2} {}^{(2)}\!{\cal C}_{i}^{(pV)}
  &=&
  - 8 \pi G a^{3} \stackrel{(0)}{C_{0}^{(p)}} \stackrel{(2)}{{\cal V}_{i}}
  - \frac{1}{2} \left(
    \partial_{\eta} + 2 {\cal H}
  \right) \stackrel{(2)}{{}^{(p)}E_{(5)i}}
  + \frac{1}{2a^{2}} \left(\Delta + 2 K\right)\stackrel{(2)}{{}^{(p)}E_{(6)i}}
  \nonumber\\
  &&
  - {\cal J}_{i} + D_{i}\Delta^{-1}D^{j}{\cal J}_{j}
  \nonumber\\
  &=& 0
  \label{eq:kouchan-19.152}
  .
\end{eqnarray}
Eqs.~(\ref{eq:kouchan-19.150}) and (\ref{eq:kouchan-19.152})
show that the second-order perturbations of the Euler equations
is consistent with the background Einstein equations and the
second-order perturbations of the Einstein equations if the
equation
\begin{eqnarray}
  {\cal J}_{j} &=& 0
  \label{eq:kouchan-19.170}
\end{eqnarray}
is satisfied under the Einstein equations of the background and
the first-order perturbations.
Actually, we can easily confirm Eq.~(\ref{eq:kouchan-19.170})
as follows.
Through Eqs.~(\ref{eq:background-Einstein-equations-3}),
(\ref{eq:kouchan-19.71})--(\ref{eq:kouchan-19.68}), we can
see the relation 
\begin{eqnarray}
  {\cal J}_{i}
  &=&
  2 \left[
    3 \partial_{\eta}\stackrel{(1)}{\Phi} D_{i}\stackrel{(1)}{v}
    - \stackrel{(1)}{\Phi} D_{i}\stackrel{(1)}{\Phi}
  \right] \left(
    3 \stackrel{(0)}{{}^{(p)}E_{(1)}}
    - \stackrel{(0)}{{}^{(p)}E_{(2)}}
  \right)
  \nonumber\\
  && \quad\quad
  - 4 \left[
      D_{i}\stackrel{(1)}{v} \left(\partial_{\eta} +  {\cal H}\right)
    - D_{i}\stackrel{(1)}{\Phi}
  \right] \stackrel{(1)}{{}^{(p)}E_{(1)}}
  \nonumber\\
  && \quad\quad
  + 4 \left[
       \stackrel{(1)}{\Phi} D_{i}
    +   D_{i}\stackrel{(1)}{\Phi}
    - 3 {\cal H} D_{i}\stackrel{(1)}{v}
  \right] \stackrel{(1)}{{}^{(p)}E_{(2)}}
  \nonumber\\
  && \quad\quad
  + 4 \left[
      D_{i}\stackrel{(1)}{v} D^{j}
    + 3 \gamma_{i}^{\;\;j} \partial_{\eta}\stackrel{(1)}{\Phi}
  \right] \stackrel{(1)}{{}^{(p)}E_{(4)j}}
  \nonumber\\
  &&
  + 6 \partial_{\eta}\stackrel{(1)}{\Psi} \stackrel{(1)}{{\cal V}_{i}} \left(
    3 \stackrel{(0)}{{}^{(p)}E_{(1)}}
    - \stackrel{(0)}{{}^{(p)}E_{(2)}}
  \right)
  - 4 \stackrel{(1)}{{\cal V}_{i}} \left[
    \partial_{\eta} + {\cal H}
  \right] \stackrel{(1)}{{}^{(p)}E_{(1)}}
  - 12 {\cal H} \stackrel{(1)}{{\cal V}_{i}} \stackrel{(1)}{{}^{(p)}E_{(2)}}
  \nonumber\\
  && \quad\quad
  + 4 \left[
    \gamma_{i}^{\;\;k} \stackrel{(1)}{\nu^{j}} D_{j}
    + \stackrel{(1)}{{\cal V}_{i}} D^{k}
    + D_{i}\stackrel{(1)}{\nu^{k}} 
  \right] \stackrel{(1)}{{}^{(p)}E_{(4)k}}
  - 3 \partial_{\eta}\stackrel{(1)}{\Psi} \stackrel{(1)}{{}^{(p)}E_{(5)i}}
  \nonumber\\
  && \quad\quad
  - \frac{2}{a^{2}} \left[
      D_{i}D^{j}\stackrel{(1)}{\Phi}
    + \gamma_{i}^{\;\;j} \Delta\stackrel{(1)}{\Phi}
    - D^{j}\stackrel{(1)}{\Phi} D_{i}
    + \frac{1}{2} \gamma_{i}^{\;\;j} \stackrel{(1)}{\Phi}\left(\Delta+2K\right)
  \right] \stackrel{(1)}{{}^{(p)}E_{(6)j}}
  \nonumber\\
  &&
  + 2 \stackrel{(1)}{\nu^{j}} D_{i}\stackrel{(1)}{\nu_{j}} \left(
    3 \stackrel{(0)}{{}^{(p)}E_{(1)}}
    - \stackrel{(0)}{{}^{(p)}E_{(2)}}
  \right)
  - \left[
    D_{i}\stackrel{(1)}{\nu^{k}}
    - \gamma_{i}^{\;\;k} \stackrel{(1)}{\nu^{j}} D_{j}
  \right] \stackrel{(1)}{{}^{(p)}E_{(5)k}}
  \nonumber\\
  && \quad\quad
  + \frac{4}{a^{2}} {\cal H} \left[
      D_{i}\stackrel{(1)}{\nu^{j}}
    + \stackrel{(1)}{\nu^{j}} D_{i}
  \right] \stackrel{(1)}{{}^{(p)}E_{(6)j}}
  \nonumber\\
  &&
  - 2 D^{j}\stackrel{(1)}{\Phi} \stackrel{(1)}{{}^{(p)}E_{(7)ij}}
  \nonumber\\
  &&
  - \frac{2}{a^{2}} \left[
    \gamma_{i}^{\;\;[j} \stackrel{(1)}{\chi^{k]m}} D_{m}D_{j}
    - D^{[j}\stackrel{(1)}{\chi_{i}^{\;\;k]}} D_{j}
    - \frac{1}{2}\left(\Delta - K\right)\stackrel{(1)}{\chi_{i}^{\;\;k}}
  \right]
  \stackrel{(1)}{{}^{(p)}E_{(6)k}}
  \nonumber\\
  &&
  + \frac{1}{2} \left[
      D_{i}\stackrel{(1)}{\chi^{kj}} \stackrel{(1)}{{}^{(p)}E_{(7)jk}}
    + 4 \stackrel{(1)}{\chi^{kj}} D_{[i}\stackrel{(1)}{{}^{(p)}E_{(7)j]k}}
  \right].
\end{eqnarray}
This represents that Eq.~(\ref{eq:kouchan-19.170}) is satisfied
due to the background Einstein equations and the first-order
perturbations of the Einstein equations, and implies that the
second-order perturbation of the Euler equation is consistent
with the set of the background, the first-order, and the
second-order Einstein equations. 
From general point of view, this is just a well-known result,
i.e., the Einstein equation includes the equations of motion for
matter field due to the Bianchi identity.
However, the above verification of the identity
(\ref{eq:kouchan-19.170}) implies that our derived second-order
perturbations of the Einstein equation and the Euler equation are
consistent.
In this sense, we may say that the derived second-order Einstein
equations, in particular, the derived formulae for the source
terms $\Gamma_{0}$, $\Gamma_{i}$, $\Gamma_{ij}$, $\Xi_{0}$, and
$\Xi_{i}$ are correct.


\subsection{Scalar field case}


Next, we consider the second-order Einstein equation
(\ref{eq:second-order-Einstein-general}) in the case of a
single scalar field.
Through the components of 
${}^{(1)}\!{\cal G}_{a}^{\;\;b}[{\cal L}]$ given by
Eqs.~(\ref{eq:kouchan-15.32-linear})--(\ref{eq:kouchan-15.35-linear})
with the replacement
(\ref{eq:replacement-linear-to-second-order}), the components of
${}^{(2)}\!{\cal G}_{a}^{\;\;b}$ given by
Eqs.~(\ref{eq:generic-2-calG-eta-eta})--(\ref{eq:generic-2-calG-i-j}),
and the components of the second-order perturbation of the energy 
momentum tensor ${}^{(2)}\!{\cal T}_{a}^{\;\;b}$ given by
Eqs.~(\ref{eq:kouchan-19.209})--(\ref{eq:kouchan-19.212}), we
can obtain the all components of the second-order perturbation
of the Einstein equation
(\ref{eq:second-order-Einstein-general}) in the case of the
universe filled with a single scalar field.
For simplicity, we used the first-order Einstein equations
(\ref{eq:kouchan-19.265}) and (\ref{eq:kouchan-19.266}).
For the scalar-mode of the second order, we obtain the equations
as follows:
\begin{eqnarray}
  \stackrel{(2)}{{}^{(s)}E_{(1)}}
  &:=&
  \left\{
                  \partial_{\eta}^{2}
    +          2  \left(
        {\cal H}
      - \frac{\partial_{\eta}^{2}\varphi}{\partial_{\eta}\varphi}
    \right)
    \partial_{\eta}
    -             \Delta
    -          4  K
    +          2  \left(
        \partial_{\eta}{\cal H}
      - \frac{\partial_{\eta}^{2}\varphi}{\partial_{\eta}\varphi} {\cal H}
    \right)
  \right\}
  \stackrel{(2)}{\Phi}
  \nonumber\\
  &&
  + \Gamma_{0}
  + \frac{1}{2} \Gamma_{k}^{\;\;k}
  - \Delta^{-1} D^{j}D^{i}\Gamma_{ij}
  \nonumber\\
  &&
  - \frac{3}{2}
  \left[
    -             \partial_{\eta}^{2}
    +             \left(
        2 \frac{\partial_{\eta}^{2}\varphi}{\partial_{\eta}\varphi}
      -   {\cal H}
    \right)
    \partial_{\eta}
  \right]
  \left( \Delta + 3 K \right)^{-1}
  \left(
    \Delta^{-1} D^{j}D^{i}\Gamma_{ij} - \frac{1}{3} \Gamma_{k}^{\;\;k}
  \right)
  \nonumber\\
  &&
  - \left(
    \partial_{\eta}
    - \frac{\partial_{\eta}^{2}\varphi}{\partial_{\eta}\varphi}
  \right)
  \Delta^{-1} D^{k}\Gamma_{k}
  \nonumber\\
  &=& 0
  \label{eq:kouchan-19.330}
  ; \\
  \stackrel{(2)}{{}^{(s)}E_{(2)}}
  &:=&
    2 \partial_{\eta}\stackrel{(2)}{\Psi}
  + 2 {\cal H} \stackrel{(2)}{\Phi}
  - 8 \pi G \partial_{\eta}\varphi \varphi_{2}
  - \Delta^{-1} D^{k}\Gamma_{k}
  \nonumber\\
  &=& 0
  \label{eq:kouchan-19.331}
  ; \\
  \stackrel{(2)}{{}^{(s)}E_{(3)}}
  &:=&
  \left(
    -             \partial_{\eta}^{2} 
    -          5  {\cal H} \partial_{\eta}
    + \frac{4}{3} \Delta
    +          4  K
  \right)
  \stackrel{(2)}{\Psi}
  -  \left(
                  {\cal H} \partial_{\eta}
    +          2  \partial_{\eta}{\cal H}
    +          4  {\cal H}^{2}
    + \frac{1}{3} \Delta
  \right)
  \stackrel{(2)}{\Phi}
  \nonumber\\
  &&
  -  8 \pi G a^{2} \varphi_{2} \frac{\partial V}{\partial\varphi}
  - \Gamma_{0} + \frac{1}{6} \Gamma_{k}^{\;\;k},
  \nonumber\\
  &=& 0
  \label{eq:kouchan-19.332}
  ; \\
  \stackrel{(2)}{{}^{(s)}E_{(4)}}
  &:=&
  \stackrel{(2)}{\Psi} - \stackrel{(2)}{\Phi}
  -
  \frac{3}{2} 
  \left( \Delta + 3 K \right)^{-1} 
  \left(
    \Delta^{-1} D^{j}D^{i}\Gamma_{ij} - \frac{1}{3} \Gamma_{k}^{\;\;k}
  \right)
  \nonumber\\
  &=& 0
  .
  \label{eq:kouchan-19.333}
\end{eqnarray}
For the vector-mode of the second order, we obtain a constraint
equation and an evolution equation as follows: 
\begin{eqnarray}
  \stackrel{(2)}{{}^{(s)}E_{(5)i}}
  &:=&
  \stackrel{(2)}{\nu_{i}}
  -
  2 \left( \Delta + 2 K \right)^{-1}
  \left\{
    D_{i}\Delta^{-1} D^{k}\Gamma_{k}
    - \Gamma_{i}
  \right\}
  \nonumber\\
  &=& 0
  \label{eq:kouchan-19.334}
  ; \\
  \stackrel{(2)}{{}^{(s)}E_{(6)i}}
  &:=&
  \partial_{\eta}
  \left(
    a^{2} \stackrel{(2)}{\nu_{i}}
  \right)
  -
  2 a^{2} 
  \left(\Delta + 2 K\right)^{-1}
  \left\{
      D_{i}\Delta^{-1}D^{k}D^{l}\Gamma_{kl}
    - D^{k}\Gamma_{ki}
  \right\}
  \nonumber\\
  &=& 0
  .
  \label{eq:kouchan-19.335}
\end{eqnarray}
For the tensor-mode of the second order, we obtain the single
evolution equation as follows:
\begin{eqnarray}
  \stackrel{(2)}{{}^{(s)}E_{(7)ij}}
  &:=&
  \left(
    \partial_{\eta}^{2} + 2 {\cal H} \partial_{\eta} + 2 K - \Delta
  \right)
  \stackrel{(2)}{\chi}_{ij}
  - 2 \Gamma_{ij}
  + \frac{2}{3} \Gamma_{k}^{\;\;k} \gamma_{ij}
  \nonumber\\
  &&
  + 3 \left(
    D_{i}D_{j} - \frac{1}{3} \gamma_{ij} \Delta
  \right) 
  \left( \Delta + 3 K \right)^{-1} 
  \left(
    \Delta^{-1}D^{k}D^{l}\Gamma_{lk} - \frac{1}{3} \Gamma_{k}^{\;\;k}
  \right)
  \nonumber\\
  &&
  - 4 \left\{
      D_{(i}\left(\Delta + 2 K\right)^{-1} D_{j)}\Delta^{-1}D^{k}D^{l}\Gamma_{kl}
    - D_{(i}\left(\Delta + 2 K\right)^{-1} D^{k}\Gamma_{j)k}
  \right\}
  \nonumber\\
  &=& 0
  \label{eq:kouchan-19.336}
  .
\end{eqnarray}
Herqe, $\Gamma_{0}$, $\Gamma_{i}$, and $\Gamma_{ij}$ in these
expressions are defined by 
\begin{eqnarray}
  \Gamma_{0}
  &:=&
    4 \pi G \left(
        (\partial_{\eta}\varphi_{1})^{2}
    +   D_{i}\varphi_{1} D^{i}\varphi_{1} 
    +   a^{2} (\varphi_{1})^{2} \frac{\partial^{2}V}{\partial\varphi^{2}}
  \right)
  \nonumber\\
  && \quad\quad
  -          4  \partial_{\eta}{\cal H} \left(\stackrel{(1)}{\Phi}\right)^{2}
  -          2  \stackrel{(1)}{\Phi} \partial_{\eta}^{2}\stackrel{(1)}{\Phi}
  -          3  D_{k}\stackrel{(1)}{\Phi} D^{k}\stackrel{(1)}{\Phi}
  -         10  \stackrel{(1)}{\Phi} \Delta\stackrel{(1)}{\Phi}
  \nonumber\\
  && \quad\quad
  -          3  \left(\partial_{\eta}\stackrel{(1)}{\Phi}\right)^{2}
  -         16  K \left(\stackrel{(1)}{\Phi}\right)^{2}
  -          8  {\cal H}^{2} \left(\stackrel{(1)}{\Phi}\right)^{2}
  \nonumber\\
  && 
  +             D_{l}D_{k}\stackrel{(1)}{\Phi} \stackrel{(1)}{\chi^{lk}}
  \nonumber\\
  && 
  + \frac{1}{8} \partial_{\eta}\stackrel{(1)}{\chi_{lk}} \partial_{\eta}\stackrel{(1)}{\chi^{kl}}
  +             {\cal H} \stackrel{(1)}{\chi_{kl}} \partial_{\eta}\stackrel{(1)}{\chi^{lk}}
  - \frac{3}{8} D_{k}\stackrel{(1)}{\chi_{lm}} D^{k}\stackrel{(1)}{\chi^{ml}}
  + \frac{1}{4} D_{k}\stackrel{(1)}{\chi_{lm}} D^{l}\stackrel{(1)}{\chi^{mk}}
  \nonumber\\
  && \quad\quad
  - \frac{1}{2} \stackrel{(1)}{\chi^{lm}} \Delta\stackrel{(1)}{\chi_{lm}}
  + \frac{1}{2} K \stackrel{(1)}{\chi_{lm}} \stackrel{(1)}{\chi^{lm}}
  \label{eq:kouchan-19.337}
  , \\
  \Gamma_{i}
  &:=&
            16  \pi G \partial_{\eta}\varphi_{1} D_{i}\varphi_{1}
  -          4  \partial_{\eta}\stackrel{(1)}{\Phi} D_{i}\stackrel{(1)}{\Phi}
  +          8  {\cal H} \stackrel{(1)}{\Phi} D_{i}\stackrel{(1)}{\Phi}
  -          8  \stackrel{(1)}{\Phi} \partial_{\eta}D_{i}\stackrel{(1)}{\Phi}
  \nonumber\\
  &&
  +          2  D^{j}\stackrel{(1)}{\Phi} \partial_{\eta}\stackrel{(1)}{\chi_{ji}}
  -          2  \partial_{\eta}D^{j}\stackrel{(1)}{\Phi} \stackrel{(1)}{\chi_{ij}}
  \nonumber\\
  &&
  - \frac{1}{2} \partial_{\eta}\stackrel{(1)}{\chi_{jk}} D_{i}\stackrel{(1)}{\chi^{kj}}
  -             \stackrel{(1)}{\chi_{kl}} \partial_{\eta}D_{i}\stackrel{(1)}{\chi^{lk}}
  +             \stackrel{(1)}{\chi^{kl}} \partial_{\eta}D_{k}\stackrel{(1)}{\chi_{il}}
  \label{eq:kouchan-19.338}
  , \\
  \Gamma_{ij}
  &:=&
    16 \pi G D_{i}\varphi_{1} D_{j}\varphi_{1}
  +  8 \pi G \left\{
      (\partial_{\eta}\varphi_{1})^{2}
    - D_{l}\varphi_{1} D^{l}\varphi_{1}
    - a^{2} (\varphi_{1})^{2} \frac{\partial^{2}V}{\partial\varphi^{2}}
  \right\} \gamma_{ij}
  \nonumber\\
  && \quad\quad
  -          4  D_{i}\stackrel{(1)}{\Phi} D_{j}\stackrel{(1)}{\Phi}
  -          8  \stackrel{(1)}{\Phi} D_{i}D_{j}\stackrel{(1)}{\Phi}
  \nonumber\\
  && \quad\quad
  + \left(
               6  D_{k}\stackrel{(1)}{\Phi} D^{k}\stackrel{(1)}{\Phi}
    +          4  \stackrel{(1)}{\Phi} \Delta\stackrel{(1)}{\Phi}
    +          2  \left(\partial_{\eta}\stackrel{(1)}{\Phi}\right)^{2}
    +          8  \partial_{\eta}{\cal H} \left(\stackrel{(1)}{\Phi}\right)^{2}
  \right.
  \nonumber\\
  && \quad\quad\quad\quad
  \left.
    +         16  {\cal H}^{2} \left(\stackrel{(1)}{\Phi}\right)^{2}
    +         16  {\cal H} \stackrel{(1)}{\Phi} \partial_{\eta}\stackrel{(1)}{\Phi}
    -          4  \stackrel{(1)}{\Phi} \partial_{\eta}^{2}\stackrel{(1)}{\Phi}
  \right) \gamma_{ij}
  \nonumber\\
  &&
  -          4  {\cal H} \partial_{\eta}\stackrel{(1)}{\Phi} \stackrel{(1)}{\chi_{ij}}
  -          2  \partial_{\eta}^{2}\stackrel{(1)}{\Phi} \stackrel{(1)}{\chi_{ij}}
  -          4  D^{k}\stackrel{(1)}{\Phi} D_{(i}\stackrel{(1)}{\chi_{j)k}}
  +          4  D^{k}\stackrel{(1)}{\Phi} D_{k}\stackrel{(1)}{\chi_{ij}}
  -          8  K \stackrel{(1)}{\Phi} \stackrel{(1)}{\chi_{ij}}
  \nonumber\\
  && \quad\quad
  +          4  \stackrel{(1)}{\Phi} \Delta\stackrel{(1)}{\chi_{ij}}
  -          4  D^{k}D_{(i}\stackrel{(1)}{\Phi} \stackrel{(1)}{\chi_{j)k}}
  +          2  \Delta \stackrel{(1)}{\Phi} \stackrel{(1)}{\chi_{ij}}
  +          2  D_{l}D_{k}\stackrel{(1)}{\Phi} \stackrel{(1)}{\chi^{lk}} \gamma_{ij}
  \nonumber\\
  &&
  +             \partial_{\eta}\stackrel{(1)}{\chi_{ik}} \partial_{\eta}\stackrel{(1)}{\chi_{j}^{\;\;k}}
  -             D^{k}\stackrel{(1)}{\chi_{il}} D_{k}\stackrel{(1)}{\chi_{j}^{\;\;l}}
  +             D^{k}\stackrel{(1)}{\chi_{il}} D^{l}\stackrel{(1)}{\chi_{jk}}
  - \frac{1}{2} D_{i}\stackrel{(1)}{\chi^{lk}} D_{j}\stackrel{(1)}{\chi_{lk}}
  \nonumber\\
  && \quad\quad
  -             \stackrel{(1)}{\chi_{lm}} D_{i}D_{j}\stackrel{(1)}{\chi^{ml}}
  +          2  \stackrel{(1)}{\chi^{lm}} D_{l}D_{(i}\stackrel{(1)}{\chi_{j)m}}
  -             \stackrel{(1)}{\chi^{lm}} D_{m}D_{l}\stackrel{(1)}{\chi_{ij}}
  \nonumber\\
  && \quad\quad
  - \frac{1}{4} \left(
      3 \partial_{\eta}\stackrel{(1)}{\chi_{lk}} \partial_{\eta}\stackrel{(1)}{\chi^{kl}}
    - 3 D_{k}\stackrel{(1)}{\chi_{lm}} D^{k}\stackrel{(1)}{\chi^{ml}}
    + 2 D_{k}\stackrel{(1)}{\chi_{lm}} D^{l}\stackrel{(1)}{\chi^{mk}}
  \right.
  \nonumber\\
  && \quad\quad\quad\quad\quad
  \left.
    - 4 K \stackrel{(1)}{\chi_{lm}} \stackrel{(1)}{\chi^{lm}}
  \right) \gamma_{ij}
  \label{eq:kouchan-19.339}
  .
\end{eqnarray}


Now, we consider the consistency check in the set of
equations (\ref{eq:kouchan-19.330})-(\ref{eq:kouchan-19.339}).
First, we consider the consistency between
Eqs.~(\ref{eq:kouchan-19.334}) and (\ref{eq:kouchan-19.335}).
Eq.~(\ref{eq:kouchan-19.334}) comes from the momentum
constraints in the Einstein equations, which is an initial value 
constraint, and should be consistent with the evolution
equations in the Einstein equations from general point of view.
In this sense, Eqs.~(\ref{eq:kouchan-19.334}) and
(\ref{eq:kouchan-19.335}) should be consistent with each other. 
Now, we explicitly check this.
Through these equations, we obtain 
\begin{eqnarray}
  &&
  \stackrel{(2)}{{}^{(s)}E_{(6)i}}
  - \partial_{\eta} \left(
    a^{2} \stackrel{(2)}{{}^{(s)}E_{(5)i}}
  \right)
  \nonumber\\
  &=&
  2 a^{2} \left( \Delta + 2 K \right)^{-1}
  \left(
    D_{i}\Delta^{-1}D^{k}
    - \gamma_{i}^{\;\;k}
  \right)
  \left(
    \partial_{\eta}\Gamma_{k}
    + 2 {\cal H} \Gamma_{k}
    - D^{l}\Gamma_{kl}
  \right)
  \nonumber\\
  &=& 0
  .
\end{eqnarray}
Therefore, the vector-part (\ref{eq:kouchan-19.334}) of the
momentum constraint is consistent with the evolution equation 
(\ref{eq:kouchan-19.335}) if the equation
\begin{eqnarray}
  \partial_{\eta}\Gamma_{k}
  + 2 {\cal H} \Gamma_{k}
  - D^{l}\Gamma_{lk} = 0
  \label{eq:kouchan-19.358}
\end{eqnarray}
is satisfied.
Actually, through Eqs.~(\ref{eq:kouchan-19.262}),
(\ref{eq:kouchan-19.263}), (\ref{eq:kouchan-19.267}), and
(\ref{eq:kouchan-17.806-first-explicit}), the left hand side of
Eq.~(\ref{eq:kouchan-19.358}) is given by  
\begin{eqnarray}
  &&
    \partial_{\eta}\Gamma_{k}
  + 2 {\cal H} \Gamma_{k}
  - D^{l}\Gamma_{lk}
  \nonumber\\
  &=&
  -         16  \pi G a^{2} D_{k}\varphi_{1} \stackrel{(1)}{{\cal C}_{(K)}}
  -          4  \stackrel{(1)}{\Phi} D_{k}\stackrel{(1)}{{}^{(s)}E_{(1)}}
  - 16 \left(
      \partial_{\eta}
    + {\cal H}
    + \frac{\partial_{\eta}^{2}\varphi}{2\partial_{\eta}\varphi} 
  \right)\stackrel{(1)}{\Phi} D_{k}\stackrel{(1)}{{}^{(s)}E_{(2)}}
  \nonumber\\
  &&
  +          2  D^{j}\stackrel{(1)}{\Phi} \stackrel{(1)}{{}^{(s)}E_{(6)jk}}
  - \frac{1}{2} \left(
    D_{k}\stackrel{(1)}{\chi^{jl}} + 2 \stackrel{(1)}{\chi^{jl}} D_{k}
  \right) \stackrel{(1)}{{}^{(s)}E_{(6)lj}}
  + \stackrel{(1)}{\chi^{jl}} D_{j}\stackrel{(1)}{{}^{(s)}E_{(6)kl}}
  .
  \label{eq:kouchan-19.364}
\end{eqnarray}
Since the first-order perturbation (\ref{eq:kouchan-19.270}) of
the Klein-Gordon equation is consistent with the Einstein
equation as shown in Eq.~(\ref{eq:kouchan-19.272}),
Eq.~(\ref{eq:kouchan-19.364}) shows that the initial value
constraint (\ref{eq:kouchan-19.334}) for the vector-mode 
of the second-order perturbation is consistent with the
evolution equation (\ref{eq:kouchan-19.335}) by virtue of the
first-order perturbations of the Einstein equations. 
This is a trivial result from general point of view,
because the Einstein equation is the first class constrained
system.
However, this trivial result implies that we have derived the
source terms $\Gamma_{i}$ and $\Gamma_{ij}$ of the second-order
Einstein equations consistently.


Next, we consider the equation (\ref{eq:kouchan-19.332}).
Through Eqs.~(\ref{eq:kouchan-19.330}) and
(\ref{eq:kouchan-19.333}), Eq.~(\ref{eq:kouchan-19.332}) is
given by
\begin{eqnarray}
  \stackrel{(2)}{{}^{(s)}E_{(3)}}
  &=&
  \left(
    -             \partial_{\eta}^{2} 
    -          5  {\cal H} \partial_{\eta}
    + \frac{4}{3} \Delta
    +          4  K
  \right)
  \stackrel{(2)}{{}^{(s)}E_{(4)}}
  - \stackrel{(2)}{{}^{(s)}E_{(1)}}
  \nonumber\\
  &&
  -          2 \left(
      2 {\cal H}
    +   \frac{\partial_{\eta}^{2}\varphi}{\partial_{\eta}\varphi}
  \right)
  \left(
    \partial_{\eta} + {\cal H}
  \right)
  \stackrel{(2)}{\Phi}
  -  8 \pi G a^{2} \varphi_{2} \frac{\partial V}{\partial\varphi}
  \nonumber\\
  &&
  -  3 \left(
      2 {\cal H}
    +   \frac{\partial_{\eta}^{2}\varphi}{\partial_{\eta}\varphi}
  \right)
  \partial_{\eta}\left( \Delta + 3 K \right)^{-1}
  \left(
    \Delta^{-1} D^{j}D^{i}\Gamma_{ij} - \frac{1}{3} \Gamma_{k}^{\;\;k}
  \right)
  \nonumber\\
  &&
  +   \Delta^{-1} D^{j}D^{i}\Gamma_{ij}
  - \left(
      \partial_{\eta}
    - \frac{\partial_{\eta}^{2}\varphi}{\partial_{\eta}\varphi}
  \right)
  \Delta^{-1} D^{k}\Gamma_{k}
  \label{eq:kouchan-19.340}
  .
\end{eqnarray}
On the other hand, from Eqs.~(\ref{eq:kouchan-19.331}) and
(\ref{eq:kouchan-19.333}), we obtain 
\begin{eqnarray}
  \stackrel{(2)}{{}^{(s)}E_{(2)}}
  - 2 \partial_{\eta}\stackrel{(2)}{{}^{(s)}E_{(4)}}
  &:=&
    2 \partial_{\eta}\stackrel{(2)}{\Phi}
  + 2 {\cal H} \stackrel{(2)}{\Phi}
  - 8 \pi G \partial_{\eta}\varphi \varphi_{2}
  \nonumber\\
  &&
  + 3 \partial_{\eta}\left( \Delta + 3 K \right)^{-1}\left(
    \Delta^{-1} D^{j}D^{i}\Gamma_{ij} - \frac{1}{3} \Gamma_{k}^{\;\;k}
  \right)
  - \Delta^{-1} D^{k}\Gamma_{k}
  \nonumber\\
  &=& 0
  .
  \label{eq:kouchan-19.342}
\end{eqnarray}
Through Eq.~(\ref{eq:kouchan-19.342}) and the background
Klein-Gordon equation (\ref{eq:background-Klein-Gordon-eq}), the
equation (\ref{eq:kouchan-19.340}) is given by  
\begin{eqnarray}
  \stackrel{(2)}{{}^{(s)}E_{(3)}}
  &=&
  \left(
    -             \partial_{\eta}^{2} 
    -          5  {\cal H} \partial_{\eta}
    + \frac{4}{3} \Delta
    +          4  K
  \right)
  \stackrel{(2)}{{}^{(s)}E_{(4)}}
  - \stackrel{(2)}{{}^{(s)}E_{(1)}}
  \nonumber\\
  &&
  -          \left(
      2 {\cal H}
    +   \frac{\partial_{\eta}^{2}\varphi}{\partial_{\eta}\varphi}
  \right)
  \left(
        \stackrel{(2)}{{}^{(s)}E_{(2)}}
    - 2 \partial_{\eta}\stackrel{(2)}{{}^{(s)}E_{(4)}}
  \right)
  -  8 \pi G a^{2} \stackrel{(0)}{C_{K}} \varphi_{2}
  \nonumber\\
  &&
  -   \Delta^{-1} \left(
        \partial_{\eta}D^{k}\Gamma_{k}
    + 2 {\cal H} D^{k}\Gamma_{k}
    -   D^{j}D^{i}\Gamma_{ij}
  \right)
  .
  \label{eq:kouchan-19.343}
\end{eqnarray}
This equation (\ref{eq:kouchan-19.343}) shows that
Eq.~(\ref{eq:kouchan-19.332}) is consistent with the set of the 
background, the first-order, and the other second-order Einstein
equations if the equation 
\begin{eqnarray}
  \left(
                  \partial_{\eta}
    +          2  {\cal H}
  \right) D^{k}\Gamma_{k}
  - D^{j}D^{i}\Gamma_{ij}
  =
  0
  \label{eq:kouchan-19.345}
\end{eqnarray}
is satisfied under the background and the first-order Einstein
equations.
Actually, we have already seen Eq.~(\ref{eq:kouchan-19.358}) is
satisfied under the background and the first-order Einstein
equation.
Taking the divergence of Eq.~(\ref{eq:kouchan-19.358}), we can
easily confirm Eq.~(\ref{eq:kouchan-19.345}).
Thus, the component (\ref{eq:kouchan-19.332}) of the Einstein
equation is not independent of the set of equations
(\ref{eq:kouchan-19.330}), (\ref{eq:kouchan-19.331}),
(\ref{eq:kouchan-19.333}), and the first-order perturbations of
the Einstein equation, i.e., Eqs.~(\ref{eq:kouchan-19.262}),
(\ref{eq:kouchan-19.263}), (\ref{eq:kouchan-19.265}). 
As seen above, the component (\ref{eq:kouchan-19.264}) of the
first-order Einstein equation is derived from the set of the
equations (\ref{eq:kouchan-19.262}), (\ref{eq:kouchan-19.263}), 
(\ref{eq:kouchan-19.265}) and the background Einstein equations.
This implies that the potential of the scalar field affects to
the evolution of the system only through the background Einstein
equations at least in the first order perturbations.
We have also seen that the situation is also same even in the
second-order perturbations, i.e., the potential of the scalar
field affects to the evolution of the system only through the
background Einstein equation even in the second-order
perturbations.


Thus, we have seen that the derive Einstein equations of the
second order
(\ref{eq:kouchan-19.330})--(\ref{eq:kouchan-19.339}) are
consistent with each other through the equation
(\ref{eq:kouchan-19.358}).
This fact implies that the derived source term $\Gamma_{i}$ and
$\Gamma_{ij}$ of the second-order perturbations of the Einstein
equations, which are defined by Eqs.~(\ref{eq:kouchan-19.338})
and (\ref{eq:kouchan-19.339}), are correct source terms of the
second-order Einstein equations.
On the other hand, for $\Gamma_{0}$, we have to consider the
consistency between the perturbative Einstein equations and the
perturbative Klein-Gordon equation as seen below.


\subsubsection{Consistency with the Klein-Gordon equation}
\label{sec:Consystency-with-the-Klein-Gordonr-equation}


Here, we consider the consistency of the second-order
perturbation of the Klein-Gordon equation and the Einstein
equations. 
As shown in KN2008\cite{kouchan-second-cosmo-matter}, the
second-order perturbation of the Klein-Gordon equation is given
by 
\begin{eqnarray}
  a^{2} \stackrel{(2)}{{\cal C}_{(K)}}
  &=&
  -    \partial_{\eta}^{2}\varphi_{2} 
  -  2 {\cal H} \partial_{\eta}\varphi_{2} 
  +    \Delta\varphi_{2} 
  +    \partial_{\eta}\stackrel{(2)}{\Phi} \partial_{\eta}\varphi
  +  3 \partial_{\eta}\stackrel{(2)}{\Psi} \partial_{\eta}\varphi
  \nonumber\\
  && 
  -  2 a^{2} \stackrel{(2)}{\Phi} \frac{\partial V}{\partial\bar{\varphi}}(\varphi)
  -    a^{2}\varphi_{2}\frac{\partial^{2}V}{\partial\bar{\varphi}^{2}}(\varphi)
  + \Xi_{(K)}
  \label{eq:second-Klein-Gordon-reduced}
  ,
\end{eqnarray}
where the term $\Xi_{(K)}$ is reduced to 
\begin{eqnarray}
  \Xi_{(K)}
  &:=&
     8 \partial_{\eta}\stackrel{(1)}{\Phi} \partial_{\eta}\varphi_{1}
  +  8 \stackrel{(1)}{\Phi} \Delta\varphi_{1}
  +  8 \stackrel{(1)}{\Phi} \partial_{\eta}\stackrel{(1)}{\Phi} \partial_{\eta}\varphi
  \nonumber\\
  &&
  -  2 \stackrel{(1)}{\chi^{ij}} D_{j}D_{i}\varphi_{1}
  +    \stackrel{(1)}{\chi^{ij}} \partial_{\eta}\stackrel{(1)}{\chi_{ij}} \partial_{\eta}\varphi
  \nonumber\\
  &&
  -  4 a^{2} \stackrel{(1)}{\Phi} \varphi_{1} \frac{\partial^{2}V}{\partial\bar{\varphi}^{2}}(\varphi)
  -    a^{2} (\varphi_{1})^{2} \frac{\partial^{3}V}{\partial\bar{\varphi}^{3}}(\varphi)
  \label{eq:kouchan-19.372-1}
  .
\end{eqnarray}
Here, we have imposed the Einstein equations
(\ref{eq:kouchan-19.265}) and (\ref{eq:kouchan-19.266}) of the
first order, the background Klein-Gordon equation
(\ref{eq:background-Klein-Gordon-eq}), and
its first-order perturbation (\ref{eq:kouchan-19.270}).


As in the case of Eq.~(\ref{eq:kouchan-19.272}) for the
first-order perturbation of the Klein-Gordon equation, we check
the consistency of the second-order perturbation 
(\ref{eq:second-Klein-Gordon-reduced}) of the Klein-Gordon
equation with the second-order perturbations
(\ref{eq:kouchan-19.330})--(\ref{eq:kouchan-19.339}) of the
Einstein equation.  
Since the vector-mode $\stackrel{(2)}{\nu_{i}}$ and the
tensor-mode $\stackrel{(2)}{\chi}_{ij}$ of the second-order do
not appear in the expressions
(\ref{eq:second-Klein-Gordon-reduced}) nor
(\ref{eq:kouchan-19.372-1}) of the second-order perturbation of
the Klein-Gordon equation, we may concentrate on the Einstein
equations for scalar-mode of the second order, i.e.,
Eqs.~(\ref{eq:kouchan-19.330})--(\ref{eq:kouchan-19.333}) with 
the definitions
(\ref{eq:kouchan-19.337})--(\ref{eq:kouchan-19.339}) of the
source terms. 
Further, as shown above, the equation (\ref{eq:kouchan-19.332})
is not independent equation from the set of equations consists
of the second-order perturbations of the Einstein equation
(\ref{eq:kouchan-19.330}), (\ref{eq:kouchan-19.331}), 
(\ref{eq:kouchan-19.333}), the first-order perturbations of the
Einstein equation (\ref{eq:kouchan-19.262}),
(\ref{eq:kouchan-19.263}), (\ref{eq:kouchan-19.265}), and the
background Einstein equations
(\ref{eq:background-Einstein-equations-scalar-1}) and
(\ref{eq:background-Einstein-equations-scalar-2}).
Moreover, as shown in
\S\S\ref{sec:Background-Equations-scalar-field} and
\ref{sec:Fist-order-scalar-field-Klein-Gordon}, the background
Klein-Gordon equation is also derived from the background 
Einstein equation, and the first-order perturbation of the
Klein-Gordon equation is also derived from the background
and the first-order perturbations of the Einstein equations.
For these reason, the second-order perturbation of the
Klein-Gordon equation should be also derived from the set of the
equations which consists of the second-order perturbations of the
Einstein equations (\ref{eq:kouchan-19.330}),
(\ref{eq:kouchan-19.331}), (\ref{eq:kouchan-19.333}), the
first-order perturbations of the Einstein equation 
(\ref{eq:kouchan-19.262}), (\ref{eq:kouchan-19.263}),
(\ref{eq:kouchan-19.265}), and the background Einstein equations
(\ref{eq:background-Einstein-equations-scalar-1}) and
(\ref{eq:background-Einstein-equations-scalar-2}).
Actually, as in the case of Eq.~(\ref{eq:kouchan-19.272}), we
can easily derive the relation
\begin{eqnarray}
  &&
  - 8 \pi G a^{2} (\partial_{\eta}\varphi) \stackrel{(2)}{{\cal C}_{(K)}}
  \nonumber\\
  &=&
  3 \left[
    \frac{\varphi_{2}}{\partial_{\eta}\varphi} \left\{
      - \partial_{\eta}^{2}
      +  \left(
             \frac{\partial_{\eta}^{2}\varphi}{\partial_{\eta}\varphi}
        +  4 {\cal H}
      \right) \partial_{\eta}
      +  2 \partial_{\eta}{\cal H}
      -  4 {\cal H}^{2}
      -  2 {\cal H} \frac{\partial_{\eta}^{2}\varphi}{\partial_{\eta}\varphi}
    \right\}
  \right.
  \nonumber\\
  && \quad\quad
  \left.
    -  2 \stackrel{(2)}{\Phi} \left(
      \partial_{\eta} - 2 {\cal H}
    \right)
  \right] \stackrel{(0)}{{}^{(s)}E_{(1)}}
  \nonumber\\
  &&
  + \left[
    \left(
           \stackrel{(2)}{\Phi}
      -  3 {\cal H} \frac{\varphi_{2}}{\partial_{\eta}\varphi}
    \right) \partial_{\eta}
    + \left(
           \partial_{\eta}\stackrel{(2)}{\Phi}
      +  3 \partial_{\eta}\stackrel{(2)}{\Psi}
      -  2 {\cal H} \stackrel{(2)}{\Phi}
    \right)
  \right.
  \nonumber\\
  && \quad\quad
  \left.
    + 3 \frac{\varphi_{2}}{\partial_{\eta}\varphi} \left(
         2 {\cal H}^{2}
      -    \partial_{\eta}{\cal H}
      +    {\cal H} \frac{\partial_{\eta}^{2}\varphi}{\partial_{\eta}\varphi}
    \right)
  \right]
  \left(
    3 \stackrel{(0)}{{}^{(s)}E_{(1)}}
    - \stackrel{(0)}{{}^{(s)}E_{(2)}}
  \right)
  \nonumber\\
  &&
  + \left[
    -    \partial_{\eta}^{2}
    +  2 \left(
      \frac{\partial_{\eta}^{2}\varphi}{\partial_{\eta}\varphi} - {\cal H}
    \right) \partial_{\eta}
    +    \Delta
    +  2 \frac{\partial_{\eta}^{3}\varphi}{\partial_{\eta}\varphi}
    +  2 \partial_{\eta}{\cal H}
    -  4 {\cal H}^{2}
  \right.
  \nonumber\\
  && \quad\quad
  \left.
    -  2 \frac{\partial_{\eta}^{2}\varphi}{\partial_{\eta}\varphi} \left(
      \frac{\partial_{\eta}^{2}\varphi}{\partial_{\eta}\varphi} - {\cal H}
    \right)
  \right] \stackrel{(2)}{{}^{(s)}E_{(2)}}
  \nonumber\\
  &&
  +  2 \left[
    \partial_{\eta}\left(
           \partial_{\eta}^{2}
      -    \Delta
    \right)
    -  2 \left(
      \frac{\partial_{\eta}^{2}\varphi}{\partial_{\eta}\varphi} - {\cal H}
    \right) \partial_{\eta}^{2}
  \right.
  \nonumber\\
  && \quad\quad\quad
  \left.
    +    \left( 
      2 \frac{(\partial_{\eta}^{2}\varphi)^{2}}{(\partial_{\eta}\varphi)^{2}}
      -  2 \frac{\partial_{\eta}^{3}\varphi}{\partial_{\eta}\varphi}
      +    \partial_{\eta}{\cal H}
      +    {\cal H}^{2}
      -  3 K
    \right) \partial_{\eta}
    -  2 {\cal H}  \frac{\partial_{\eta}^{2}\varphi}{\partial_{\eta}\varphi}
  \right] \stackrel{(2)}{{}^{(s)}E_{(4)}}
  \nonumber\\
  &&
  +  2 \left(
    \partial_{\eta} + {\cal H}
  \right) \stackrel{(2)}{{}^{(s)}E_{(1)}}
  \nonumber\\
  &&
  - 2 \left(
    \partial_{\eta} + {\cal H}
  \right) \Gamma_{0}
  -    {\cal H} \Gamma_{k}^{\;\;k}
  +    D^{k}\Gamma_{k}
  - 8 \pi G (\partial_{\eta}\varphi)^{3} \Xi_{(K)}
  \nonumber\\
  &&
  + \Delta^{-1}\left[
    \left(
      \partial_{\eta} - 2 {\cal H} 
    \right)
    D^{k}\left\{
           \partial_{\eta}\Gamma_{k}
      +  2 {\cal H} \Gamma_{k}
      -    D^{l}\Gamma_{lk}
    \right\}
  \right]
  ,
  \label{eq:kouchan-19.373}
\end{eqnarray}
where we have used
Eqs.~(\ref{eq:background-Einstein-equations-scalar-1}),
(\ref{eq:background-Einstein-equations-scalar-3}),
(\ref{eq:kouchan-19.331}), (\ref{eq:kouchan-19.333}),
and (\ref{eq:kouchan-19.330}). 
Equation (\ref{eq:kouchan-19.373}) shows that the
second-order perturbation of the Klein-Gordon equation is
consistent with the background, the second-order Einstein
equations if the last two lines in Eq.~(\ref{eq:kouchan-19.373})
vanish.
Further, since $\Gamma_{k}$ and $\Gamma_{ij}$ satisfy
Eq.~(\ref{eq:kouchan-19.358}), the last line in
Eq.~(\ref{eq:kouchan-19.373}) vanishes due to
Eq.~(\ref{eq:kouchan-19.358}).
Therefore, we may say that the second-order perturbation of the
Klein-Gordon equation is consistent with the background and the
second-order Einstein equations if the equation 
\begin{eqnarray}
  2 \left(
    \partial_{\eta} + {\cal H}
  \right) \Gamma_{0}
  -    D^{k}\Gamma_{k}
  +    {\cal H} \Gamma_{k}^{\;\;k}
  + 8 \pi G \partial_{\eta}\varphi \Xi_{(K)}
  = 0
  \label{eq:kouchan-19.374}
\end{eqnarray}
is satisfied under the background and first-order Einstein
equations. 
Actually, we can derive the relation
\begin{eqnarray}
  &&
  2 \left(
    \partial_{\eta} + {\cal H}
  \right) \Gamma_{0}
  -    D^{k}\Gamma_{k}
  +    {\cal H} \Gamma_{k}^{\;\;k}
  +  8 \pi G (\partial_{\eta}\varphi) \Xi_{(K)}
  \nonumber\\
  &=&
  4 \stackrel{(1)}{\Phi} \left[
        \stackrel{(1)}{\Phi} \partial_{\eta}
    + 2 \left(\partial_{\eta} + 2 {\cal H}\right)\stackrel{(1)}{\Phi}
  \right] \left(
    3 \stackrel{(0)}{{}^{(s)}E_{(1)}}
    - \stackrel{(0)}{{}^{(s)}E_{(2)}}
  \right)
  \nonumber\\
  && \quad\quad
  -  16 \pi G a^{2} \left[
    - 2 \partial_{\eta}\varphi \stackrel{(1)}{\Phi}
    +   \partial_{\eta}\varphi_{1}
  \right] \stackrel{(1)}{{\cal C}_{(K)}}
  \nonumber\\
  && \quad\quad
  - 8 \left[
    \stackrel{(1)}{\Phi} \left\{
          \partial_{\eta}^{2}
      -   \left(
        \frac{\partial_{\eta}^{2}\varphi}{\partial_{\eta}\varphi}
        + 4 {\cal H}
      \right) \partial_{\eta}
      +   \Delta
      + \frac{\partial_{\eta}^{2}\varphi}{\partial_{\eta}\varphi} \left(
        \frac{(\partial_{\eta}^{2}\varphi)}{(\partial_{\eta}\varphi)}
        - 4 {\cal H}
      \right)
      -   \frac{\partial_{\eta}^{3}\varphi}{\partial_{\eta}\varphi}
    \right\}
  \right.
  \nonumber\\
  && \quad\quad\quad\quad\quad
  \left.
    + 4 \partial_{\eta}\stackrel{(1)}{\Phi} \left\{
      \partial_{\eta}
      - \frac{\partial_{\eta}^{2}\varphi}{\partial_{\eta}\varphi}
    \right\}
  \right]
  \stackrel{(1)}{{}^{(s)}E_{(2)}}
  \nonumber\\
  && \quad\quad
  + 4 \left[
    \stackrel{(1)}{\Phi} \partial_{\eta}
    +  4 \left( \partial_{\eta} + {\cal H} \right) \stackrel{(1)}{\Phi}
  \right] \stackrel{(1)}{{}^{(s)}E_{(1)}}
  \nonumber\\
  &&
  + 4  \stackrel{(1)}{\chi^{ij}} D_{i}D_{j}\stackrel{(1)}{{}^{(s)}E_{(2)}}
  \nonumber\\
  &&
  - \stackrel{(1)}{\chi^{jl}} \partial_{\eta}\stackrel{(1)}{\chi_{lj}} \left(
    3 \stackrel{(0)}{{}^{(s)}E_{(1)}}
    - \stackrel{(0)}{{}^{(s)}E_{(2)}}
  \right)
  + \frac{1}{2} \left(\partial_{\eta} + 4 {\cal H}\right)\stackrel{(1)}{\chi^{lm}} \stackrel{(1)}{{}^{(s)}E_{(6)lm}}
  \nonumber\\
  &=&
  0,
  \label{eq:kouchan-19.374-2}
\end{eqnarray}
where we have used the background Einstein equation
(\ref{eq:background-Einstein-equations-scalar-4}), the
first-order perturbation (\ref{eq:kouchan-19.270}) of the
Klein-Gordon equation, the scalar-part of the first-order
perturbation of the momentum constraint
(\ref{eq:kouchan-19.263}), the evolution equation
(\ref{eq:kouchan-19.262}) of the scalar-mode in the first-order
perturbation of the Einstein equation, and the evolution
equation (\ref{eq:kouchan-19.267}) of the tensor-mode in the
first-order perturbation of the Einstein equation.


As shown in
\S\ref{sec:Fist-order-scalar-field-Klein-Gordon}, the
first-order perturbation of the Klein-Gordon equation is derived
from the background and the first-order perturbations
of the Einstein equation.
In the case of the second-order perturbation of the Klein-Gordon
equation (\ref{eq:second-Klein-Gordon-reduced}), we have derived
the relation (\ref{eq:kouchan-19.373}) from the background
Einstein equations
(\ref{eq:background-Einstein-equations-scalar-1}) and
(\ref{eq:background-Einstein-equations-scalar-2}), the
scalar-part of the second-order perturbation of the Einstein
equation (\ref{eq:kouchan-19.330}), (\ref{eq:kouchan-19.331}),
and (\ref{eq:kouchan-19.333}).
These equations include the source terms $\Gamma_{0}$,
$\Gamma_{i}$, $\Gamma_{ij}$, and $\Xi_{(K)}$ due to the
mode-coupling of the linear-order perturbations.
The equation (\ref{eq:kouchan-19.373}) gives the relation
(\ref{eq:kouchan-19.374}) between the source terms $\Gamma_{0}$, 
$\Gamma_{i}$, $\Gamma_{ij}$, $\Xi_{(K)}$ and we have also
confirmed that the equation (\ref{eq:kouchan-19.374}) is
satisfied due to the background, the first-order perturbation of
the Einstein equations, and the Klein-Gordon equation.
Thus, the second-order perturbation of the Klein-Gordon equation
is not independent equation of the Einstein equations if we
impose on each order perturbations of the Einstein equation at
any conformal time $\eta$.
This also implies that the derived formulae of the source terms
$\Gamma_{0}$, $\Gamma_{i}$, $\Gamma_{ij}$, and $\Xi_{(K)}$ are
consistent with each other.
In this sense, we may say that the formulae
(\ref{eq:kouchan-19.337})--(\ref{eq:kouchan-19.339}) and
(\ref{eq:kouchan-19.372-1}) for these source terms are correct.


\section{Summary and Discussion}
\label{sec:summary}


In summary, we derived the all components of the second-order
perturbation of the Einstein equation without ignoring any modes
of perturbation in the case of a perfect fluid and a scalar
field.
The derivation is based on the general framework of the
second-order gauge-invariant perturbation theory developed in
the paper KN2003\cite{kouchan-gauge-inv} and
KN2005\cite{kouchan-second}. 
In this formulation, any gauge fixing is not necessary and we
can obtain any equation in the gauge-invariant form which is
equivalent to the complete gauge fixing.
In other words, our formulation gives complete gauge fixed
equations without any gauge fixing.
Therefore, equations which are obtained in gauge-invariant
manner cannot be reduced without physical restrictions any more. 
In this sense, the equations shown here are irreducible.
This is one of the advantages of the gauge-invariant perturbation
theory.


The resulting Einstein equations of the second order shows that
any types of mode-coupling appears as the quadratic terms of the
linear-order perturbations due to the non-linear effect of the
Einstein equations, in principle.
Perturbations in cosmological situations are classified into
three types: scalar-; vector-; tensor-types.
In the second-order perturbations, we also have these three
types of perturbations as in the case of the first-order
perturbations.
Further, in the equations for the second-order perturbations,
there are many quadratic terms of linear-order perturbations due
to the nonlinear effects of the system.
Due to these nonlinear effects, the above three types of
perturbations couple with each other.


Actually, the source terms $\Gamma_{0}$, $\Gamma_{i}$,
$\Gamma_{ij}$, $\Xi_{0}$, and $\Xi_{i}$ defined by 
Eqs.~(\ref{eq:kouchan-19.117})--(\ref{eq:kouchan-19.119}),
(\ref{eq:kouchan-19.133}), and (\ref{eq:kouchan-17.682}) in the
perfect fluid case include all types of mode-coupling, i.e.,
the scalar-scalar; the scalar-vector; the scalar-tensor; the
vector-vector; the vector-tensor; the tensor-tensor types.
Since we concentrate only on the case of a single perfect fluid,
there is no anisotropic stress in the energy momentum tensor and
we have Eq.~(\ref{eq:kouchan-19.66}).
This equation is imposed in the definitions
(\ref{eq:kouchan-19.117})--(\ref{eq:kouchan-19.119}).
For this reason, the resulting Einstein equations are simpler
than those in the case of the fluid with anisotropic stress and
the source terms in
Eqs.~(\ref{eq:kouchan-19.117})--(\ref{eq:kouchan-19.119}) are
not generic form in this sense.
Even in this simple case,
Eqs.~(\ref{eq:kouchan-19.117})--(\ref{eq:kouchan-19.119}), 
(\ref{eq:kouchan-19.133}), and (\ref{eq:kouchan-17.682}) include
all types of mode-coupling.
Thus, we should keep in mind that all types of mode-coupling may 
occur in some situations. 
However, we may neglect vector- and tensor-modes of the linear
order in many realistic situations because these modes rapidly
decay due to the expansion of the universe.
If we take these behaviors of each mode into account, the
source terms
Eqs.~(\ref{eq:kouchan-19.117})--(\ref{eq:kouchan-19.119}),  
(\ref{eq:kouchan-19.133}), and (\ref{eq:kouchan-17.682}) become
simpler.


In the case of the single scalar field, the vector-mode of the
linear order vanishes due to the first-order perturbation of the
momentum constraint.
Further, we also have Eq.~(\ref{eq:kouchan-19.265}).
Due to these two facts, the source terms $\Gamma_{0}$,
$\Gamma_{i}$, $\Gamma_{ij}$, and $\Xi_{(K)}$, which are defined 
Eqs.~(\ref{eq:kouchan-19.337})--(\ref{eq:kouchan-19.339}) and
(\ref{eq:kouchan-19.372-1}), are simpler than those in the case
of a perfect fluid.
As a result, the source terms
(\ref{eq:kouchan-19.337})--(\ref{eq:kouchan-19.339}) shows the
mode-coupling of the scalar-scalar; the scalar-tensor; and the
tensor-tensor types.
Since the tensor-mode of the linear order is also generated due
to quantum fluctuations during the inflationary phase, the
mode-coupling of the scalar-tensor and the tensor-tensor types
may appear in the inflation.
If these mode-coupling occur during the inflationary phase,
these effects will depend on the scalar-tensor ratio $r$.
If so, there is a possibility that the accurate observations of
the second-order effects in the fluctuations of the scalar-type
in our universe also restrict the scalar-tensor ratio $r$ or
give some consistency relations between the other observations
such as the measurements of the B-mode of the polarization of
CMB.
This is a new effect which gives some information of the
scalar-tensor ratio $r$.


Further, we have also checked the consistency between the
second-order perturbations of the equation of motion of matter
field and the Einstein equations. 
In the case of a perfect fluid, we considered the consistency
between the second-order perturbations of the energy continuity
equation, the Euler equation, and the Einstein equations.
As a result, we obtain the consistency relation between the
source terms in these equations $\Gamma_{0}$, $\Gamma_{i}$,
$\Gamma_{ij}$, $\Xi_{0}$, and $\Xi_{i}$ which are given by
Eqs.~(\ref{eq:kouchan-19.135}) and (\ref{eq:kouchan-19.170})
with Eq.~(\ref{eq:kouchan-19.151}).
We also showed that these consistency relations between the
source terms are satisfied through the background and the
first-order perturbation of the Einstein equations.
This implies that the set of all equations are self-consistent
and the derived source terms $\Gamma_{0}$, $\Gamma_{i}$,
$\Gamma_{ij}$, $\Xi_{0}$, and $\Xi_{i}$ are correct.
We also note that these results are independent of the equation
of state of the perfect fluid.


In the case of a scalar field, we checked the consistency
between the second-order perturbations of the Klein-Gordon
equation and the Einstein equations.
As in the case of a perfect fluid, we have also obtained the
consistency relation between the source terms in these equations
$\Gamma_{0}$, $\Gamma_{i}$, $\Gamma_{ij}$, and $\Xi_{(K)}$ which
are given by Eqs.~(\ref{eq:kouchan-19.358}) and
(\ref{eq:kouchan-19.374}).
We note that the relation (\ref{eq:kouchan-19.358}) comes
from the initial value constraint in the Einstein equations of
the second order by itself, while the relation
(\ref{eq:kouchan-19.374}) comes from the second-order
perturbation of the Klein-Gordon equation.
We also showed that these relations between the
source terms are satisfied through the background and the
first-order perturbation of the Einstein equations.
This implies that the set of all equations are self-consistent
and the derived source terms $\Gamma_{0}$, $\Gamma_{i}$,
$\Gamma_{ij}$, and $\Xi_{(K)}$ are correct.
We also note that these relations are independent of the
details of the potential of the scalar field.


Thus, we have derived the self-consistent set of equations of
the second-order perturbation of the Einstein equations and the
evolution equation of matter fields in the cases of a perfect
fluid and a scalar field, respectively.
Therefore, in the case of the single matter field, we may say
that we have been ready to clarify the physical behaviors of the
second-order cosmological perturbations.
The physical behavior of the second-order perturbations in the
universe filled with a single matter field will be instructive
to clarify the physical behaviors of the second-order
cosmological perturbations in more realistic situations.
We leave these issues as future works.


\section*{Acknowledgements}
The author acknowledges participants of the international
workshop on ``11th Capra meeting'', which was held at CNRS in
France on June 2008, for valuable discussions in this workshop,
in particular, to Prof. S.~Detweiler, Prof. A.~Higuchi, and
Prof. D.~Galsov for valuable discussions and encouragements
during the workshop.
The author also thanks to Prof. J.~Nester for his continuous
encouragement to write this paper.
The author deeply thanks members of Division of Theoretical
Astronomy at NAOJ and my family for their continuous
encouragement.


\appendix
\section{Perturbations of energy momentum tensors}
\label{sec:Perturbations-of-energy-momentum-tensors}


Since we consider the perturbations of the Einstein equations,
we summarize the components of the perturbations of the energy
momentum tensor for a perfect fluid and a scalar field.
Though the ingredients of this section are already given in
KN2007\cite{kouchan-second-cosmo} and
KN2008\cite{kouchan-second-cosmo-matter}, we show again these
components to summarize the explicit definitions of the
perturbative variables which are necessary in the main text of
this paper.


\subsection{Perfect fluid}
\label{sec:Perfect-fluid-generic}


Here, we consider the perturbative expressions of the energy
momentum tensor for a perfect fluid.
The total energy momentum tenor of the fluid is characterized by
the energy density $\bar{\epsilon}$, the pressure $\bar{p}$, and
the four-velocity $\bar{u}^{a}$, and it is given by 
\begin{equation}
  {}^{(p)}\!\bar{T}_{a}^{\;\;b} 
  = (\bar{\epsilon} + \bar{p}) \bar{u}_{a} \bar{u}^{b} 
  + \bar{p} \delta_{a}^{\;\;b}.
  \label{eq:MFB-5.2-again}
\end{equation}
Since $\bar{\epsilon}$, $\bar{p}$, and $\bar{u}_{a}$ are
variables on the physical spacetime ${\cal M}$, these variables
are pulled back to the background spacetime through an
appropriate gauge choice ${\cal X}_{\lambda}$ to evaluate these
variables on the background spacetime.
We expand the fluid components $\bar{\epsilon}$, $\bar{p}$, and
$\bar{u}_{a}$ in Eq.~(\ref{eq:MFB-5.2-again}) as follows:
\begin{eqnarray}
  \bar{\epsilon}
  &:=&
  \epsilon 
  + \lambda \stackrel{(1)}{\epsilon}
  + \frac{1}{2} \lambda^{2} \stackrel{(2)}{\epsilon} 
  + O(\lambda^{3})
  \label{eq:energy-density-expansion}
  ; \\
  \bar{p}
  &:=&
  p
  + \lambda\stackrel{(1)}{p}
  + \frac{1}{2} \lambda^{2} \stackrel{(2)}{p}
  + O(\lambda^{3})
  \label{eq:pressure-expansion}
  ; \\
  \bar{u}_{a} 
  &:=& 
  u_{a}
  + \lambda \stackrel{(1)}{(u_{a})}
  + \frac{1}{2} \lambda^{2} \stackrel{(2)}{(u_{a})}
  + O(\lambda^{3})
  .
  \label{eq:four-velocity-expansion}
\end{eqnarray}
Following to Eqs.~(\ref{eq:matter-gauge-inv-decomp-1.0})
and (\ref{eq:matter-gauge-inv-decomp-2.0}), we define the
gauge-invariant variable for the perturbations of the fluid
components $\bar{\epsilon}$, $\bar{p}$, and $\bar{u_{a}}$: 
\begin{eqnarray}
  \label{eq:kouchan-16.13}
  &&
  \stackrel{(1)}{{\cal E}} 
  :=
  \stackrel{(1)}{\epsilon}
  - {\pounds}_{X}\epsilon;
  \quad
  \stackrel{(1)}{{\cal P}}
  :=
  \stackrel{(1)}{p}
  - {\pounds}_{X}p;
  \quad
  \stackrel{(1)}{{\cal U}_{a}}
  :=
  \stackrel{(1)}{(u_{a})}
  - {\pounds}_{X}u_{a}; \\
  &&
  \stackrel{(2)}{{\cal E}} 
  := \stackrel{(2)}{\epsilon} 
  - 2 {\pounds}_{X} \stackrel{(1)}{\epsilon}
  - \left\{
    {\pounds}_{Y}
    -{\pounds}_{X}^{2}
  \right\} \epsilon
  ; \quad
  \label{eq:kouchan-16.17}
  \stackrel{(2)}{{\cal P}}
  := \stackrel{(2)}{p}
  - 2 {\pounds}_{X} \stackrel{(1)}{p}
  - \left\{
    {\pounds}_{Y}
    -{\pounds}_{X}^{2}
  \right\} p
  ; \\
  &&
  \stackrel{(2)}{{\cal U}_{a}}
  :=
  \stackrel{(2)}{(u_{a})}
  - 2 {\pounds}_{X} \stackrel{(1)}{u_{a}}
  - \left\{
    {\pounds}_{Y}
    -{\pounds}_{X}^{2}
  \right\} u_{a}
  ,
  \label{eq:kouchan-16.18}
\end{eqnarray}
where the vector fields $X_{a}$ and $Y_{a}$ are the gauge-variant
parts of the first- and second-order metric perturbations,
respectively, and these vector fields are defined in
Eqs.~(\ref{eq:linear-metric-decomp}) and
(\ref{eq:second-metric-decomp}).


Components of the background value and the first- and the
second-order perturbations of the fluid four-velocity are
summarized as
\begin{eqnarray}
  \label{eq:kouchan-17.378}
  u_{a} &=& - a (d\eta)_{a}, \\
  \label{eq:kouchan-17.380}
  \stackrel{(1)}{{\cal U}_{a}}
  &=&
  - a \stackrel{(1)}{\Phi} (d\eta)_{a}
  + a \left(
    D_{i} \stackrel{(1)}{v} 
    + 
    \stackrel{(1)}{{\cal V}_{i}}
  \right) (dx^{i})_{a}
  , \\
  \label{eq:kouchan-17.399}
  \stackrel{(2)}{{\cal U}_{a}} 
  &=&
  \stackrel{(2)}{{\cal U}_{\eta}} (d\eta)_{a}
  + a \left(
    D_{i} \stackrel{(2)}{v} 
    + 
    \stackrel{(2)}{{\cal V}_{i}}
  \right) (dx^{i})_{a}
  ,
\end{eqnarray}
where
\begin{eqnarray}
  D^{i}\stackrel{(1)}{{\cal V}_{i}} &=& 0, \quad
  D^{i}\stackrel{(2)}{{\cal V}_{i}} = 0
  , \\
  \label{eq:kouchan-17.398}
  \stackrel{(2)}{{\cal U}_{\eta}} 
  &=&
  a \left\{
        \left(\stackrel{(1)}{\Phi}\right)^{2}
    -   \stackrel{(2)}{\Phi}
    - \left(
        D_{i}\stackrel{(1)}{v}
      + \stackrel{(1)}{{\cal V}_{i}}
      - \stackrel{(1)}{\nu_{i}}
    \right)
    \left(
        D^{i}\stackrel{(1)}{v}
      + \stackrel{(1)}{{\cal V}^{i}}
      - \stackrel{(1)}{\nu^{i}}
    \right)
  \right\}
  .
\end{eqnarray}
Here, we have used the normalization conditions of the
four-velocity 
$\bar{g}^{ab}\bar{u}_{a}\bar{u}_{a} = g^{ab}u_{a}u_{b} = -1$,
and its perturbations.


The perturbative expansion of the energy momentum tensor
(\ref{eq:MFB-5.2-again}) is given by 
\begin{eqnarray}
  {}^{(p)}\!\bar{T}_{a}^{\;\;b}
  &=:& 
  {}^{(p)}\!T_{a}^{\;\;b}
  + \lambda \stackrel{(1)}{{}^{(p)}\!T_{a}^{\;\;b}}
  + \frac{1}{2} \lambda^{2} \stackrel{(2)}{{}^{(p)}\!T_{a}^{\;\;b}}
  + O(\lambda^{3}).
  \label{eq:energy-momentum-tensor-expansion-perfect}
\end{eqnarray}
The background energy momentum tensor for a perfect fluid is
given by  
\begin{eqnarray}
  \label{eq:energy-momentum-perfect-fluid}
  T_{a}^{\;\;b}
  &=&
  \epsilon u_{a} u^{b}
  + p (\delta_{a}^{\;\;b} + u_{a} u^{b}) 
  \\
  &=&
  - \epsilon (d\eta)_{a} \left(\frac{\partial}{\partial\eta}\right)^{b}
  + p \gamma_{a}^{\;\;b},  
  \label{eq:energy-momentum-perfect-fluid-homogeneous}
\end{eqnarray}
where we have used
\begin{eqnarray}
  \delta_{a}^{\;\;b}
  = (d\eta)_{a}\left(\frac{\partial}{\partial\eta}\right)^{b}
  + \gamma_{a}^{\;\;b},
  \label{eq:energy-momentum-for-velocity-homogeneous}
\end{eqnarray}
and $\gamma_{ab}:=\gamma_{ij}(dx^{i})_{a}(dx^{j})_{b}$,
$\gamma_{a}^{\;\;b}:=\gamma_{i}^{\;\;j}(dx^{i})_{a}(\partial/\partial
x^{j})^{b}$.


The first- and the second-order perturbations
$\stackrel{(1)}{{}^{(p)}\!T_{a}^{\;\;b}}$ and
$\stackrel{(2)}{{}^{(p)}\!T_{a}^{\;\;b}}$ of the energy 
momentum tensor are also decomposed into the form as
Eqs.~(\ref{eq:matter-gauge-inv-decomp-1.0}) and
(\ref{eq:matter-gauge-inv-decomp-2.0}), respectively, i.e., 
\begin{eqnarray}
  \label{eq:first-energy-momentum-tensor-perfect-decomposed}
  \stackrel{(1)}{{}^{(p)}\!T_{a}^{\;\;b}}
  &=:&
  \stackrel{(1)}{{}^{(p)}\!{\cal T}_{a}^{\;\;b}}
  + {\pounds}_{X}{}^{(p)}\!T_{a}^{\;\;b}
  ,\\
  \label{eq:second-energy-momentum-tensor-perfect-decomposed}
  \stackrel{(2)}{{}^{(p)}\!T_{a}^{\;\;b}}
  &=:&
  \stackrel{(2)}{{}^{(p)}\!{\cal T}_{a}^{\;\;b}}
  + 2 {\pounds}_{X} \stackrel{(1)}{{}^{(p)}\!T_{a}^{\;\;b}}
  + \left\{
    {\pounds}_{Y}
    - {\pounds}_{X}^{2}
  \right\} {}^{(p)}\!T_{a}^{\;\;b}
  .
\end{eqnarray}
Here, the components of the gauge-invariant parts
$\stackrel{(1)}{{}^{(p)}\!{\cal T}_{a}^{\;\;b}}$ of the
first order are given by 
\begin{eqnarray}
  \stackrel{(1)}{{}^{(p)}\!{\cal T}_{\eta}^{\;\;\eta}}
  &=&
  - \stackrel{(1)}{{\cal E}}
  \label{eq:kouchan-19.22}
  , \\
  \stackrel{(1)}{{}^{(p)}\!{\cal T}_{\eta}^{\;\;i}}
  &=&
  - \left( \epsilon + p \right) \left(
      D^{i} \stackrel{(1)}{v} 
    + \stackrel{(1)}{{\cal V}^{i}}
    - \stackrel{(1)}{\nu^{i}}
  \right)
  \label{eq:kouchan-19.23}
  , \\
  \stackrel{(1)}{{}^{(p)}\!{\cal T}_{i}^{\;\;\eta}}
  &=&
  \left( \epsilon + p \right) \left(
      D_{i}\stackrel{(1)}{v} 
    + \stackrel{(1)}{{\cal V}_{i}}
  \right)
  \label{eq:kouchan-19.24}
  , \\
  \stackrel{(1)}{{}^{(p)}\!{\cal T}_{i}^{\;\;j}}
  &=&
  \stackrel{(1)}{{\cal P}} \delta_{i}^{\;\;j}
  \label{eq:kouchan-19.25}
  ,
\end{eqnarray}
and the components of the gauge-invariant part
$\stackrel{(2)}{{}^{(p)}\!{\cal T}_{a}^{\;\;b}}$ of the second
order are given by 
\begin{eqnarray}
  \stackrel{(2)}{{}^{(p)}\!{\cal T}_{\eta}^{\;\;\eta}}
  &=&
  -   \stackrel{(2)}{{\cal E}} 
  - 2 \left( \epsilon + p \right) \left(
      D_{i}\stackrel{(1)}{v} 
    + \stackrel{(1)}{{\cal V}_{i}}
  \right)
  \left(
      D^{i} \stackrel{(1)}{v} 
    + \stackrel{(1)}{{\cal V}^{i}}
    - \stackrel{(1)}{\nu^{i}}
  \right)
  \label{eq:kouchan-19.34}
  , \\
  \stackrel{(2)}{{}^{(p)}\!{\cal T}_{i}^{\;\;\eta}}
  &=&
    2 
  \left(
    \stackrel{(1)}{{\cal E}} + \stackrel{(1)}{{\cal P}} 
  \right) 
  \left(
      D_{i}\stackrel{(1)}{v}
    + \stackrel{(1)}{{\cal V}_{i}}
  \right)
  \nonumber\\
  && 
  + \left( \epsilon + p \right) \left(
      D_{i}\stackrel{(2)}{v}
    + \stackrel{(2)}{{\cal V}_{i}}
    - 2 \stackrel{(1)}{\Phi} D_{i}\stackrel{(1)}{v}
    - 2 \stackrel{(1)}{\Phi} \stackrel{(1)}{{\cal V}_{i}}
  \right)
  \label{eq:kouchan-19.36}
  , \\
  \stackrel{(2)}{{}^{(p)}\!{\cal T}_{\eta}^{\;\;i}}
  &=&
  - 2 \left( 
    \stackrel{(1)}{{\cal E}} + \stackrel{(1)}{{\cal P}} 
  \right) 
  \left(
      D^{i}\stackrel{(1)}{v}
    + \stackrel{(1)}{{\cal V}^{i}}
    - \stackrel{(1)}{\nu^{i}}
  \right)
  \nonumber\\
  && 
  + \left( \epsilon + p \right) \left\{
    -   D^{i}\stackrel{(2)}{v}
    -   \stackrel{(2)}{{\cal V}^{i}}
    +   \stackrel{(2)}{\nu^{i}}
    - 2 \stackrel{(1)}{\Phi} \left(
        D^{i}\stackrel{(1)}{v}
      + \stackrel{(1)}{{\cal V}^{i}}
    \right)
  \right.
  \nonumber\\
  && \quad\quad\quad\quad\quad
  \left.
    + 2 \left(
      - 2 \stackrel{(1)}{\Psi} \gamma^{ij}
      +   \stackrel{(1)}{\chi^{ij}}
    \right) 
    \left(
        D_{j}\stackrel{(1)}{v}
      + \stackrel{(1)}{{\cal V}_{j}}
      - \stackrel{(1)}{\nu_{j}}
    \right)
  \right\}
  \label{eq:kouchan-19.35}
  , \\
  \stackrel{(2)}{{}^{(p)}\!{\cal T}_{i}^{\;\;j}}
  &=&
    2 \left( \epsilon + p \right) \left(
      D_{i}\stackrel{(1)}{v}
    + \stackrel{(1)}{{\cal V}_{i}}
  \right)
  \left(
      D^{j}\stackrel{(1)}{v}
    + \stackrel{(1)}{{\cal V}^{j}}
    - \stackrel{(1)}{\nu^{j}}
  \right)
  +   \stackrel{(2)}{{\cal P}} \delta_{i}^{\;\;j}
  \label{eq:kouchan-19.37}
  .
\end{eqnarray}


\subsection{Scalar fluid}
\label{sec:scalar-field-generic}


Next, we summarize the gauge-invariant variables for the
perturbations of a scalar field.
The energy momentum tensor of the single scalar field
$\bar{\varphi}$ is given by 
\begin{eqnarray}
  \bar{T}_{a}^{\;\;b} = 
  \bar{g}^{bc} \bar{\nabla}_{a}\bar{\varphi} \bar{\nabla}_{c}\bar{\varphi} 
  - \frac{1}{2} \delta_{a}^{\;\;b}
  \left(
    \bar{g}^{cd} \bar{\nabla}_{c}\bar{\varphi}\bar{\nabla}_{d}\bar{\varphi}
    + 2 V(\bar{\varphi})
  \right),
  \label{eq:MFB-6.2-again}
\end{eqnarray}
where $V(\varphi)$ is the potential of the scalar field
$\varphi$.
Since we shall consider a homogeneous and isotropic universe with
small perturbations, the scalar field must also be approximately
homogeneous.
In this case, the scalar field $\bar{\varphi}$ can be expanded
as 
\begin{eqnarray}
  \bar{\varphi}
  =
  \varphi
  + \lambda \hat{\varphi}_{1}
  + \frac{1}{2} \lambda^{2} \hat{\varphi}_{2}
  + O(\lambda^{3}),
  \label{eq:scalar-field-expansion-second-order}
\end{eqnarray}
where $\varphi$ is the homogeneous function on the homogeneous
isotropic universe, i.e.,
\begin{eqnarray}
  \varphi = \varphi(\eta).
  \label{eq:kouchan-19.181}
\end{eqnarray}
The background energy momentum tensor for the scalar field on
the homogeneous isotropic universe is given by
\begin{eqnarray}
  T_{a}^{\;\;b}
  &=&
  \nabla_{a}\varphi\nabla^{b}\varphi -
  \frac{1}{2}\delta_{a}^{\;\;b}\left(\nabla_{c}\varphi\nabla^{c}\varphi +
    2V(\varphi)\right)
  \label{eq:energy-momentum-single-scalar}
  \\
  &=&
  -
  \left(
      \frac{1}{2a^{2}} (\partial_{\eta}\varphi)^{2}
    + V(\varphi)
  \right)
  (d\eta)_{a} \left(\frac{\partial}{\partial\eta}\right)^{b}
  +
  \left(
    \frac{1}{2a^{2}} (\partial_{\eta}\varphi)^{2}
    - V(\varphi)
  \right)
  \gamma_{a}^{\;\;b}.
  \label{eq:energy-momentum-single-scalar-homogeneous}
\end{eqnarray}
The energy momentum tensor (\ref{eq:MFB-6.2-again}) can be also
decomposed into the background, the first-order perturbation,
and the second-order perturbation:
\begin{eqnarray}
  \bar{T}_{a}^{\;\;b} = T_{a}^{\;\;b} 
  + \lambda {}^{(1)}\!\left(T_{a}^{\;\;b}\right)
  + \frac{1}{2} \lambda^{2} {}^{(2)}\!\left(T_{a}^{\;\;b}\right)
  + O(\lambda^{3}),
\end{eqnarray}
where ${}^{(1)}\!\left(T_{a}^{\;\;b}\right)$ is linear in matter
and metric perturbations $\hat{\varphi}_{1}$ and $h_{ab}$, and 
${}^{(2)}\!\left(T_{a}^{\;\;b}\right)$ includes the second-order
metric and matter perturbations $l_{ab}$ and $\hat{\varphi}_{2}$
and quadratic terms of the first-order perturbations
$\hat{\varphi}_{1}$ and $h_{ab}$.


As in the case of the perfect fluid, each order perturbations of
the scalar field $\varphi$ is decomposed into the
gauge-invariant part and gauge-variant part: 
\begin{eqnarray}
  \label{eq:varphi-1-def}
  \hat{\varphi}_{1} &=:& \varphi_{1} + {\pounds}_{X}\varphi, \\
  \label{eq:varphi-2-def}
  \hat{\varphi}_{2} &=:& \varphi_{2} 
  + 2 {\pounds}_{X}\hat{\varphi}_{1} 
  + \left( {\pounds}_{Y} - {\pounds}_{X}^{2} \right) \varphi, 
\end{eqnarray}
where $\varphi_{1}$ and $\varphi_{2}$ are the first-order and
the second-order gauge-invariant perturbation of the scalar
field.


The perturbed energy momentum tensor of each order also
decomposed into the gauge-invariant and gauge-variant parts as
(\ref{eq:matter-gauge-inv-decomp-1.0}) and
(\ref{eq:matter-gauge-inv-decomp-2.0}).
Through Eqs.~(\ref{eq:linear-metric-decomp}),
(\ref{eq:second-metric-decomp}), (\ref{eq:varphi-1-def}), and
(\ref{eq:varphi-2-def}), we can decompose the perturbations
${}^{(1)}\!\left(T_{a}^{\;\;b}\right)$ and
${}^{(2)}\!\left(T_{a}^{\;\;b}\right)$ of the energy momentum
tensor as
\begin{eqnarray}
  {}^{(1)}\!\left(T_{a}^{\;\;b}\right)
  &=:&
  {}^{(1)}\!{\cal T}_{a}^{\;\;b} + {\pounds}_{X}T_{a}^{\;\;b}
  \label{eq:first-order-energy-momentum-scalar-decomp}
  , \\
  {}^{(2)}\!\left(T_{a}^{\;\;b}\right)
  &=:&
  {}^{(2)}\!{\cal T}_{a}^{\;\;b}
  + 2 {\pounds}_{X}{}^{(1)}\!\left(T_{a}^{\;\;b}\right)
  + \left( {\pounds}_{Y} - {\pounds}_{X}^{2}\right) T_{a}^{\;\;b}.
  \label{eq:second-order-energy-momentum-scalar-decomp}
\end{eqnarray}


Further, through the components (\ref{eq:components-calHab}) of
the gauge-invariant part of the first-order metric perturbation,
the definition (\ref{eq:varphi-1-def}) of the first-order
perturbation of the scalar field, and the homogeneous condition
(\ref{eq:kouchan-19.181}) for the background field, the
components of the first-order perturbation of the
energy-momentum tensor of the scalar field are given by
\begin{eqnarray}
  {}^{(1)}\!{\cal T}_{\eta}^{\;\;\eta}
  &=& 
  -   \frac{1}{a^{2}} \left\{
      \partial_{\eta}\varphi \partial_{\eta}\varphi_{1} 
    - \stackrel{(1)}{\Phi} (\partial_{\eta}\varphi)^{2}
    + a^{2} \frac{\partial V}{\partial\varphi} \varphi_{1}
  \right\}
  \label{eq:kouchan-19.196}
  , \\
  {}^{(1)}\!{\cal T}_{\eta}^{\;\;i}
  &=&
  \frac{1}{a^{2}} \partial_{\eta}\varphi \left(
      D^{i}\varphi_{1}
    + \stackrel{(1)}{\nu^{i}} \partial_{\eta}\varphi
  \right)
  \label{eq:kouchan-19.197}
  , \\
  {}^{(1)}\!{\cal T}_{i}^{\;\;\eta}
  &=& 
  - \frac{1}{a^{2}} D_{i}\varphi_{1} \partial_{\eta}\varphi 
  \label{eq:kouchan-19.198}
  , \\
  {}^{(1)}\!{\cal T}_{i}^{\;\;j}
  &=& 
  \frac{1}{a^{2}} \gamma_{i}^{\;\;j}
  \left\{
        \partial_{\eta}\varphi \partial_{\eta}\varphi_{1}
    -   \stackrel{(1)}{\Phi} (\partial_{\eta}\varphi)^{2}
    -   a^{2} \frac{\partial V}{\partial\varphi} \varphi_{1}
  \right\}
  .
  \label{eq:kouchan-19.199}
\end{eqnarray}


Finally, we summarize the components of the gauge-invariant part
of the second-order perturbation of the energy momentum tensor
for a scalar field.
Through the components (\ref{eq:components-calHab}) and
(\ref{eq:components-calLab}) of the gauge-invariant parts of the
first- and the second-order metric perturbations, the
definitions (\ref{eq:varphi-1-def}) and (\ref{eq:varphi-2-def})
of the gauge-invariant variables for the first- and the
second-order perturbations of the scalar field, and the  
homogeneous background condition (\ref{eq:kouchan-19.181}), the
components of the second-order perturbation of the
energy-momentum tensor for a single scalar field are given by 
\begin{eqnarray}
  {}^{(2)}\!{\cal T}_{\eta}^{\;\;\eta}
  &=&
  -   \frac{1}{a^{2}} \left\{
        \partial_{\eta}\varphi \partial_{\eta}\varphi_{2}
    -   (\partial_{\eta}\varphi)^{2} \stackrel{(2)}{\Phi}
    +   a^{2} \varphi_{2}\frac{\partial V}{\partial\varphi}
    - 4 \partial_{\eta}\varphi \stackrel{(1)}{\Phi} \partial_{\eta}\varphi_{1}
    + 4 (\partial_{\eta}\varphi)^{2} (\stackrel{(1)}{\Phi})^{2}
  \right.
  \nonumber\\
  && \quad\quad\quad
  \left.
    -   (\partial_{\eta}\varphi)^{2} \stackrel{(1)}{\nu^{i}} \stackrel{(1)}{\nu_{i}}
    +   (\partial_{\eta}\varphi_{1})^{2}
    +   D_{i}\varphi_{1} D^{i}\varphi_{1} 
    +   a^{2} (\varphi_{1})^{2} \frac{\partial^{2}V}{\partial\varphi^{2}}
  \right\}
  \label{eq:kouchan-19.209}
  , \\
  {}^{(2)}\!{\cal T}_{i}^{\;\;\eta}
  &=&
  -   \frac{1}{a^{2}} \left\{
        \partial_{\eta}\varphi \left(
          D_{i}\varphi_{2}
      - 4 D_{i}\varphi_{1} \stackrel{(1)}{\Phi}
    \right)
    + 2 D_{i}\varphi_{1} \partial_{\eta}\varphi_{1}
  \right\}
  \label{eq:kouchan-19.211}
  , \\
  {}^{(2)}\!{\cal T}_{\eta}^{\;\;i}
  &=&
      \frac{1}{a^{2}} \left[
        \partial_{\eta}\varphi D^{i}\varphi_{2}
    + 2 \partial_{\eta}\varphi_{1} D^{i}\varphi_{1}
    + 2 \partial_{\eta}\varphi \left(
        2 \stackrel{(1)}{\nu^{i}} \partial_{\eta}\varphi_{1}
      + 2 \stackrel{(1)}{\Psi} D^{i}\varphi_{1}
      -   \stackrel{(1)}{\chi^{il}} D_{l}\varphi_{1}
    \right)
  \right.
  \nonumber\\
  && \quad\quad
  \left.
    +   (\partial_{\eta}\varphi)^{2} \left(
          \stackrel{(2)}{\nu^{i}}
      - 4 \stackrel{(1)}{\Phi} \stackrel{(1)}{\nu^{i}}
      + 4 \stackrel{(1)}{\Psi} \stackrel{(1)}{\nu^{i}}
      - 2 \stackrel{(1)}{\chi^{ik}} \stackrel{(1)}{\nu_{k}}
    \right)
  \right]
  \label{eq:kouchan-19.210}
  , \\
  {}^{(2)}\!{\cal T}_{i}^{\;\;j}
  &=&
  \frac{2}{a^{2}} \left[
       D_{i}\varphi_{1} D^{j}\varphi_{1}
    +  D_{i}\varphi_{1} \stackrel{(1)}{\nu^{j}} \partial_{\eta}\varphi 
  \right.
  \nonumber\\
  && \quad\quad
  \left.
    + \frac{1}{2} \gamma_{i}^{\;\;j}
    \left\{
      +   \partial_{\eta}\varphi \left(
            \partial_{\eta}\varphi_{2}
        - 4 \stackrel{(1)}{\Phi} \partial_{\eta}\varphi_{1} 
        - 2 \stackrel{(1)}{\nu_{l}} D^{l}\varphi_{1} 
      \right)
    \right.
  \right.
  \nonumber\\
  && \quad\quad\quad\quad\quad\quad
  \left.
    \left.
      +   (\nabla_{\eta}\varphi)^{2} \left(
          4 (\stackrel{(1)}{\Phi})^{2}
        -   \stackrel{(1)}{\nu^{l}} \stackrel{(1)}{\nu_{l}}
        -   \stackrel{(2)}{\Phi}
      \right)
      +   (\partial_{\eta}\varphi_{1})^{2}
      -   D_{l}\varphi_{1} D^{l}\varphi_{1} 
    \right.
  \right.
  \nonumber\\
  && \quad\quad\quad\quad\quad\quad
  \left.
    \left.
      -   a^{2} \varphi_{2} \frac{\partial V}{\partial\varphi}
      -   a^{2} (\varphi_{1})^{2} \frac{\partial^{2}V}{\partial\varphi^{2}}
    \right\}
  \right]
  \label{eq:kouchan-19.212}
  .
\end{eqnarray}




\begin{thebibliography}{99}
\bibitem{WMAP}
  C.L.~ Bennett et al., Astrophys. J. Suppl. Ser. {\bf 148},
  (2003), 1.
\bibitem{Non-Gaussianity-observation-WMAP}
  E.~Komatsu et al., preprint arXiv:0803.0547 [astro-ph], (2008).
\bibitem{Non-Gaussianity-inflation}
  V.~Acquaviva, N.~Bartolo, S.~Matarrese, and A.~Riotto,
  Nucl. Phys. B {\bf 667} (2003), 119; \\ 
  J.~Maldacena, JHEP, {\bf 0305} (2003), 013; \\
  K.~A.~Malik and D.~Wands, Class. Quantum Grav. {\bf 21} (2004), L65; \\
  N.~Bartolo, S.~Matarrese and A.~Riotto, Phys. Rev. D {\bf 69}
  (2004), 043503; \\
  N.~Bartolo, S.~Matarrese and A.~Riotto, JHEP {\bf 0404}
  (2004), 006; \\
  D.H.~Lyth and Y.~Rodr\'iguez, Phys. Rev. D {\bf 71} (2005), 123508; \\
  F.~Vernizzi, Phys. Rev. D {\bf 71} (2005), 061301R.
\bibitem{Non-Gaussianity-in-CMB}
  N.~Bartolo, S.~Matarrese and A.~Riotto, JCAP {\bf 0401} (2004), 003; \\
  N.~Bartolo, S.~Matarrese and A.~Riotto, Phys. Rev. Lett. 
  {\bf 93} (2004), 231301; \\  
  N.~Bartolo, E.~Komatsu, S.~Matarrese and A.~Riotto,
  Phys. Rept. {\bf 402} (2004), 103; \\
  N.~Bartolo, S.~Matarrese, and A.~Riotto, [arXiv:astro-ph/0512481].
\bibitem{kouchan-gauge-inv}
  K.~Nakamura, Prog.~Theor.~Phys. {\bf 110}, (2003), 723.
\bibitem{kouchan-second} 
  K.~Nakamura, Prog. Theor. Phys. {\bf 113} (2005), 481.
\bibitem{kouchan-second-cosmo} 
  K.~Nakamura, Phys. Rev. D {\bf 74} (2006), 101301(R);\\
  K.~Nakamura, Prog. Theor. Phys. {\bf 117} (2007), 17.
\bibitem{kouchan-second-cosmo-matter} 
  K.~Nakamura, preprint (arXiv:0804.3840 [gr-qc]).
\bibitem{Bardeen-1980}
  J.~M.~Bardeen, Phys.~Rev.\ D\ {\bf 22} (1980), 1882.
\bibitem{Kodama-Sasaki-1984}
  H.~Kodama and M.~Sasaki, Prog.~Theor.~Phys.~Suppl.\ No.78 (1984), 1.
\bibitem{Mukhanov-Feldman-Brandenberger-1992}
  V.~F.~Mukhanov, H.~A.~Feldman and R.~H.~Brandenberger,
  Phys.~Rep.\ {\bf 215} (1992), 203. 
\bibitem{Bel-Damour-Deruelle-Ibanez-Martine-1981}
  L.~Bel, T.~Damour, N.~Deruelle, J.~Ibanez, and J.~Martin,
  Gen. Rel. and Grav. {\bf 13}, 963, (1981).
\bibitem{Wald-book}
  R.M.~Wald, {\it General Relativity} (Chicago, IL: University of
  Chicago Press, 1984).
\end{thebibliography}
\end{document}